\documentclass[iop,twocolappendix,numberedappendix]{emulateapj}
\pdfoutput=1
\usepackage{epsfig}
\usepackage{hyperref}
\usepackage{amsmath}
\usepackage{amssymb}
\usepackage{natbib}
\usepackage{setspace}
\usepackage{graphicx}
  \usepackage[flushleft]{threeparttable}

\begin{document}
\defcitealias{meidt2018}{Paper~1}
\defcitealias{usero}{U15}
\defcitealias{veh16}{VEH16}
\title{A model for the onset of self-gravitation and star formation in molecular gas governed by galactic forces: II. the bottleneck to collapse set by cloud-environment decoupling} 
\author{Sharon E. Meidt\altaffilmark{1,2}}
\author{Simon C.~O.~Glover\altaffilmark{3}}
\author{J.~M.~Diederik Kruijssen\altaffilmark{4}}
\author{Adam K. Leroy\altaffilmark{5}}
\author{Erik Rosolowsky\altaffilmark{6}}
\author{Andreas Schruba\altaffilmark{7}}
\author{Annie Hughes\altaffilmark{8,9}}
\author{Eva Schinnerer\altaffilmark{1}}
\author{Antonio Usero\altaffilmark{10}}
\author{Frank Bigiel\altaffilmark{11}}
\author{Guillermo Blanc\altaffilmark{12,13,14}}
\author{M\'{e}lanie Chevance\altaffilmark{4}}
\author{Jerome Pety\altaffilmark{15}}
\author{Miguel Querejeta\altaffilmark{10,16}}
\author{Dyas Utomo\altaffilmark{5}}


\altaffiltext{1}{Max-Planck-Institut f\"ur Astronomie, K\"{o}nigstuhl 17, D-69117 Heidelberg, Germany}
\altaffiltext{2}{Sterrenkundig Observatorium, Universiteit Gent, Krijgslaan 281 S9, B-9000 Gent, Belgium}
\altaffiltext{3}{Institut f\"{u}r theoretische Astrophysik, Zentrum f\"{u}r Astronomie der Universit\"{a}t Heidelberg, Albert-Ueberle Str. 2, D-69120 Heidelberg, Germany}
\altaffiltext{4}{Astronomisches Rechen-Institut, Zentrum f\"{u}r Astronomie der Universit\"{a}t Heidelberg, M\"{o}nchhofstra\ss e 12-14, 69120 Heidelberg, Germany} 
\altaffiltext{5}{Department of Astronomy, The Ohio State University, 140 W. 18th Ave., Columbus, OH 43210, USA} 
\altaffiltext{6}{Department of Physics, University of Alberta, Edmonton, AB, Canada} 
\altaffiltext{7}{Max-Planck-Institut f\"ur extraterrestrische Physik, Giessenbachstra\ss e 1, 85748 Garching, Germany} 
\altaffiltext{8}{CNRS, IRAP, 9 av. du Colonel Roche, BP 44346, F-31028 Toulouse cedex 4, France}
\altaffiltext{9}{Universit\'{e} de Toulouse, UPS-OMP, IRAP, F-31028 Toulouse cedex 4, France}
\altaffiltext{10}{Observatorio Astron\'{o}mico Nacional - (IGN), Observatorio de Madrid Alfonso XII, 3, 28014 - Madrid, Spain}
\altaffiltext{11}{Argelander-Institut f\"{u}r Astronomie, Universit\"{a}t Bonn, Auf dem H\"{u}gel 71, 53121 Bonn, Germany}
\altaffiltext{12}{Departamento de Astronom\'{i}a, Universidad de Chile, Casilla 36-D, Santiago, Chile}
\altaffiltext{13}{Centro de Astrof\'{i}sica y Tecnolog\'{i}as Afines (CATA), Camino del Observatorio 1515, Las Condes, Santiago, Chile}
\altaffiltext{14}{Visiting Astronomer, Observatories of the Carnegie Insti- tution for Science, 813 Santa Barbara St, Pasadena, CA, 91101, USA}
\altaffiltext{15}{Institut de Radioastronomie Millim\'etrique, 300 Rue de la Piscine, F-38406 Saint Martin d'H\`eres, France}
\altaffiltext{16}{European Southern Observatory, Karl-Schwarzschild-Stra\ss e. 2, D-85748 Garching, Germany}

\date{\today}
\begin{abstract}
In \cite{meidt2018}, we showed that gas kinematics on the scale of individual molecular clouds are not dominated by self-gravity but also track a component that originates with orbital motion in the potential of the host galaxy. This agrees with observed cloud line widths, which show systematic variations from virial motions with environment, pointing at the influence of the galaxy potential. In this paper, we hypothesize that these motions act to slow down the collapse of gas and so help regulate star formation. Extending the results of \cite{meidt2018}, we derive a dynamical collapse timescale that approaches the free-fall time only once the gas has fully decoupled from the galactic potential.  Using this timescale we make predictions for how the fraction of free-falling, strongly self-gravitating gas varies throughout the disks of star-forming galaxies.  We also use this collapse timescale to predict variations in the molecular gas star formation efficiency, which is lowered from a maximum, feedback-regulated level in the presence of strong coupling to the galactic potential.  
Our model implies that gas can only decouple from the galaxy to collapse and efficiently form stars deep within clouds.  We show that this naturally explains the observed drop in star formation rate per unit gas mass in the Milky Way's CMZ and other galaxy centers. The model for a galactic bottleneck to star formation also agrees well with resolved observations of dense gas and star formation in galaxy disks and the properties of local clouds. \end{abstract}

\section{Introduction\label{sec:intro}}
\setcounter{footnote}{0}
Uncovering the conditions for the onset of star formation within molecular gas is one of the principal outstanding issues at the intersection of modern star formation theory and observation.   Where and how stars form (or do not form) is a key ingredient in models of galaxy formation and evolution that must be tightly constrained in order to properly calibrate models of (stellar and AGN) feedback and its impact on the cold gas reservoirs of galaxies over cosmic time.   

One of the cornerstones of star formation theory is the observed inefficiency of the process. It takes $2{-}3$ orders of magnitude longer than the free-fall time for the molecular gas in a typical galaxy to be consumed by star formation \citep{zuck}.  There are two concepts to explain this low global efficiency. In the first, the star-forming medium is organized into cold and dense, roughly virialized clouds, each of which forms stars but with an overall low efficiency \citep[i.e., ][]{elmegreen2002,krumholz05,krumholz12}.  
In the second, star formation is rare but intrinisically efficient, while only a small portion of the cold gas reservoir is ever in a state to undergo star formation \citep[i.e.,][]{padoan,fk12,hennebelle13,semenov17,semenov18}.  

In both cases, feedback from newly formed stars is thought to play a pivotal role either by restricting the conversion of gas into stars or by limiting the star-forming reservoir.  Many observational efforts to distinguish between them have therefore leveraged the subtle differences in gas properties predicted in the two scenarios \citep[c.f.,][]{hopkins13}.

Initially, surveys of extragalactic cloud populations, first from within the Local Group, indicated that clouds obey a well-defined size--linewidth relationship \citep[e.g.,][]{larson, solomon, bolatto} and thus appear approximately virialized \citep{fukuiKawamura2010}.  Based on these observations, clouds have been treated---much like virialized stellar clusters---as ballistic objects whose internal kinematics are largely decoupled from the large-scale motions of material orbiting in the host galaxy potential.\footnote{The internal velocity dispersion of clouds has been treated as a 
consequence of the collisions of clouds as they orbit the galaxy \citep{JO, gammie}, but the orbital motions in this scenario do not apply to the motions of material within the clouds.}  

In this context, stellar feedback provides a key source of internal motions that can maintain the near-equilibrium virial state, as it acts to replenish these motions in the face of rapid dissipation of turbulence (e.g., \citealt{zuck}). Other interpretations for the origin of the size--linewidth relation include pure incompressible or shock-dominated turbulence (\citealt{elmscalo}; \citealt{mckeeOst}). 

More recent studies have emphasized that virialized clouds and gravitationally collapsing clouds are hard to distinguish observationally \citep{vs08, ballesteros, ibanez}. These studies support a more dynamic view of the star-forming reservoir in which collapse is pervasive \citep{BurkertHartmann,elmegreencoll}. In this context, feedback once again plays a critical role in limiting the efficiency of star formation by acting across a range of spatial scales: it directly influences gas at the highest densities where star formation occurs \citep{hopkins13}, limits the evolution toward those high densities \citep{elmegreencoll}, and disperses clouds preventing further star formation \citep{semenov17, semenov18, kruijssen19b,chevance19,rahner19}.  

From this perspective, the dynamical state of the gas on different scales is a sensitive predictor of the onset of collapse and star formation \citep{dobbs11,padoan12,padoanRev,semenov17,semenov18}. Modern probes of the physical properties of molecular clouds across a diversity of galactic environments are beginning to reveal such a link \citep{meidt, leroy2017m51, colombo2018, utomo, schruba19}, starting with the observation that molecular gas in some environments is not always organized entirely into long-lived, virialized clouds \citep[e.g.,][]{hughesI, colombo2014a, meidt15, kruijssen19b,chevance19}. Deviations from approximate virialization consistently occur in environments with high shear, short orbital times, deep stellar potential wells but also low pressure environments \citep{kruijssen14,leroy2017m51, sun} and may be partially linked to local hydrostatic midplane pressure in the gas disk \citep[][J.~Sun et al., in prep.]{oka, rosow05, heyer, field, schruba19}.

Here we take the view that variations in the gas dynamical state arise as part of the dynamic nature of the star-forming gas reservoir.  
With our treatment of three-dimensional, cloud-scale gas motions in the first paper of this series (\citealt{meidt2018}; hereafter Paper I) the observed line widths and virial parameters of clouds are described as reflecting a combination of motions in the galactic potential and the cloud's self-gravitational potential. In this picture, departures from virialization indicate a systematic imbalance of gravitational energies on cloud scales, signifying weakly self-gravitating gas. The balance is altered on small scales at high density within the cloud interior, where self-gravity dominates and the gas decouples from the galactic potential.  

The coupling of gas motions to the galaxy potential in this framework resonates with the dynamically evolving molecular clouds in the high resolution, full disk simulations of \citet{dobbsPringle} that capture the thermal evolution of the gas and feedback from star formation down to the cloud scale.  These simulations show molecular gas passing smoothly between bound and unbound states, with the cloud boundary a constantly evolving surface that appears and disappears as clouds interact with their surroundings.   

The impact that this cycling has on the efficiency of star formation has been examined more explicitly in the cloud-scale simulations of \citet{semenov17,semenov18} and in the statistical formalism of \citealt{kruijsLong14} and \citealt{kruijssen18b}. In their framework, the cycling of gas between star-forming and non-star-forming states is set by the time gas spends in a star-forming state, which is limited either by feedback or by dynamics acting to disperse clouds \citep[also see][]{jeffreson18}.  
In this paper, we use the framework developed in \citetalias{meidt2018} to advance an additional regulatory mechanism and consider how cycling is affected by the time spent {\it before} gas reaches a star-forming state. The model we introduce describes a bottleneck to self-gravitation and collapse imposed by orbital motions in the galactic potential.

This paper is organized as follows.  We begin by summarizing our three-dimensional model for internal cloud motions that originate both with the self-gravity of the gas cloud and with orbital motions arranged by the `external' potential defined by the large-scale distribution of gas, stars and dark matter ($\S$~\ref{sec:themodel}).  Using this framework, we derive the dynamical timescale for collapse when self-gravity is opposed by the energy in galactic orbital motions ($\S$~\ref{sec:collapsetime}) and identify regimes in which collapse is either inhibited or progressing near the free-fall rate.  Then in $\S$~\ref{sec:SFmodel} we introduce a model for star formation proceeding at the rate given by the environmentally-dependent collapse timescale.  

We use empirical cloud and galaxy models (introduced in Paper~I) to explore how the properties of the host galaxy help regulating the onset of collapse ($\S$~\ref{sec:predictions}) and the efficiency with which gas is observed to form stars ($\S$~\ref{sec:Obscomparison}).  In order to highlight the degree to which cloud-scale variations in the star formation efficiency (SFE) reflect the bottleneck imposed by the decoupling of gas kinematics from galactic orbital motions, we assume a universal, dimensionless conversion efficiency that we calibrate from observations of local clouds in the MW disk ($\S$~\ref{sec:anchor}). 

We close by discussing in $\S$~\ref{sec:discussion} how the galactic bottleneck to star formation contributes to the observed, long molecular gas depletion times on large scales in galaxies.  In that section, we also discuss how gravitationally-induced, turbulent motions coupled with star formation feedback lead to a picture in which the galaxy participates in the regulation of star formation.  

Additional material to supplement the predictions of the model given in the main text is included in two Appendices.  To estimate the star formation rate (SFR) in gas with a given density distribution, in Appendix~\ref{sec:appendixdeltad}, we calculate the scaling factor that relates the integrated SFR of a cloud to the SFR estimated from properties measured on some scale $R_c$, which depends on the distribution of material in the cloud \citep[see][]{tan06,burk18}.  In Appendix~\ref{sec:appendix1}, we present a prediction for how the scale at which gravitational collapse and star formation occurs varies with galactic environment given the balance of gravitational energies in the gas.  In Appendix~\ref{sec:appSFEveldisp}, we present scale-dependent expressions for the link between SFE and gas velocity dispersion depending on the strength of self-gravity.   

\section{The model}
\label{sec:themodel}

\subsection{The dynamical coupling of clouds to their galactic environment}
In light of the recent observational challenges to the virialized `isolated cloud' view of molecular gas structure \citep{hughesI, colombo2014a, meidt, leroy2017m51, sun, schruba19}, in Paper~I we revisited the question of the coupling of clouds to their surroundings.  

The idea that the gas on cloud scales should be decoupled from its galactic environment is largely based on the expectation that regions smaller than the Toomre scale are able to collapse whereas regions larger than the Toomre scale are stabilized by rotation.  In this framework, rotation is assumed to be restricted to the disk plane, as its stabilizing influence applies on scales much larger than the disk thickness. 

With the framework adopted in Paper~I, however, we aim to describe the motions of gas embedded within the disk, where the disk itself is embedded within an external potential.\footnote{The influence of an external potential on disk stability as parameterized by the Toomre criterion has been examined by \cite{jog2014}.  Note that a number of other stability criteria have been introduced for multiple-component (star and gas) disks \citep[e.g.,][]{jogsolomon,rafikov,elmegreen2011}, mostly to quantify the destabilizing influence of gas on the combined system.  In many of these criteria, the influence of disk thickness \citep[][]{toomre,bertinromeo} and vertical motions in the gas \citep{romeo} are also included.}  We therefore adopt a picture in which the orbital motions framed by the background galactic gravitational potential are distributed within three dimensions.  These motions reflect the present distribution of orbital energies in the gas that is assumed to be constantly evolving, set by an initial accretion level and a history of dissipation, torques and advection (see e.g. \citealt{krumholz17}), and presumably also shaped by the energy injected by stellar feedback.  
 
From this perspective, we argued that internal cloud motions reflect galactic orbital motions (unlike in the limit of only tidal effects that largely applies to dense stellar clusters).  The kinematic response in this case (in the absence of self-gravity or other non-gravitational forces) is identical to basic epicyclic motions as material orbits the galaxy potential.

Our description includes both in-plane motions as well as vertical epicyclic motions. The latter describe orbits not entirely restricted to the mid-plane, which make an important contribution given the spatial extent of molecular clouds compared to the typical vertical height of galaxy disks.  In this scenario, only the non-intersecting orbits will be populated by gas, eventually helped to settle into the plane over long timescales (many orbital periods) by turbulent and collisional viscosity (e.g. \citealt{steiman}; \citealt{katz}).  Thus we envision that the epicyclic motions in the gas describe motions about non-intersecting non-circular orbits, such as those configured by well-defined bar and spiral arm patterns (though with the potential for overlap restricted to dynamical resonances, where orbit geometries are altered).  

Observations of ordered motions on cloud scales throughout molecular gas disks (P.~Lang et al., ApJ subm.) do indeed suggest that gas is populating non-intersecting orbits to lowest order, since pervasive shocking and viscous and gravitational torques would otherwise considerably rearrange the gas into a more centrally-concentrated distribution, and virial or collapse motions would be conspicuous.  More quantitatively, in Paper~I we found that observed velocity dispersions in excess of what is expected from virialized or self-gravitating clouds are consistent with a contribution from unresolved, ordered motions predicted by our model.    

The picture of 3D galactic motions hypothesized in Paper~I thus applies the same motions responsible for stabilizing gas on large scales, as described by the Toomre criterion, to the 3D kinematics of gas at and below the cloud scale.  Although greatly reduced in magnitude on the scales of GMCs, these 3D orbital motions remain large enough that they are comparable to the motions needed to support gas against its own self-gravity on cloud scales (see Figure~1).  

Our estimation of these motions is as follows.  As described in Paper~I, we are interested in accessing the contribution of coherent orbital motions to the internal motions of clouds through their observed velocity dispersions.  For the models of cloud structure examined in Paper~I, the density-weighted second moment of the velocity distribution across a cloud of size $R_{\rm c}$ in the plane and vertical extent $Z_{\rm c}$ yields 
 \begin{equation}
3\sigma_{\rm gal}^2\approx(\kappa R_{\rm c})^2+2(\Omega R_{\rm c})^2+(\nu Z_{\rm c})^2\, ,\label{eq:siggalnet}
\end{equation}
where $\sigma_{\rm gal}$ denotes the one-dimensional velocity dispersion associated with gas motions in the galactic potential on the scale of a cloud. We have ignored factors of order unity that account for the internal density distribution.
Here the frequency of vertical oscillations
\begin{equation}
\nu^2=\frac{\partial^2\Phi(z)}{\partial z^2}\approx2\pi G\Sigma_{\rm tot}z_0^{-1} \label{eq:nu}
\end{equation}
generally exceeds the frequency of radial oscillations in the plane within the main disk environment of typical nearby galaxies (except within galaxy centers; see Paper~I). The expression for the radial oscillations in polar coordinates is
\begin{eqnarray}
\kappa^2&=&\frac{\partial^2\Phi(R,\phi)}{\partial R^2}=4\Omega^2+R\frac{d\Omega^2}{dR}\\
&=&2\Omega^2(\beta+1)\, ,
\end{eqnarray}
with the logarithmic derivative of the rotation curve $\beta=\partial (\ln V_{\rm rot})/\partial(\ln R_{\rm gal})$ measuring rotation curve shear.  

As discussed in Paper~I, the dissipative and turbulent nature of gas is not explicitly incorporated into the model and is expected to lead to deviations from the purely gravitational kinematics described here.  
Shocks and dissipation and/or instabilities in the gas \citep[e.g.,][]{sbalbus, huber, wada, kim03, ko06, vs06, krumBurkert, krumkruij15, sormani17} will transform ordered, galaxy-driven motions into turbulent motions.  The formalism presented here provides an estimate of the magnitudes of these turbulent motions under the assumption that they are driven continuously from the orbital energy distribution of the gas framed by the background galaxy potential.   

\subsection{Decoupling from the galactic potential as a bottleneck to star formation}\label{sec:decoupling}
If some non-negligible part of the internal motions of clouds reflects motions in the host galaxy potential, as hypothesized in Paper~I, then the decoupling of molecular gas from the environment is potentially a key bottleneck for the process of star formation.  The idea that star formation in molecular gas is influenced by motions in the galactic potential has so far been most clearly inspired by observations of galaxy centers, where orbital times are short, tides are strong and circular velocities vary rapidly \citep[e.g.,][]{downes, kruijssen19}. The formalism presented in the previous section offers a description of this influence and extends it also to the normal disk environment.  

As will be described in more detail in the upcoming sections, in this paper we use our model of gas kinematics to relate the rate at which gas forms stars to the strength of its coupling to the galactic potential.  We do this by describing a smooth transition between two regimes, one in which the collapse of gas is regulated by self-gravity and one in which collapse is slowed (or even prevented) by motions in the galactic potential.  In the model, star formation sets in with a characteristic time set by the free-fall time of the gas only once the cloud decouples from its environment.  

As an element fundamental to the star formation process, the collapse in our model is meant to resemble the pervasive collapse envisioned by \citet{BurkertHartmann} and \citet{elmegreencoll}.  In our 3D framework, the galactic motions that are most influential on the largest scales have a slowing influence on collapse of gas clouds.  Thus the collapse that occurs in clouds in the present scenario is not exclusively free-fall and it sets in at densities that depend on location in the galaxy.  This leads to variations in the efficiency of star formation to levels lower than predicted in the case of free-fall collapse. 

At the basis of the description we introduce here is the idea that once gas decouples from the galactic gravitational potential it gains the ability to collapse.  We will call this collapse 'free-fall collapse' to distinguish it from the 'inhibited collapse' characteristic of weakly self-gravitating gas.  However, free-fall is far from guaranteed as a result of other (non-gravitational) factors that can still oppose collapse.  We chose a convention in which the action of these factors is parameterized by a star formation efficiency per free-fall time $\epsilon$ that is much less than unity, and use the free-fall time as the characteristic star formation timescale.  We find that this approach offers a straightforward way to isolate the role of galactic motions.  Later we will use this approach to investigate the degree to which the environmental variations in cloud-scale star formation efficiencies detected by observations and simulations \citep[e.g.,][]{dobbsPringle, leroy2017, utomo, schruba19,chevance19} can be attributed to the galactic bottleneck.  

\subsection{The relative strengths of gravitational potential energies within clouds}\label{sec:forcebalance}
Based on the model of 3D cloud-scale gas motions introduced in Paper~I and summarized above, we expect the galactic potential to induce differential motions across gas structures whose sizes are of the order of tens of parsecs. By contrasting these motions with those needed to support a cloud against its own self-gravity, we consider how the galactic potential slows the collapse of gas and thus the rate at which it forms stars.  

We express the decoupling of cloud material from the galactic potential (and the onset of strong self-gravitation) as a comparison of the strengths of the gravitational potential energies within clouds in three dimensions, 
\begin{equation}
\gamma^2=\frac{\Phi_{\rm c}}{\Phi_{\rm gal}}\, .\nonumber \\ 
\end{equation}
This can be expressed as a ratio of kinetic energies
\begin{equation}
\gamma^2=\frac{3\sigma_{\rm sg}^2 }{3 \sigma_{\rm gal}^2} \label{eq:gamma1}
\end{equation}
in terms of the kinematic response of the gas to the galactic potential, $\sigma_{\rm gal}$ in Eq.~\eqref{eq:siggalnet} as hypothesized in Paper~I, and using $\sigma_{\rm sg}$ to represent the one-dimensional cloud velocity dispersion on scale $R$ associated with self-gravity that we assume obeys 
\begin{equation}
\sigma_{\rm sg}=\sqrt{2\pi (a_k/5) G\Sigma R}\label{eq:sigSG}
\end{equation}
in the case of a spherical cloud with volume density profile $\rho\propto R^{-k}$ and surface density $\Sigma$ on scale $R$.  
Here, the geometric factor is
\begin{equation}
a_k=\frac{(1-k/3)}{(1-2k/5)}
\end{equation}  
following \cite{bertoldimckee}.  In the case of a homogeneous (uniform density) cloud, $a_k=1$, while isothermal clouds with $\rho\propto R^{-2}$ have $a_k=5/3$. 

For the model of gas motions constructed in Paper~I, the galactic motions $\sigma_{\rm gal}$ combine with the velocity dispersion $\sigma_{\rm sg}$, with $\sigma_{\rm sg}$ envisioned in two different ways.  In the first, motions represent the collapse response of the gas in energy equipartition (following \citealt{vs08}, \citealt{ballesteros} and \citealt{ibanez}), in which gas self-gravity is converted into kinetic energy during collapse.  In the second, $\sigma_{\rm sg}$ represents the turbulent velocity dispersion in the gas (arising from a variety of sources) assuming that the turbulent energy balances the gas self-gravity in an equilibrium scenario.  In this paper, $\sigma_{\rm sg}$ is used exclusively as a measure of gravo-turbulent collapse motions set by the strength of self-gravity.  

Likewise, in estimating the relative strengths of gravitational energies, we use $\sigma_{\rm gal}$ in the denominator of Eq.~\eqref{eq:gamma1}.  We further replace $\sigma_{\rm gal}$ by its equivalent expansion in the epicyclic approximation (Eq.~\eqref{eq:siggalnet}) assuming that clouds are approximately spherical.  This yields 
\begin{eqnarray}
\gamma^2&\approx&\frac{2\pi( a_k/5)G \Sigma_{\rm c}}{R_{\rm c}(\kappa^2 +2\Omega^2+\nu^2)/3}\nonumber \\
&\approx&\frac{3\sigma_{\rm sg}^2}{R_{\rm c}^2(\kappa^2+2\Omega^2 +\nu^2)}\, , \label{eq:balance}
\end{eqnarray}
where the subscript `c' denotes quantities defined on a scale $R_{\rm c}$ in the interior of the cloud.   

The flatness of galaxy disks, which yields non-isotropy in the background potential on cloud scales, leads to non-isotropic motions in the spherical cloud case that provide a good match to observed cloud velocity dispersions (Section 3.3, Paper~I).  We note, however, that galactic motions are roughly isotropic for highly flattened clouds so that 
\begin{equation}
\gamma^2\approx\left(\frac{\sigma_{\rm sg}}{\kappa R_{\rm c} }\right)^2\, . \label{eq:balanceOut2}
\end{equation}
This may also be relevant on the smallest scales within the deep interiors of clouds (independent of cloud geometry), where it approximates the limit in which the galactic potential is effectively isotropic.

In general, when $\gamma\lesssim1$ we expect 
galactic motions to make an increasingly important contribution to observed gas motions within the cloud.  
Clouds begin to decouple from the background potential when local rotation is matched to  internal motions due to self-gravity, i.e., $\gamma\sim1$.  For larger values of $\gamma$, the energy in (collapse) motions associated with gas self-gravity dominates over any other energy present until eventually the cloud fully decouples from the external potential and becomes strongly self-gravitating.  

\subsection{The timescale for collapse in the presence of galactic orbital motions}\label{sec:collapsetime}
In Paper~I, we presented evidence suggesting that the orbital motions of gas in the galactic potential constitute a source of motion across molecular clouds that is comparable in magnitude to (and in some environments, slightly larger than) motions associated with self-gravitating gas on the scales of individual clouds.  This suggests that almost everywhere throughout molecule-rich gas disks (within ${\sim}4 R_{\rm e}$) gas is not strongly self-gravitating on the scales of cloud envelopes but only starting to decouple from the background galactic potential.  As a consequence, in order for collapse to proceed efficiently and lead to star formation, the gas must overcome the non-negligible and continuous source of energy associated with orbital motions in the galactic potential. The required imbalance is possible only within the deep interiors of clouds, where the energy in galactic motions becomes negligible compared to the strengthening gas self-gravity at increasingly high densities.  This implies that only a small fraction of a cloud will actually collapse in a (local) free-fall time, with the remainder of the cloud undergoing much slower collapse.  This in turn acts to lower the efficiency with which gas forms stars.

We can estimate the collapse timescale in the presence of galactic motions assuming that the energy in these motions opposes gas self-gravity.  Our derivation adopts the equation of motion for a spherical shell at position $r$ in a uniform density cloud with potential $\Phi_{\rm c}$.  In the absence of any other sources of energy, this equation of motion can be integrated to yield the standard free-fall time $t_{\rm ff} = \sqrt{3\pi/(32 G \rho_{\rm c})}$ where $\rho_{\rm c}$ is the gas density.  The modification we introduce is to place the cloud at location $R_{\rm gal}$ in the  potential of the host galaxy so that
\begin{equation}
\frac{d^2 r}{dt^2}=-\nabla_r \Phi_{\rm c}(r) +\left[F_{\rm ng}(R_{\rm gal},r)+3\frac{\sigma_{\rm gal}^2(R_{\rm gal},r) }{r}\right]\, . \label{eq:collapseeqn}
\end{equation} 
The total opposing force in square brackets on the right is separated into two parts, the first due to non-gravitational factors (including feedback-driven turbulence and magnetic fields) and the second due to galactic motions, which are approximated using that $\Phi_{\rm gal} = 3\sigma_{\rm gal}^2/2$ with $3\sigma_{\rm gal}^2\propto r^2$ as given by Eq.~\eqref{eq:siggalnet} in the spherical cloud case ($Z_c$=$R_c$; see Paper~I) with the replacement $r=R_c$.  We use this to write the opposing force as $\nabla_r\Phi_{\rm gal}$=3$\sigma_{\rm gal}^2/r$.

For the present exercise, we will drop $F_{\rm ng}$ with the understanding that the collapse timescale we derive is a lower limit to the true collapse timescale.  Without a precise analytical model for how the non-gravitational factors acting within the gas should vary with location in a galaxy, we prefer to incorporate their influence at a later stage, which we assume results in a star formation efficiency per free-fall time that falls below unity.  

\begin{figure}[t]
\begin{center}
\begin{tabular}{c}
\includegraphics[width=0.85\linewidth]{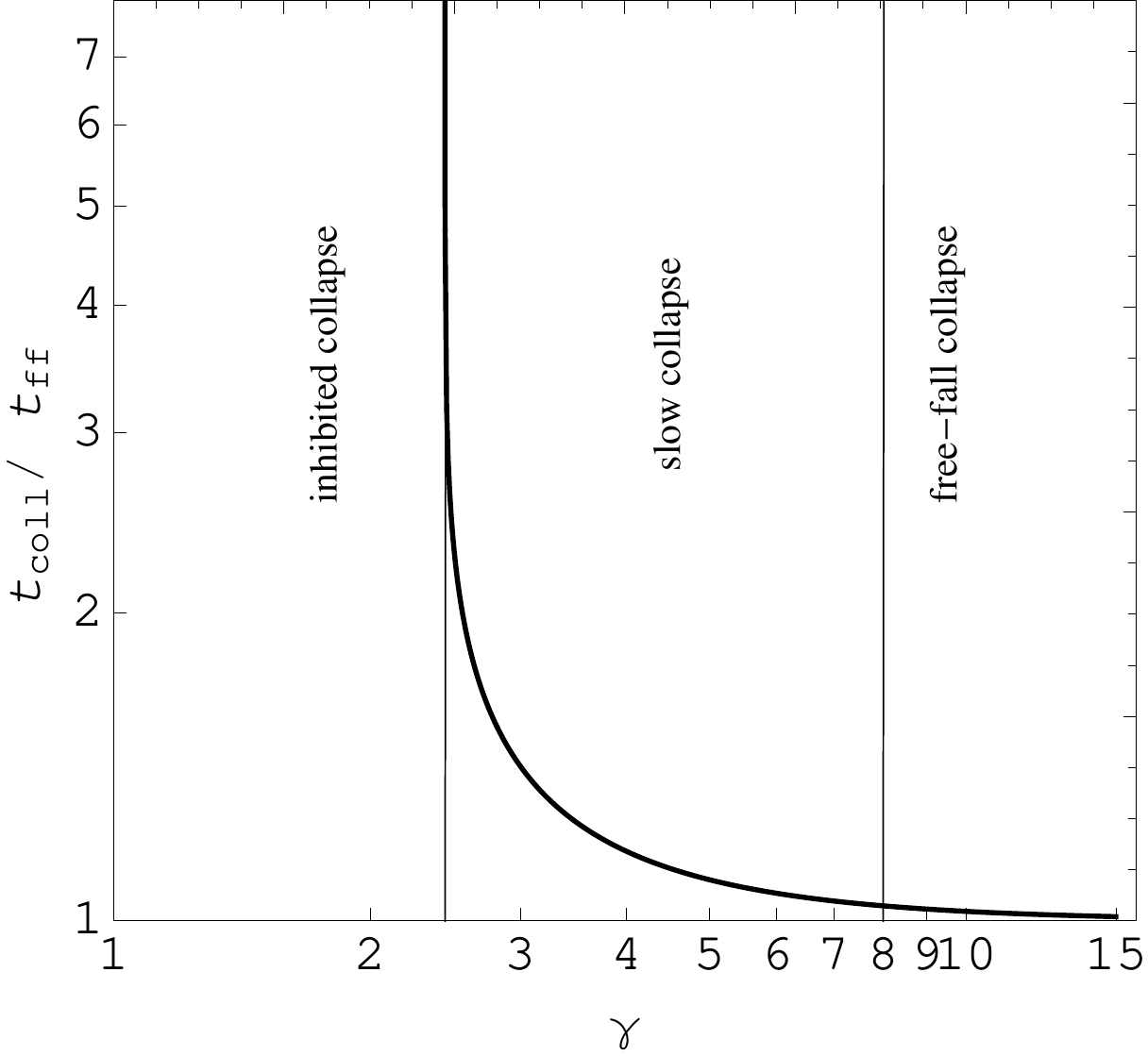}
\end{tabular}
\end{center}
\caption{The timescale for collapse when gas self-gravity is opposed by the energy in galactic orbital motions.  The ratio of the strength of self-gravity to the galactic potential is quantified by $\gamma$.  The collapse timescale is shown in units of the free-fall time $t_{\rm ff}$ for unopposed collapse.  
 }
\label{fig:tcoll}
\end{figure}

Figure~\ref{fig:tcoll} shows the behavior of the gravitational collapse timescale in units of the free-fall time determined through numerical integration of Eq.~\eqref{eq:collapseeqn} with $F_{\rm ng}$ set to zero.  In the limit of large $\gamma\gg1$ it can be shown that the time for the shell to reach $r=0$ is
\begin{equation}
t_{\rm coll}=t_{\rm ff} \left(1+\frac{11}{16}\frac{3}{\gamma_0^2}\right)\, ,
\label{eq:slowcoll}
\end{equation}
where $t_{\rm ff}$ is the free-fall time in the absence of all forces besides the force of self-gravity and $\gamma_0^2=\gamma^2/(a_k/5)$, given our definition of $\gamma$ in the case of non-uniform gas. This timescale approaches the free-fall time when the gas becomes more strongly self-gravitating (i.e., with $\gamma\gg1$).  

As $\gamma$ decreases, collapse slows considerably more until $\gamma_0\approx2.45$ is reached.  Beyond this point, the collapse timescale becomes infinite.  As we show in the next section, the behavior of $t_{\rm coll}$ depending on $\gamma$ leads to variations in the efficiency of star formation that depend on location in the galaxy. 

\subsection{A model for inefficient star formation}\label{sec:SFmodel}
As in most theories of star formation, we hypothesize that the timescale for gas to collapse sets the characteristic time for gas to convert a fixed fraction of its mass into stars, i.e.,
\begin{equation}
\dot{M}_{\rm stars}=\int\frac{\epsilon}{t_{\rm coll}}dM_{\rm gas}\, ,
\label{eq:sfmodel}
\end{equation}
where $\epsilon$ is the dimensionless star formation efficiency (discussed further below).  
In our convention the collapse timescale is explicitly the free-fall time only when $\gamma\gg1$, i.e., when the galactic potential is negligible compared to gas self-gravity.  
In this case, star formation occurs at the free-fall rate.  

For other circumstances, the rate of star formation is set by the slower collapse timescale ($t_{\rm coll}$) derived in the previous section.  The behavior of $t_{\rm coll}$ suggests that the star formation process regulated by galactic dynamics occurs in three smoothly-connected regimes corresponding to three characteristic stages of collapse:  
the onset of collapse, accelerated (but still slower than free-fall) collapse, and the transition to free-fall collapse. 

\subsubsection{The approach to free-fall collapse}
Gas that passes the threshold $\gamma_0=\gamma_{\rm coll}\sim 2.5$ undergoes collapse at a rate that is initially considerably slower than the free-fall rate for small $\gamma\gtrsim2.5$ but approaches the free-fall rate as $\gamma$ increases.  For large $\gamma\gg1$, the energy in galactic motions constitutes an increasingly negligible factor relative to self-gravity.  In this strongly self-gravitating regime, the efficiency of star formation per unit time $\mathit{SFE}$=$\dot{M}_{\rm stars}/M_{\rm gas}$ is written as
\begin{equation}
\mathit{SFE}_{\gamma >\gamma_{\rm coll}}\approx\frac{\epsilon}{t_{\rm ff}}\left(1+\frac{11}{16}\frac{3(a_k/5)}{\gamma^2}\right)^{-1}
\end{equation}
using the collapse timescale when $\gamma\gg\gamma_{\rm coll}$ in Eq.~\eqref{eq:slowcoll}.  In this regime, the star formation efficiency per free-fall time $\epsilon_{\rm ff} = \mathit{SFE}\, t_{\rm ff}$ is a weak function of $\gamma$.  Note that this approximation ignores the dependence of $t_{\rm coll}$ (and $t_{\rm ff}$ and $\gamma$) on density to pull it out of the integral in eq. (\ref{eq:sfmodel}) and takes $\epsilon$ to be approximately universal, yielding a lower bound on the SFE in non-uniform density gas \citep[see, e.g.,][and Appendix \ref{sec:appendixdeltad}]{tan06,burk18,parm}.  

\subsubsection{The onset of collapse}\label{sec:onsetofcollapse}
When galactic motions contribute significantly to internal cloud motions, so that $\gamma_0\leq2.5$, $t_{\rm coll}$ is infinite and collapse is inhibited.   Gas that falls below $\gamma_0\leq2.5$ on cloud scales is thus prevented from forming stars.  For star formation to occur, material must be present within the cloud with $\gamma_0>2.5$ so that collapse can commence.  Our estimate of the star formation rate therefore incorporates an additional factor related to the collapsing fraction $f_{\rm coll} = M_{\rm coll}/M_{\rm c}$ of a cloud with total mass $M_{\rm c}$, where $M_{\rm coll}$ measures the amount of material present within the cloud with $\gamma_0>\gamma_{\rm coll}$ and the ability to collapse to form stars.  Thus, we write the star formation efficiency of the cloud as 
\begin{equation}
\mathit{SFE}_{\gamma\leq\gamma{\rm coll}}\approx\frac{\epsilon}{t_{\rm coll,on}}\delta_{\rm d} f_{\rm coll}\, ,
\label{eq:weakSFE}
\end{equation}
where $t_{\rm coll,on}$ is the collapse timescale at the densities when collapse turns on and $\delta_{\rm d}$ is a factor related to the distribution of material within the cloud (e.g. \citealt{tan06}; \citealt{parm}).  For a power-law density distribution $\rho\propto r^{-k}$, for example, $\delta_d$=$2/3(3-k)(2-k)^{-1}$, assuming that $\epsilon$ is universal (see Appendix \ref{sec:appendixdeltad}).  This approximation slightly underestimates the SFE, since the collapse time $t_{\rm coll}$ has been taken as a fixed multiple of the free-fall time at all densities, whereas the derivation in Figure 1 suggests that $t_{\rm coll}/t_{\rm ff}$ continues to vary considerably in decoupled gas with $\gamma_0\gtrsim 2.5$.  The true SFE would sit between this level and the upper bound that assumes that collapsing material collapses at exactly the free-fall time, i.e. $\mathit{SFE}$=$\epsilon\delta_d f_{\rm coll} t_{\rm ff,coll}^{-1}$. 

In what follows we adopt $\gamma_{\rm coll,on}=2.5$ as the onset of collapse and use the timescale $t_{\rm coll,on} = 2.4 t_{\rm ff,coll}$ that is matched to this level (see Figure~\ref{fig:tcoll}), where $t_{\rm ff,coll}$ is the free-fall time specifically at the onset of collapse.  Given the rapid rise in $t_{\rm coll}$ as $\gamma$ decreases, this choice is arbitrary to the extent that a slightly smaller value of $\gamma$ would be associated with a considerably higher $t_{\rm coll}$.  On the other hand, this matched pair of $t_{\rm coll,on} = 2.4 t_{\rm ff,coll}$ and $\gamma_{\rm coll,on}=2.5$ yields star formation rates that are consistent (to within a factor of $1.5$) with pairs chosen up to the level $\gamma=3$.     

The collapse fraction $f_{\rm coll}$ is determined by the balance of gravitational energies $\gamma$ in the cloud.  
Writing Eq.~\eqref{eq:balance} in terms of  $\gamma_{\rm coll,on}=2.5$, we define the density required for collapse to proceed as 
\begin{equation}
\rho_{\rm coll}=\frac{(\kappa^2+2\Omega^2+\nu^2)\gamma_{{\rm coll}}^2}{2\pi (3 a_k/5)G}
\label{eq:volthresh}
\end{equation}
using our adopted model for gas self-gravity (see Eq.~\ref{eq:sigSG}).  Note that this is the {\it minimum} threshold for the onset of collapse, since other non-gravitational forces may be present that prevent the gas above this threshold from collapsing.  

The density threshold for collapse given by Eq.~\eqref{eq:volthresh} is largely insensitive to the properties of the gas but strongly dependent on the galactic potential.  In the main disk environment, it is essentially proportional to the mid-plane density $\rho_{\rm gal}$ of the host galaxy itself according to Poisson's equation, which we write as 
$4\pi G\rho_{\rm gal} \approx \kappa^2-2\Omega^2+\nu^2$ in our model.  

It should be emphasized that the scale associated with $\rho_{\rm coll}$, which we derive in Appendix~\ref{sec:appendix1}, is smaller than the Toomre scale that marks the size of the region that can be stabilized by rotation in the plane.  As argued earlier, the assumption of rotation restricted to the plane applies on scales larger than the disk scale height.  For the view of gas structure envisioned in this work, in which the gas is embedded within a disk that is itself embedded in an external potential, stability must be assessed in three dimensions.  In this case, there can be a component of orbital motion in the vertical direction that has an important stabilizing influence.  For the molecular gas situated in a galactic disk, the external galactic potential varies more rapidly in the vertical direction than in the plane, making the energy in the vertical orbital component dominant over radial epicyclic motions in the competition against self-gravity.  This limits the collapse to scales below the Toomre scale.  

In this regard, the picture of collapse described by our model is not quite as pervasive as envisioned by \citet{elmegreencoll} (or \citealt{BurkertHartmann}).  However, once collapse sets in, the view of star formation is the same: it occurs with no explicit threshold proceeding smoothly at the collapse rate.  

As we show in $\S$~\ref{sec:rhosgpred}, the collapse scale is typically located below the cloud scale and may thus give the appearance of a threshold for star formation.  However, in line with the above view of the star formation process, we prefer to describe this limit to collapse as a bottleneck.  Our accounting of star formation on cloud scales and larger is mostly a reflection of this bottleneck, as we show below.  

\subsubsection{Star formation in weakly self-gravitating gas coupled to the galactic potential}\label{sec:coupledSFE}
With the threshold density for gravitational collapse (Eq.~\eqref{eq:volthresh}), the $\mathit{SFE}$ becomes a strong function of the galactic potential.  Consider the case of a basic power-law density distribution $\rho\propto r^{-k}$, for which $f_{\rm coll} = (\rho_{\rm coll} / \rho_{\rm c})^{(k-3)/k}$.  This allows us to write the cloud-scale star formation efficiency in Eq.~\eqref{eq:weakSFE} to 
\begin{equation}
\mathit{SFE}_{\gamma\leq\gamma_{\rm coll}} \approx  \left(\frac{\gamma_{\rm coll}^{2(k-3)/k}}{t_{\rm coll,on}/t_{\rm ff,coll}}\right)\left(\frac{\delta_d\epsilon}{t_{\rm ff,coll}}\right)\gamma^{2(3-k)/k}\, .
\label{eq:genericSFEc}
\end{equation}
The first term in parentheses on the right is a constant factor that amounts to ${\sim}1/6$ when $k=2$ and ${\sim}1/15$ when $k=1.5$.  The second term in parentheses measures the star formation efficiency characteristic of the onset of collapse and is later referred to as $\delta_d\mathit{SFE}_{\rm coll}$.  

Using that $t_{\rm coll,on}=2.4 t_{\rm ff,coll}$,  $\rho_{\rm coll}/\rho_{\rm c} = (\gamma_{\rm coll}/\gamma)^{2}$ and $t_{\rm ff,coll}=\sqrt{3\pi/(32G\rho_{\rm coll})}$, we simplify this further to
\begin{equation}
\mathit{SFE}_{\gamma\leq\gamma_{\rm coll}} \approx  \left(\frac{\gamma_{\rm coll}^{3-6/k}}{2.4}\right)\left(\frac{\delta_d\epsilon}{t_{\rm ff}}\right)\gamma^{-3+6/k}\, . \label{eq:SFEweak}
\end{equation} 
From this we see that, even for a fixed internal density distribution, the efficiency per free-fall time in the regime of weakly self-gravitating gas varies strongly, with variations driven primarily by  $\gamma$, i.e. 
\begin{equation}
\epsilon_{\rm ff}\approx\epsilon \delta_d\left(\frac{\gamma_{\rm coll}^{3-6/k}}{2.4}\right)\gamma^{-3+6/k}.\label{eq:epweak}
\end{equation}

\subsubsection{The division between star-forming and non-star-forming gas}
The factor $\gamma$ that determines collapse is a ratio of timescales, namely the period of the local epicycle $t_{\rm epic} \sim 2\pi/\kappa$ divided by the local free-fall time $t_{\rm ff}$.  Our model of the bottleneck to collapse can thus be viewed as allowing star formation when the local free-fall time is considerably shorter than the circulation time for material in the gas as set by local galaxy dynamics.   The bottleneck model thus separates gas into star-forming and non-star-forming components.  

Note that the action of the bottleneck can resemble star formation limited by shear-regulated cloud dispersal.  However, the two pictures are conceptually different as the action of shear in the bottleneck model is responsible for \emph{preventing} star formation, rather than \emph{stopping} it (which is instead implicitly attributed to feedback).

Stellar dynamical bar and spiral arm features, which locally enhance $\rho_{\rm coll}$, may provide an opportunity to distinguish between these pictures as gas orbits the galaxy, since the fraction of non-star-forming gas should peak at the over-density where the bottleneck is narrowest.  In shear-limited star formation, the non-star-forming reservoir might be expected to become maximal downstream of the density maximum (where shear tends to be reduced), given the flow of gas back to an environment where shear is raised to the high background differential level.  The accounting of non-star-forming gas recently introduced by \citet{piechart} has the potential to make this distinction.  

\subsection{An inverse relation between SFE and gas velocity dispersion on cloud scales}\label{sec:SFEvdisp}
A distinguishing feature of the bottleneck model is a dependence of the $\mathit{SFE}$ on the three-dimensional motions in the gas, not just shear in the plane.  On large scales, the observed gas kinematics are expected to be dominated by galactic motions $\sigma_{\rm gal} \approx R_{\rm c}(2\kappa^2+\nu^2)^{-1/2}$ (in the flat part of the rotation curve), which increase linearly with spatial scale $R_{\rm c}$  while the gas self-gravity falls off away from the cloud center.  The expressions presented in the previous section therefore notably encode, for a given $\epsilon$, an inverse relation between the $\mathit{SFE}$ and the gas velocity dispersion in weakly self-gravitating gas, or a dependence on the  boundedness of the gas as measured by the virial parameter $\alpha_{\rm vir}=5\sigma^2R/(GM)$ \citep[][see Appendix~\ref{sec:appSFEveldisp}]{bertoldimckee}.   

The predicted behavior distinguishes our model from most other theories of star formation applied on the cloud scale (e.g., \citealt{krumholz05, ostrikerShetty, fk12}; but see \citealt{padoan12, padoan17, burkmocz} and the discussion in $\S$~\ref{sec:SFEvdisppred}).  In many of these theories, high Mach numbers, which lead to compressive shocks and the build-up of high density material, raise $\epsilon$ and thus the $\mathit{SFE} = \epsilon/t_{\rm ff}$.  Assuming that high Mach numbers emerge through the turbulent cascade  from elevated cloud-scale velocity dispersions (raised by the contribution from galactic motions), star formation might be expected to become more efficient with increasing velocity dispersion rather than less, as observed (see $\S$~\ref{sec:SFEvdisppred}).  In our model, motions on large scales that keep the envelopes of clouds coupled to the galaxy reduce the star formation efficiency by limiting the fraction of star-forming material.  

\section{Quantitative predictions of the galactic bottleneck using semi-empirical cloud and galaxy models}\label{sec:predictions}
\begin{figure}[t]
\begin{center}
\begin{tabular}{c}
\includegraphics[width=0.985\linewidth]{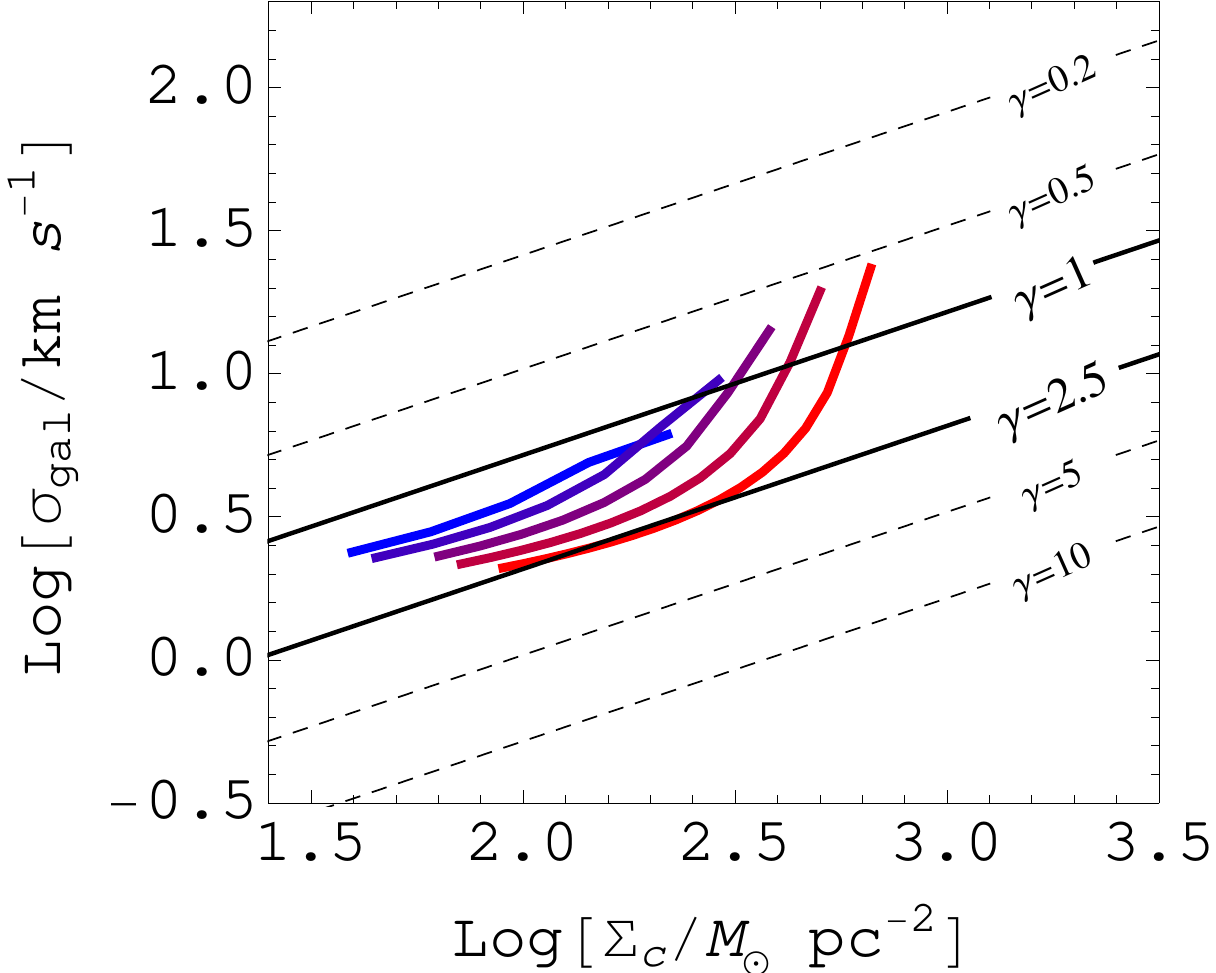}
\end{tabular}
\end{center}
\caption{The magnitude of gas motions on cloud scales $\sigma_{\rm gal}$ as function of cloud-scale surface density $\Sigma_{\rm c}$ for clouds with radius $R_{\rm c}=30$~pc. The thick dotted black line highlights when motions balance self-gravity so that $\gamma$=1 and $\sigma_{\rm  gal}$ = $\sigma_{\rm sg}$ as given by  Eq.~\eqref{eq:sigSG}. The thick solid black line shows when $\gamma$=2.5 and galactic motions can prevent collapse. Thinner dashed lines highlight offsets from these trends, for different levels of $\gamma$ in Eq.~\eqref{eq:balance} assuming that clouds are spherical.  Colored curves illustrate the contribution of cloud-scale orbital motions in the galactic potential to the internal motions of clouds in five model disk galaxies.  The models assume a rotation curve and cloud surface density distribution based on the galaxy mass, which varies here from $9.25~{\rm (blue)} < \log M/{\rm M}_\odot < 10.75$ (red) in steps of $0.25 \log M/{\rm M}_\odot$. Motions associated with the galactic potential have the same order of magnitude as those due to self-gravity throughout all disks and become very strong at galaxy centers, which are characterized by the high $\Sigma_{\rm c}$ and high $\sigma_{\rm gal}$ end of each curve.  }
\label{fig:paramspace}
\end{figure}
In this section we use the formalism introduced in the previous section to investigate how galactic orbital motions in the gas introduce environmental variations in the onset of collapse within molecular gas.  

\subsection{A model of gas self-gravity and orbital motions throughout galaxies}\label{sec:globalgalaxymodels}
The cloud-scale measurements needed to examine how several of the parameters in our model of star formation are related and how they vary together throughout the disks of real star-forming galaxies, are currently being assembled by the PHANGS collaboration (A.~K.~Leroy et al., in prep.; see also \citealt{gallagher}; \citealt{sun}; \citealt{utomo}; \citealt{piechart}; \citealt{chevance19}; P.~Lang et al., ApJ subm.; E.~Rosolowsky et al., in prep.).  In the near future, it will be possible to place fundamental constraints on how gas structure and kinematics are organized across spatial scales and how this impacts star formation across the local galaxy population.

In the meantime, to capture how the strength of gas self-gravity varies in relation to the galactic potential throughout real galaxies, in this section we introduce generic `global galaxy models' that build on the typical structure and dynamical properties of galaxies as well as the characteristic distributions and properties of their molecular disks. We adopt the same physically-motivated models for galactocentric rotation and cloud-scale surface density as used in Paper~I. In brief, at a given galaxy stellar mass, empirical scaling relations specify the shape and maximum of the rotation curve, the mass in molecular gas, and the shape of its mass distribution (its variation in the plane and in the vertical direction).  The distribution of the gas on cloud scales is assigned by assuming an exponential distribution of gas across the disk and a representative cloud size (e.g., set to a fixed value at all locations in the disk or varying in the case of a fixed cloud mass).  The disk's molecular hydrogen gas surface density $\Sigma_{\rm H_2}$ is then increased by a clumping factor $c$ to generate a model for the cloud surface density $\Sigma_{\rm c} = c\Sigma_{\rm H_2}$ at all locations.  For our fiducial $R_{\rm c}=30$~pc case, we adopt $c=2$ chosen to match observations on $60$~pc scales \citep{leroy2016}.

 Altogether, our empirically-motivated `global galaxy models' provide a prediction for the radial variation in $\sigma_{\rm gal}$ and $\Sigma_{\rm c}$ at fixed spatial scale throughout a given galaxy disk.  Models are typically extended out to the observed edge of molecule-bright emission near $2.5R_{\rm e}$ given the typical scale length of the molecular disk $R_{\rm e} \approx 0.2R_{25}$ in nearby galaxies (\citealt{schruba}; in terms of the customary isophotal radius $R_{25}$ at which the stellar surface brightness reaches $25$ mag arcsec$^{-2}$ in the $B$-band).  Typically, half of the total CO flux tracing molecular hydrogen is enclosed within a radius $R^{\rm CO}_{50} \approx 1.5 R_{\rm e}$ (very near the transition from HI to H$_2$) and $90$\% is within $R^{\rm CO}_{90} \approx 4 R_{\rm e}$ \citep{schruba}.  

\subsubsection{Gravitational motions on cloud scales in global galaxy models}\label{sec:motions}
The set of colored curves in Figure~\ref{fig:paramspace} highlight the region of parameter space occupied by molecular clouds in our empirically-motivated `global galaxy models' (Appendix~A) at a fiducial scale of $R_{\rm c} = 30$~pc, typical of the measured sizes of clouds in the MW \citep{heyer, mv17} and in external galaxies \citep[e.g.,][]{bolatto, hughesI, leroy2016, schruba19}.  In these models, the galactic potential falls off more rapidly than gas self-gravity in the main disk environment, yielding a characteristic increase in the contribution of orbital motions to the internal cloud motions toward galaxy centers (and toward the far outer, atomic-dominated disk beyond $R_{\rm gal} = 8R_{\rm e}$; not shown in Figure~\ref{fig:paramspace}), as discussed at greater length in $\S$~\ref{sec:massfrac}. 

The location of models in this parameter space implies that the energies associated with cloud self-gravity and the local galactic potential are comparable on average at the scale of typical molecular clouds where $\gamma\approx1{-}2$.  This implies that gas begins to decouple from the galactic potential and becomes weakly self-gravitating on the cloud scale. It thus also suggests that much higher densities within the cloud are required for gas self-gravity to become effectively unopposed so that collapse can proceed at the free-fall rate, as explored later.  In this context, the ratio of bound, self-gravitating gas to unbound, molecular material can be expected to vary strongly with galactic environment.  

We note that the precise balance of gravitational energies in any given cloud depends on the properties of that cloud and its galactic environment (discussed in greater detail in Paper~I).  Systematic variation in cloud sizes throughout a global cloud population (as we consider in Appendix~\ref{sec:appendix1}) alter the average cloud-scale proportion of $\sigma_{\rm sg}$ relative to $\sigma_{\rm gal}$ from the level indicated in Figure~\ref{fig:paramspace}.  Including the change to orbital motions characteristic in the presence of bars and spiral arm perturbations, $\sigma_{\rm gal}$ can also be locally enhanced (see Paper~I).  Compared to the basic axisymmetric disk models portrayed in Figure~\ref{fig:paramspace}, we therefore expect observations to show a greater degree of variety.   

\subsubsection{A model of internal cloud structure}
\label{sec:internalstruc}
In the previous section, we demonstrated that the motions due to self-gravity and the galactic potential are comparable on the scales of molecular clouds throughout the disks of typical galaxies, preventing gas from becoming strongly self-gravitating on the cloud-scale.  We can expect this balance to change in the cloud interior, where increasingly high densities provide the opportunity for gas self-gravity to overcome the energy in galactic motions. 

\begin{figure}[t]
\includegraphics[width=0.965\linewidth]{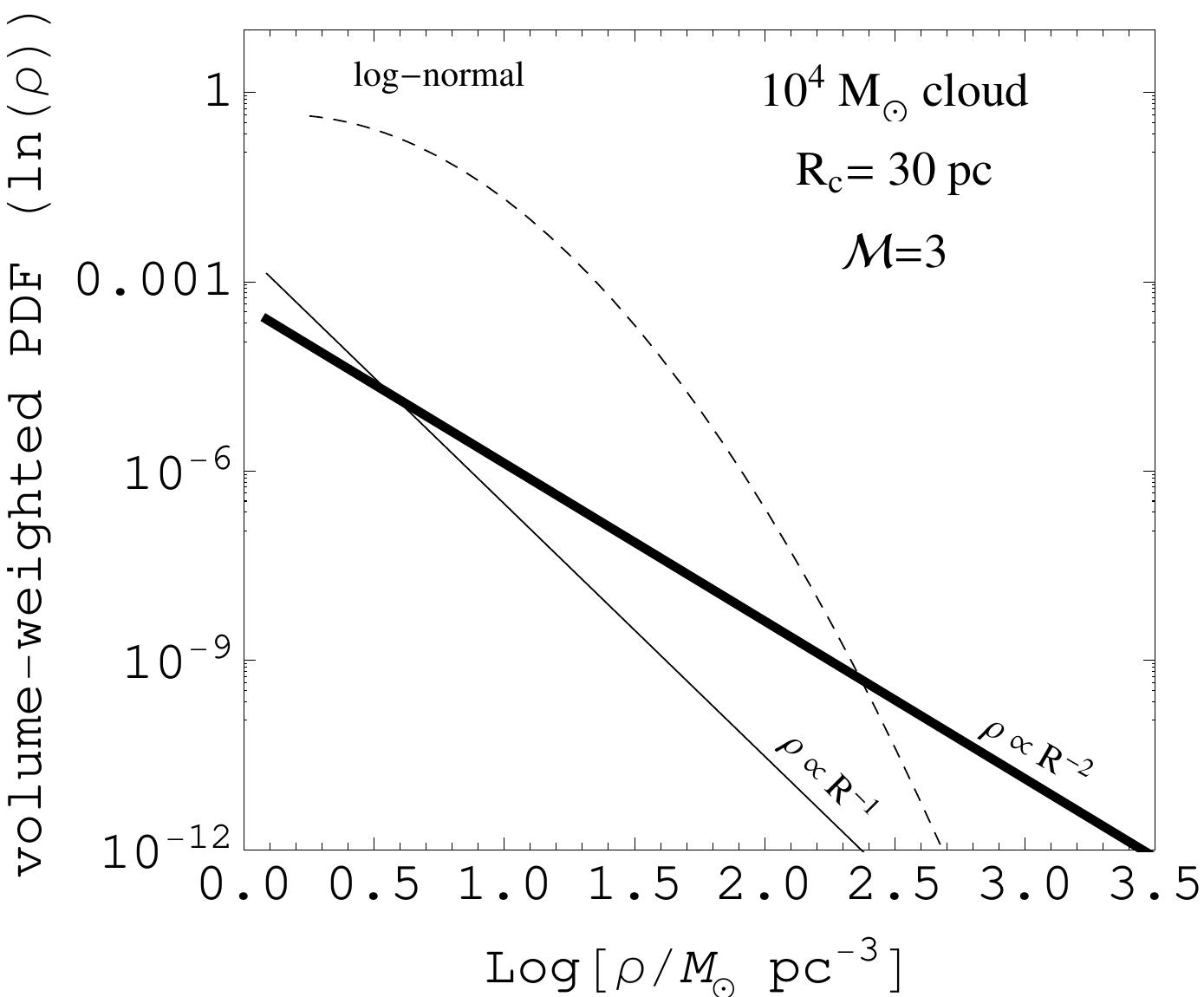}
\caption{Example volume-weighted log-density PDFs for a $10^4~ {\rm M}_\odot$ cloud with radius $R_{\rm c}=30$~pc.  The thick solid line shows the PDF for our nominal case of a power-law density distribution $\rho\propto R^{-2}$.  The thin solid line assumes a shallower density profile $\rho\propto R^{-1}$ associated with lower dense gas fractions.  The dashed line represents a log-normal PDF with density variance set by the characteristics of driven turbulence, assuming solenoidal driving \citep{federrath10} and a conservative Mach number $\mathcal{M}=3$. \vspace*{0.35cm} }
\label{fig:PDFs}
\end{figure}
To make concrete predictions for the scales and densities at which gas decouples from galactic orbital motions, we must make an assumption about the way material is distributed within clouds.  One option would be a log-normal density PDF (see Figure~\ref{fig:PDFs}), which describes the density structure that develops in the presence of isothermal MHD turbulence \citep[e.g.][]{vazSem94, padoan97, scalo98, klessen, ostriker01, vsg01}.  

Since our model is meant to apply to molecular clouds in normal star-forming disk galaxies, we adopt a power-law density distribution, which arguably best captures the distribution of material throughout the bulk of clouds with observationally reconstructed PDFs, from high density cores to the gas at the cloud edge, near the H$_2$--HI transition \citep{lombardi}.  We further assume that the density follows $\rho\propto r^{-k}$ and the material is arranged (spherically) symmetrically across most of the cloud.  This should provide a reasonable description over the range of densities that we are interested in here, as the gas motions that would be expected from this density distribution approximately reproduce the observed velocity dispersions of clouds \citep[e.g.,][see Eq.~\eqref{eq:sigSG}]{heyer}. 
As we use it later in $\S$~\ref{sec:massfrac}, this power-law model offers a straightforward analytical connection between densities at large and small scales within the cloud.  

In order to highlight how dense gas fractions in equal mass clouds are impacted by the shape of the density distribution, we choose to normalize densities to the value at the cloud edge rather than the volume-weighted mean density.  Thus, we note that clouds with power-law density PDFs will tend to have fractionally more mass at high densities than equivalent-mass clouds with log-normal PDFs and modestly transonic Mach numbers (see Figure~\ref{fig:PDFs}).  Steeper power-law profiles also contain moderately more high density material compared to flatter power-law profiles.  

Observations of molecular gas in local samples are consistent with power-law density PDFs with a range of power-law slopes ($k=1{-}2$; see \citealt{mclaren, abreu, lombardi, meidt16}).  For the sake of generality, we cast most of our model predictions in terms of the generic power-law profile $\rho\propto r^{-k}$.  However, for making direct comparisons to observational results, we select $k=2$ and $k=1$ as our nominal density profiles (unless noted otherwise).  This choice is not meant to favor a particular origin scenario (i.e., pressure equilibrium vs.\ dynamical collapse, e.g., \citealt{larson69, whitworth, foster, naranjo, li}), but is chosen for consistency with the observational results that we use for comparison.  Existing observations suggest that either density profile is a reasonable choice. In the Galaxy, probes of cloud material at high and low density appear consistent with $k=1{-}1.5$, as we find in $\S$ \ref{sec:MWSFR}.  On the other hand, a $k=2$ density profile is compatible with the local clouds analyzed by \cite{lombardi}.  The observed trend in extragalactic dense gas fractions $f_{\rm d}$ with increasing molecular gas surface density are also in agreement with a $k=2$ density profile 
(\citealt{usero}, henceforth \citetalias{usero}; \citealt{gallagher}), since shallower density profiles would lead to a steeper increase in $f_{\rm d}$ with gas surface density than what is observed \citep[][and see Eq.~\eqref{eq:eqRatiosGen} below]{meidt16}. Nevertheless, we caution that the steep $k=2$ profile inferred from the \citetalias{usero} results may only apply to massive, high surface density clouds that dominate in extragalactic surveys. 

Appendix \ref{sec:appendixdeltad} presents a derivation of the  factor $\delta_d$ introduced in $\S$ \ref{sec:SFmodel} that is associated with our chosen density profile.  There we incorporate a central core rather than strictly power-law behavior at all densities.  
\begin{figure*}[t]
\begin{center}
\begin{tabular}{cc}
\includegraphics[width=0.5\linewidth]{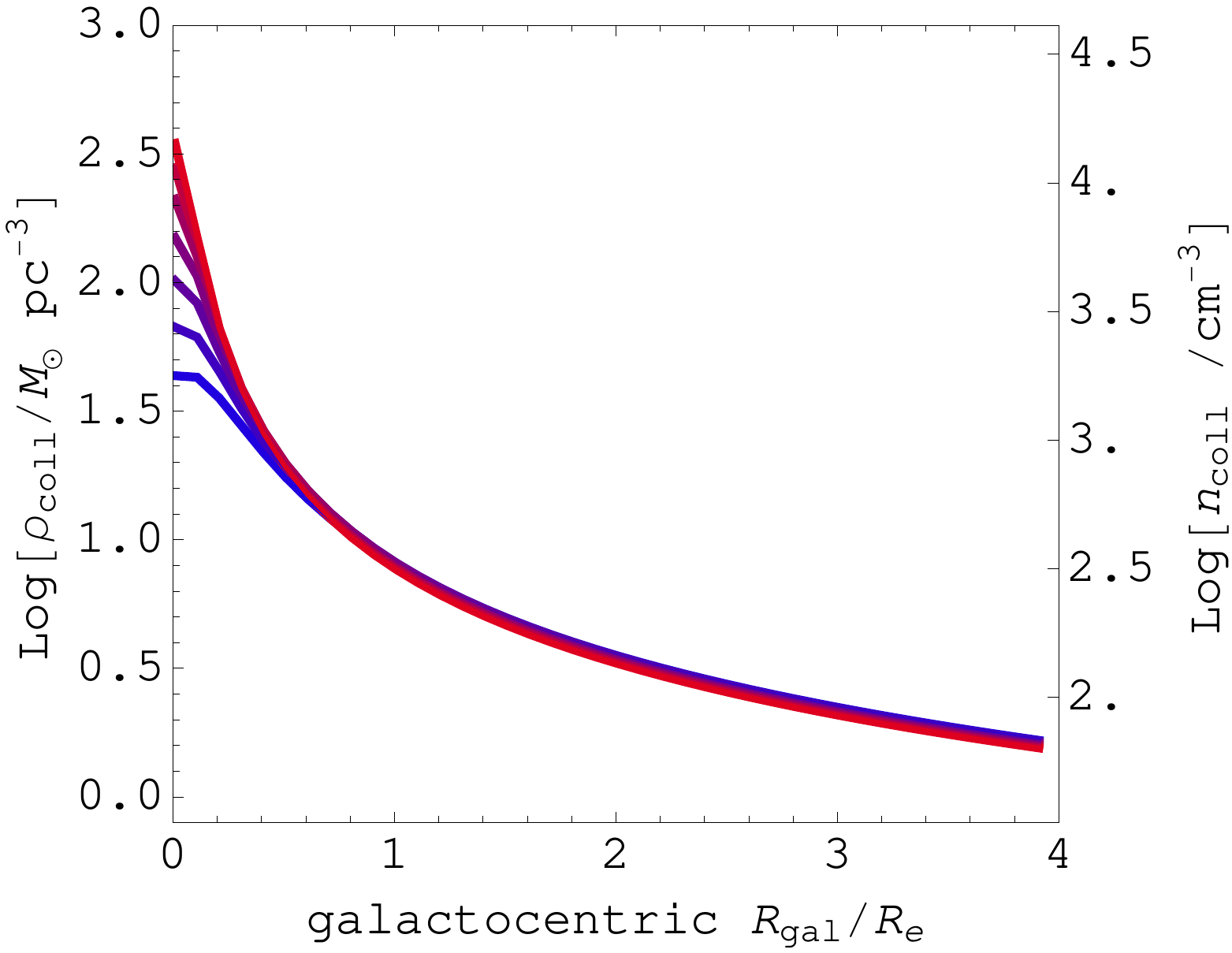}&\includegraphics[width=0.5\linewidth]{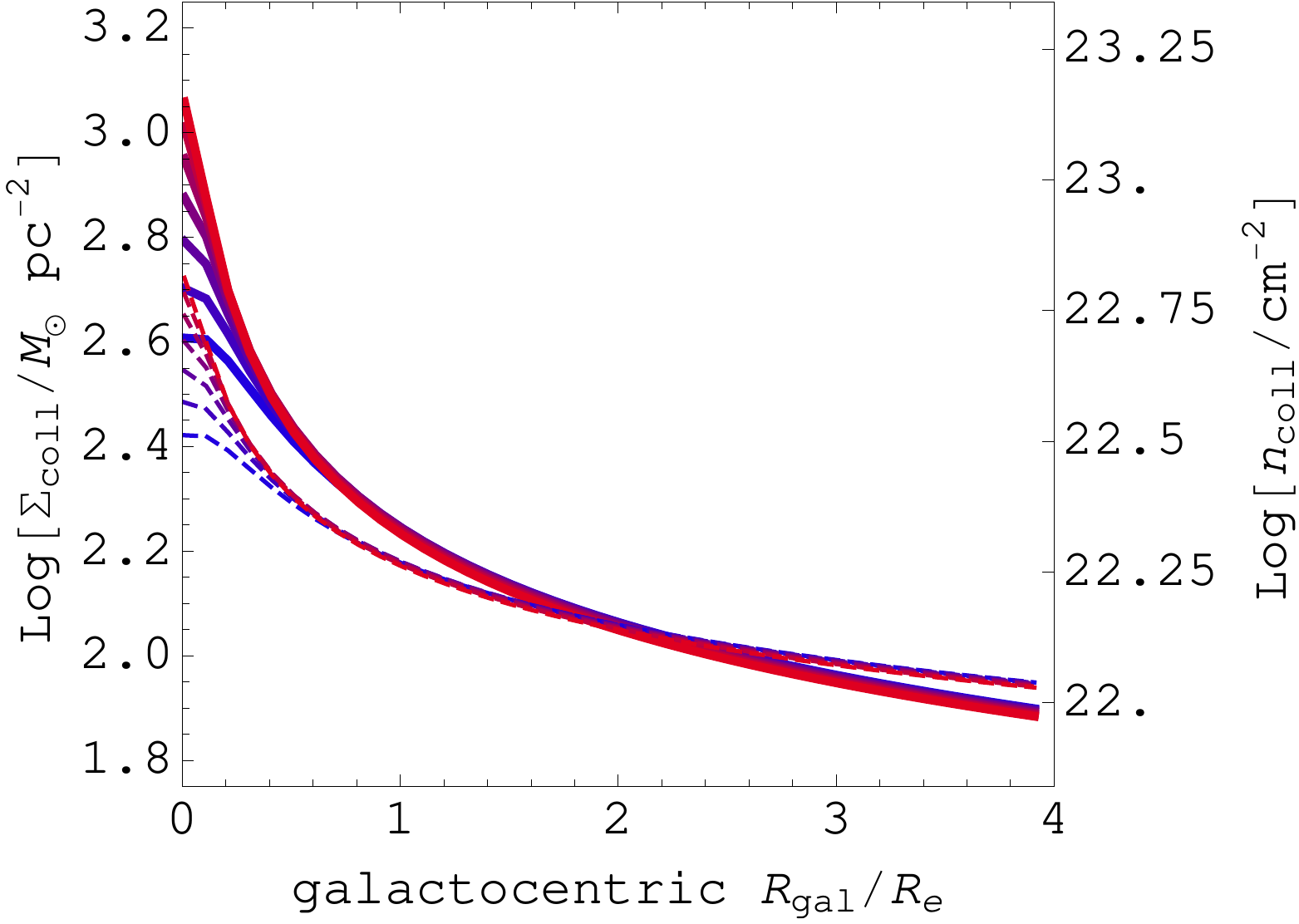}
\end{tabular}
\end{center}
\caption{(Left) The volume densities at which gas decouples from the galactic potential and begins to collapse (Eq.~\eqref{eq:volthresh}), measured from where self-gravity dominates the galactic potential by the derived factor $\gamma_{\rm coll}=2.5$ assuming $k=2$ (see $\S$~\ref{sec:collapsetime}).  In this illustration we adopt the empirical galaxy models described in $\S$~\ref{sec:globalgalaxymodels} with stellar masses in the range $9.25<\log M/{\rm M}_\odot<10.75$ (from blue to red; shown in steps of $0.25\log M/{\rm M}_\odot$).  (Right) An estimate for the surface densities associated with collapse in typical clouds, using the volume density in the left panel and assuming the gas has surface density $\Sigma_{\rm c}=60$ M$_\odot$~pc$^{-2}$ on scale $R_{\rm c}=30$~pc \citep[typical of clouds in the disk of the Milky Way,][]{mv17} below which we assume the density to be distributed as power-law with index $k=2$ (solid) or $k=1.5$ (dashed). \vspace*{0.35cm} }
\label{fig:volthresh}
\end{figure*}

\subsection{The volume densities at which cloud material decouples from the galactic environment }\label{sec:rhosgpred}
\subsubsection{Predicted variation with galactic environment}
In this section, we use our suite of semi-empirical cloud and galaxy models introduced in $\S$~\ref{sec:motions} to 
estimate the densities at which gas decouples from the galactic potential so that it can collapse and form stars.  

The left panel of Figure~\ref{fig:volthresh} shows the dependence of the volume density $\rho_{\rm coll}$ given by Eq.~\eqref{eq:volthresh} on the properties of a given galaxy potential.  For reference, the right panel shows an estimate of the surface density $\Sigma_{\rm coll}$ associated with collapse.  This estimate uses our nominal internal power-law density distribution to express the collapse scale in terms of the surface density $\Sigma_{\rm c}$ on some larger scale $R_{\rm c}$.  In this illustration we adopt $R_{\rm c}=30$~pc and $\Sigma_{\rm c}=60$ M$_{\odot}$~pc$^{-2}$, to match the average properties of clouds in the disk of the Milky Way \citep{mv17}. 

In the main disk environment (from $1\lesssim R_{\rm gal}/R_{\rm e} \lesssim 4$), where rotation curves flatten out, $\rho_{\rm coll}$ decreases along with $\nu^2\approx R_{\rm gal}^{-2}$, which dominates the numerator in Eq.~\eqref{eq:volthresh}.  There is very little variation from galaxy to galaxy, although this is mostly a product of our adopted semi-empirical galaxy model.  Global scaling relations suggest that the increase in stellar scale height with stellar mass alters $\nu$ in a way that is cancelled by the increase in $V_{\rm c}$ with stellar mass.  In real galaxies the level of $\rho_{\rm coll}$ is expected to show greater diversity, given the specific density distributions of their stellar disks and non-axisymmetric features therein.

By the outer radii, where the Solar Neighborhood is located ($R_{\rm gal} \approx 2R_{\rm e}$), the predicted $\rho_{\rm coll}$ in all models falls to ${\sim}10^2$~cm$^{-3}$.  We estimate that this corresponds to $\Sigma_{\rm coll} \approx 100$ M$_{\odot}$~pc$^{-2}$ or $A_K=1$~mag (adopting $A_K=0.112 A_V$ and $N_{\rm H}/A_V=1.37\times10^{21}~{\rm cm}^{-2}~{\rm mag}^{-1}$; \citealt{evans09}) in typical Milky Way clouds.  This falls near the level of $A_K=0.2$~mag at which high dynamic range density PDFs (observationally reconstructed by \citealt{kain09} and \citealt{lombardi}) develop clear power-law behavior, indicating the onset of self-gravitation.  This moreover coincides with the apparent density threshold above which local clouds are observed to form stars, as further discussed below. 

Moving inwards toward galaxy centers, the threshold density increases rapidly.  For a Milky Way-mass galaxy, the model predicts an increase in $\rho_{\rm coll}$ by $1{-}2$ orders of magnitude from the disk to the center.  In contrast to the main disk environment, the threshold density varies more from galaxy to galaxy at small galactocentric radii, where galaxy mass distributions and rotation curve shapes can differ substantially.  As we will show in the next sections, the rapid increase in $\rho_{\rm coll}$ toward small $R_{\rm gal}$ leads to characteristic variations in the rate at which gas can form stars. 

\subsubsection{Relation to a `critical density' for star formation}
\label{sec:motivation}
Studies of molecular clouds in the Solar Neighborhood of the Milky Way suggest that star formation is strongly correlated with high column density gas. Specifically, they find that there is a tight correlation between the mass of gas along sight lines with dust extinctions $A_V\geq8$~mag (corresponding to $\Sigma_{\rm crit} = 120{-}200$ M$_{\odot}$~pc$^{-2}$) and the star formation rate \citep{evans14}.  This has been interpreted as evidence for a star formation threshold at $A_V=8$~mag \citep{Johnstone, heid, lada}, although several authors have argued that this should be interpreted as a sharp but gradual decline in the star formation rate below this value, rather than an absolute cut-off \citep{Gutermuth, BurkertHartmann}. 

In the bottleneck model, this putative critical threshold is a consequence of both a sharp decline in star formation (passing from free-fall to the slow collapse regime) and a baseline threshold for collapse.  For the Solar Neighborhood, our model predicts that this baseline sits near $\Sigma_{\rm coll} \sim 100$ M$_\odot$~pc$^{-2}$, independent of the properties of the gas.  The apparent threshold would then sit above this level to varying degrees, depending on subtle differences in how rapidly star formation rises to the free-fall rate, which is sensitive to the local gas distribution (i.e., the strength of self-gravity; see Figure~\ref{fig:tcoll}).  

The model also describes variation in the critical star formation threshold both within and between galaxies, following environmental changes in the baseline collapse threshold illustrated in Figure~\ref{fig:volthresh}.  A non-universal threshold like this would help explain the existence of clouds that have little obvious evidence of ongoing (massive) star formation but surface densities far in excess of the critical value $\Sigma_{\rm crit} = 120{-}200$ M$_\odot$~pc$^{-2}$ determined for Solar Neighborhood clouds. Such high surface density, non-star-forming clouds exist within the Central Molecular Zone (CMZ) of the Milky Way \citep{longmore, kruijsLong, johnston, kruijssen14} and along the inner, high column density portion of the spiral arms of the nearby galaxy M51 \citep{meidt, leroy2017, q19}.  The variations in dense gas star formation efficiency in extragalactic surveys that probe a wide variety of cloud environments \citep{usero, bigiel16, gallagher,jd19} also imply that the critical threshold for star formation is not universal.  As we demonstrate later in $\S$~\ref{sec:sfedenseobs}, the galactic bottleneck model unifies the large range of thresholds inferred from observations under one framework.  

\subsection{The fraction of collapsing, decoupled gas}\label{sec:massfrac}
The volume density threshold given in Eq.~\eqref{eq:volthresh} determines the mass of material in a given cloud where collapse (i.e., strong self-gravitation) is possible.  This can be expressed as a fraction of the cloud's total mass when the internal distribution of material within a cloud is known.  With our assumed $\rho\propto r^{-k}$ density profile, 
the fraction of the total mass $M_{\rm c}$ above a volume density threshold $\rho_{\rm i}$ can be easily expressed as
\begin{equation}
\frac{M_{\rm i}}{M_{\rm c}} = \left(\frac{\rho_{\rm i}}{\rho_{\rm c}}\right)^{\frac{k-3}{k}} = \left(\frac{\Sigma_{\rm i}}{\Sigma_{\rm c}}\right)^{\frac{3-k}{1-k}} = \left(\frac{R_{\rm i}}{R_{\rm c}}\right)^{3-k}\, .
\label{eq:eqRatiosGen}
\end{equation}
Here $R_{\rm i}$ is the scale  probed by $\rho_{\rm i}$ and its associated column density $\Sigma_{\rm i}$, $M_{\rm i}$ is the mass above $\rho_{\rm i}$, and $\rho_{\rm c}$ is the volume density above which $M_{\rm c}$ on scale $R_{\rm c}$ is measured.  In the nominal case with $k=2$, for example, the self-gravitating mass fraction can be written as
\begin{equation}
\frac{M_{\rm coll}}{M_{\rm c}} = \left(\frac{\rho_{\rm c}}{\rho_{\rm coll}}\right)^{1/2}\, .
\label{eq:eqRatios}
\end{equation}

\begin{figure*}[t]
\begin{tabular}{cc}
\includegraphics[width=0.45\linewidth]{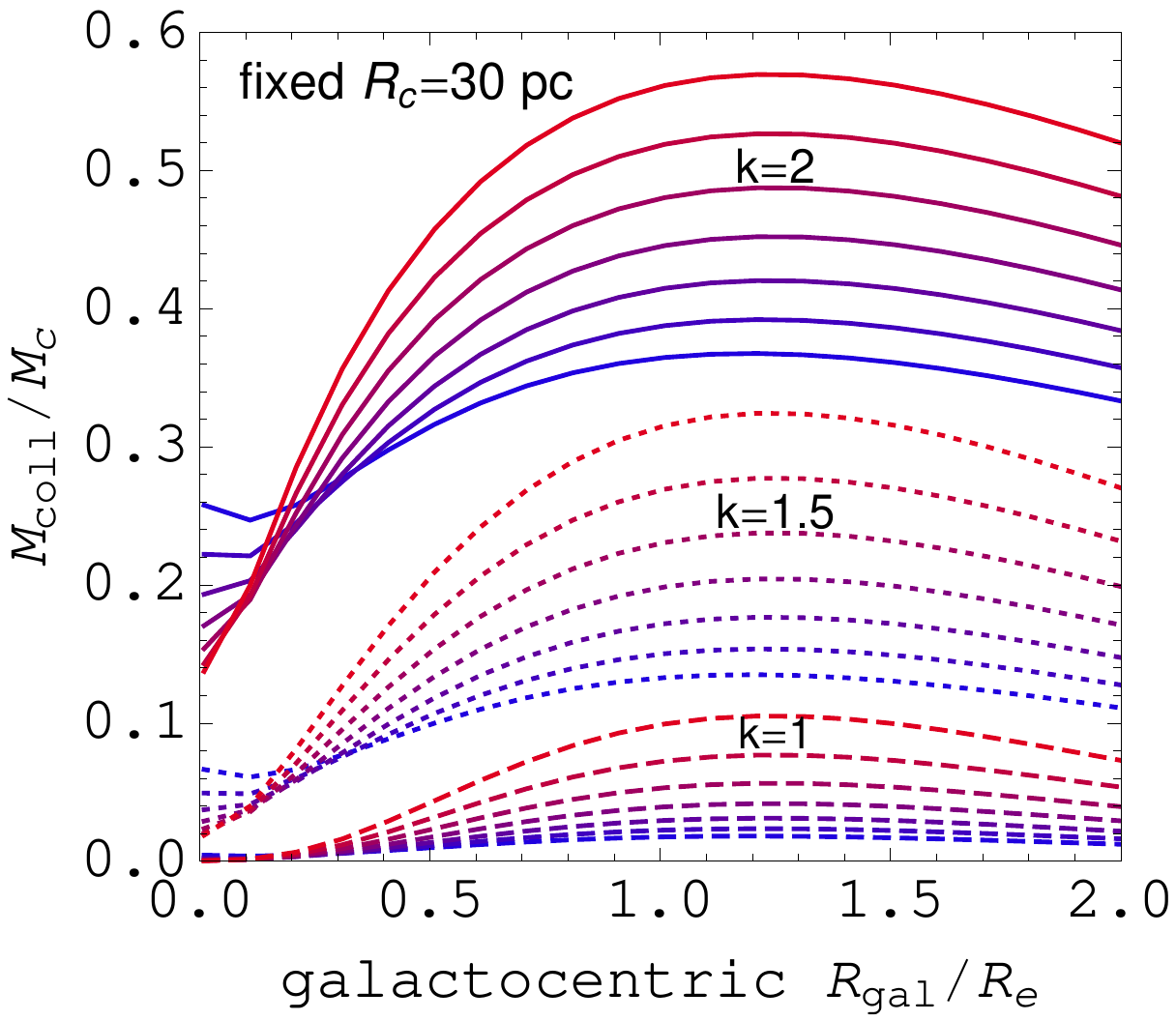}&\includegraphics[width=0.45\linewidth]{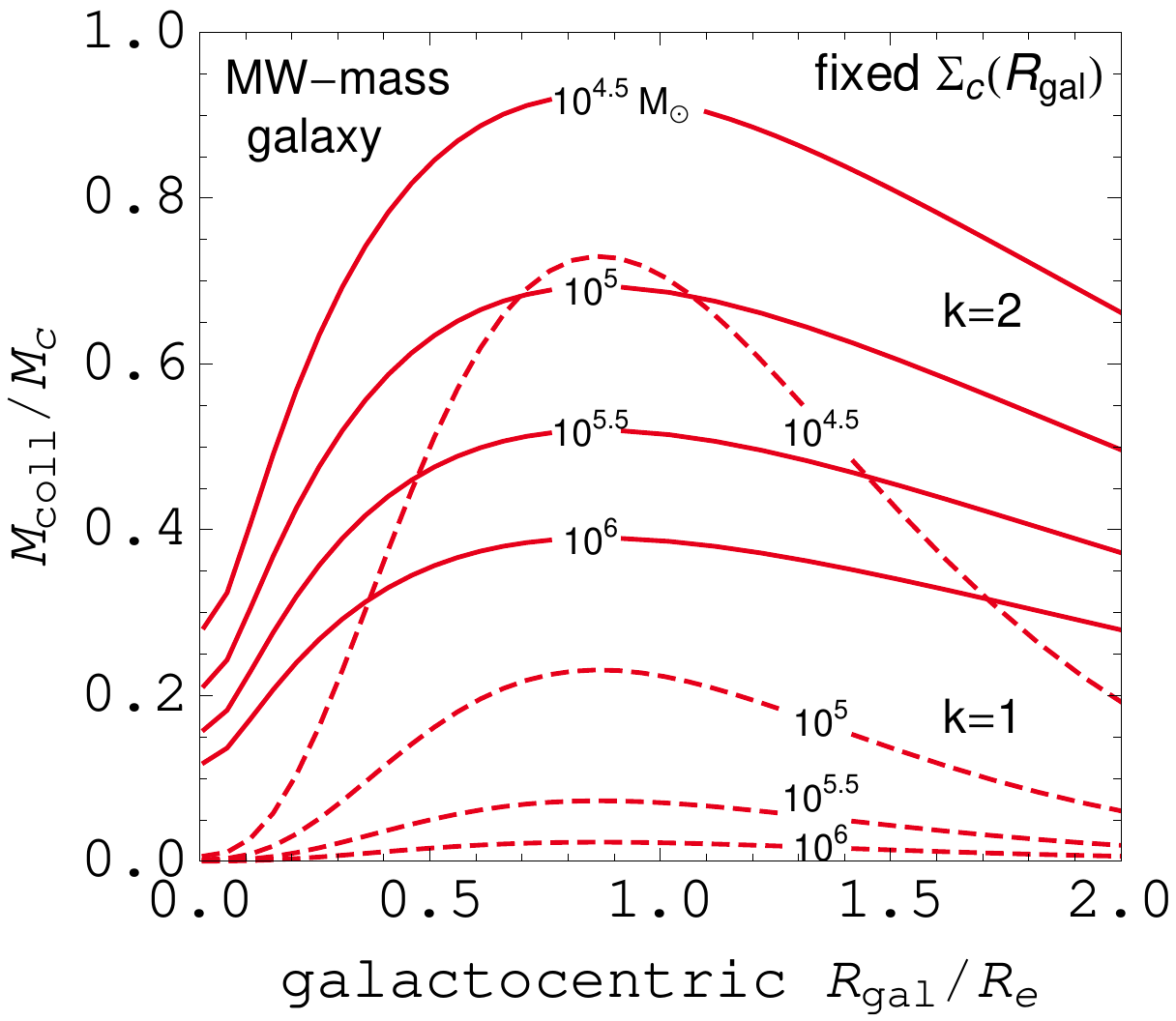}
\end{tabular}
\caption{(Left) {\bf (Needs to be updated!)} The fraction of cloud mass $M_{\rm coll}/M_{\rm c}$ where self-gravity dominates the galaxy potential for clouds of three different sizes: $R_{\rm c}=10$~pc (top light gray region), $R_{\rm c}=30$~pc (middle black region), and $R_{\rm c}=50$~pc (bottom dark gray region) situated in a Milky Way-mass galaxy.  The curves illustrate the radial dependence of $M_{\rm coll}$ across $0.01R_{\rm e}$ to $2.5R_{\rm e}$, extending out to the typical disk scale length corresponding to the edge of the bright molecular emission in nearby star-forming disk galaxies \citep{schruba}.  The dotted black line shows the value of $M_{\rm coll}/M_{\rm c}$ predicted at fixed volume density $n({\rm H}_2)=275$~cm$^{-3}$ at all radii. 
(Right) Hatched regions indicate the spread in $M_{\rm coll}/M_{\rm c}$ associated with a range in volume density that decreases on average exponentially with galactocentric radius (see text).  The three regions assume three different cloud sizes and the empirically-motivated model for the distribution of cloud-scale surface densities introduced in Paper~I, described in $\S$~\ref{sec:globalgalaxymodels}.
The ratio $M_{\rm coll}/M_{\rm c}$ for a cloud of the same size in the inner and outer disk varies by a factor of ${\sim}5$.  
 \vspace*{0.35cm} }
\label{fig:rsgplot0}
\end{figure*}

According to Eq.~\eqref{eq:eqRatios}, the fraction $M_{\rm coll}/M_{\rm c}$ of the cloud, in which self-gravity dominates the energy in galactic motions by the amount $\gamma_{\rm coll}$, will vary throughout a cloud population both due to variation in cloud volume density and the threshold $\rho_{\rm coll}$ determined by the cloud's location in the galactic potential.  This is illustrated in Figure~\ref{fig:rsgplot0}, which shows the collapsing mass fraction $M_{\rm coll}/M_{\rm c}$ predicted according to Eq.~\eqref{eq:eqRatios} throughout a set of empirically-based `global galaxy' models introduced in $\S$~\ref{sec:motions}.  

In the left panel, predictions assume a power-law distribution of gas densities below a fixed scale $R_{\rm c}=30$~pc with three different values of $k$ adopted.  A fixed cloud size is chosen to facilitate direct comparisons to observations obtained at fixed beam size.  In all cases, the nominal cloud volume density model at scale $R_{\rm c}=30$~pc assumes the exponential disk surface density model for the molecular gas distribution $\Sigma_{\rm H_2}(R_{\rm gal})$ developed in Paper~I that we increase by a clumping factor $c=2$ to the cloud scale to match observations on $60$~pc scales (Leroy et al. 2016), i.e., $\Sigma_{\rm c} (R_{\rm gal}) = c\Sigma_{\rm H_2}(R_{\rm gal})$.  For reference, the trend assuming a constant cloud volume density at all galactocentric radii $R_{\rm gal}$ is illustrated by the dashed line.  

The right panel of Figure~\ref{fig:rsgplot0} highlights the variations predicted throughout a cloud population in a single, Milky Way-mass galaxy, adopting a cloud size $R_{\rm c}(R_{\rm gal})$ that varies with position in the galaxy according to the assumed cloud mass $M_{\rm c}$ and fixed surface density model $\Sigma_{\rm c} (R_{\rm gal})$, i.e., $R_{\rm c} = \sqrt{M_{\rm c} / (\pi\Sigma_{\rm c}(R_{\rm gal}))}$.  The density distribution below the cloud scale in all cases is also assumed to be a power-law with either $k=2$ or $k=1$.

\subsubsection{Characteristic trends in the collapsing mass fraction throughout galaxy disks}
Figure~\ref{fig:rsgplot0} demonstrates that the collapse fraction of any given region or cloud depends strongly on its size or mass and its internal distribution of material.  In the highest-mass clouds, which entail larger cloud sizes at fixed $\Sigma_{\rm c}$, more of the gas is distributed in the weakly self-gravitating cloud envelope, reducing the overall collapse fraction.  Shallower profiles, which can accommodate more mass in the cloud envelope, also contain smaller fractions of collapsing gas.  

Figure~\ref{fig:rsgplot0} also demonstrates that the collapse fraction in a cloud of a given size or mass depends strongly on location in the galaxy. The increase in $M_{\rm coll}/M_{\rm c}$ with increasing galactocentric radius at fixed scale, highlighted in the left panel of Figure~\ref{fig:rsgplot0}, is a trend characteristic of our adopted semi-empirical galaxy models based on the observed properties of galaxies. 
Exponentially decreasing gas surface densities imply that the cloud-scale gas density distribution also decreases with galactocentric radius $R_{\rm gal}$ at fixed size scale (or disk height; here modeled as fixed cloud size), leading to a weakening of self-gravity with increasing $R_{\rm gal}$.  This decrease is typically less rapid than the weakening of the background galaxy potential with increasing  $R_{\rm gal}$ given the properties (surface density, scale length and height) of stellar disks.  As a result, $M_{\rm coll}/M_{\rm c}$ exhibits a characteristic increase from the inner to the outer disk. Were the gas self-gravity to remain high at large $R_{\rm gal}$, such as is possible when the cloud volume density stays larger than in the modeled exponential decline, the self-gravitating fraction would exhibit an even larger increase from small to large $R_{\rm gal}$ at fixed cloud size (as indicated by the dashed line).  

The general trend of increasing $M_{\rm coll}/M_{\rm c}$ with increasing $R_{\rm gal}$ is also a feature of other models for how the gas density varies across galaxies, i.e., assuming varying disk scale heights or varying cloud sizes in a realistic molecular cloud population.  In the Milky Way, cloud sizes are relatively larger in the disk compared to those in the Galactic Center analyzed in $\S$~\ref{sec:MWSFR}, for example.  In this case we would predict higher gas densities at small $R_{\rm gal}$ than assumed in Figure~\ref{fig:rsgplot0}.  However, this does not substantially change the variation in the collapsing mass fraction, which is mostly driven by $\rho_{\rm coll}$ according to our empirical galaxy models.   

At any given location in a galaxy, though, the precise value of $M_{\rm coll}$ predicted for a set of clouds depends on several factors.  Most directly, $M_{\rm coll}$ depends on the level $\gamma_{\rm coll}$ at which gas collapses, which we have tied to the level when gas decouples from the galactic potential.  If (non-equilibrium) turbulent motions prevent the gas from collapsing when it fully decouples from galactic orbital motions, then the collapse threshold would increase so that $M_{\rm coll}$ constitutes a smaller portion of the cloud.  

\subsubsection{Relation to the dense gas mass fraction}\label{sec:corrDG}
The volume density threshold introduced in $\S$~\ref{sec:onsetofcollapse} can be used to predict the galaxy-decoupled, collapsing mass fraction above any (arbitrary) density, such as a `dense gas' volume density threshold $\rho_{\rm d}$ that is much higher than the typical gas density at the cloud boundary.  In practice, a threshold of interest might correspond to the effective densities ${\sim}3\times 10^4 {-} 1\times 10^5$~cm$^{-3}$ probed by commonly used extragalactic dense gas tracers (e.g., HCN or HNC; see for instance \citealt{shirley15}, \citealt{leroy2017}).  Following Eq.~\eqref{eq:eqRatios}, the collapsing fraction above $\rho_{\rm d}$ for clouds with $\rho\propto R^{-2}$ is 
\begin{eqnarray}
\frac{M_{\rm coll}}{M_{\rm d}} = \left(\frac{\rho_{\rm coll}}{\rho_{\rm d}}\right)^{1/2} = \left(\frac{\Sigma_{\rm d}}{\Sigma_{\rm coll}}\right) = \frac{R_{\rm coll}}{R_{\rm d}}\, ,
\end{eqnarray}
where $M_{\rm d}$ is the total dense gas mass and $\Sigma_{\rm d}$ is the dense gas surface density threshold associated with $\rho_{\rm d}$ on scale $R_{\rm d}$.  

In the context of our model, we infer that the onset of collapse occurs near (or within) the `dense gas' (i.e., $\rho_{\rm coll}\gtrsim\rho_{\rm d}$ so that $M_{\rm coll}/M_{\rm d}\lesssim1$).  In disk galaxies, the dense gas ratio is observed to be $0.03<\Sigma_{\rm dense}/\Sigma_{{\rm H}_2}<0.1$ (U15, \citealt{gallagher}).  Comparable (but slightly higher) mass fractions $0.1\lesssim M_{\rm coll}/M_{\rm c}\lesssim 0.5$ are implied by our empirically-based `global galaxy' models introduced in $\S$~\ref{sec:motions} (see Figure~\ref{fig:rsgplot0}), assuming that collapse occurs when self-gravity exceeds the energy in galactic motions by a factor $\gamma_{\rm coll}\sim2.5$.  Thus we expect that dense gas may be close to forming stars at the free-fall rate.  As explored in $\S$~\ref{sec:MWSFR}, this arguably leads to the observation that the star formation efficiency in local clouds is approximately uniform above an apparently universal density threshold 
\citep[i.e.,][]{lada,evans14}.  

As described in the following sections, however, one of the features of this model is that the collapsing mass and the `dense gas' mass can systematically differ, especially in environments where the galactic potential varies strongly.  This can lead to characteristic trends in the fraction of the dense gas that goes on to form stars (i.e., the dense gas star formation efficiency).   

\section{Comparing predicted and observed star formation efficiencies}\label{sec:Obscomparison}
In this section, we use the formalism introduced in the previous two sections to investigate how the bottleneck to collapse imposed by galactic orbital motions in the gas introduces variations in the SFEs of parcels of gas or whole clouds.  We are especially concerned with whether the bottleneck would introduce clear signatures (e.g., trends with galactocentric radius and galaxy type) in the current generation of observations of the cloud-scale SFE.  

\subsection{Anchoring the model: star formation in galaxy-decoupled, free-falling gas}\label{sec:anchor}
To use the model for the decoupling of gas from the galactic potential to investigate environmental variations in the SFEs of molecular clouds, we must assume a model for the rate at which decoupled, free-falling gas forms stars.  

Broadly following \citet{maclow04} and \citet{krumholz05}, we envision star formation as being regulated by the interplay of self-gravity with magnetic fields, turbulence, and energy- and momentum-driven feedback from star formation (i.e., in the form of supernovae (SNe), stellar winds, photoionization, and radiation pressure).  As already noted in $\S$~\ref{sec:decoupling}, the bottleneck model incorporates a heuristic description of these factors by assuming that they regulate the dimensionless star formation efficiency $\epsilon$ in the model to a value much less than unity. 
We will further assume that this value is universal on average, i.e., not strongly dependent on the location of a cloud within its host galaxy, although this has yet to be clearly demonstrated by simulations. 

In practice, we anchor the model by incorporating an empirical calibration of $\epsilon$ that lets us assign a value to the efficiency $\epsilon/t_{\rm ff,coll}$ at the onset of collapse, in particular.  As used in Eq.~\eqref{eq:genericSFEc}, this factor allows us to predict star formation rates in clouds that are partially coupled to their environment (see $\S$~\ref{sec:coupledSFE}).  

Our calibration leverages the observation that SFEs in the star-forming 
gas at the highest densities in local clouds vary only slightly \citep{evans14}.  As we show in the next section, at these densities, gas is decoupled from the galaxy and thus our calibration for $\epsilon$ should be appropriate for gas collapsing near the free-fall rate. The alternative would be to adopt a recalibration of this SFE that applies to not only the collapsing gas, but all gas directly down to the cloud edge.  Such cloud-scale molecular gas efficiencies exhibit large variations both within and between galaxies \citep[e.g.,][]{schruba10, leroy2017m51, utomo17, schruba19}, part of which we aim to describe with our model.  We show later how the first approach can be used to make predictions for the recalibrated SFE of the second approach.

We emphasize that cloud-scale variations in SFE due to variations in $\epsilon$---arising from changes in the balance between the small-scale physical processes that regulate star formation within clouds or cloud evolutionary effects \citep{lee16, grudic}---can be incorporated into the model presented here.  For now, however, our main goal is to determine whether the influence of the galaxy on cloud scales has a discernible effect on cloud SFEs that could be distinguished observationally from these other effects. Our assumption of a universal efficiency is a choice that lets us explore the degree to which the decoupling of molecular gas from the galactic potential acts as a bottleneck to star formation. Later in $\S$~\ref{sec:Obscomparison} we show that this appears to offer a good match to observations over a wide range of scales.

\begin{figure*}[ht]
\begin{center}
\begin{tabular}{cc}
\includegraphics[width=0.425\linewidth]{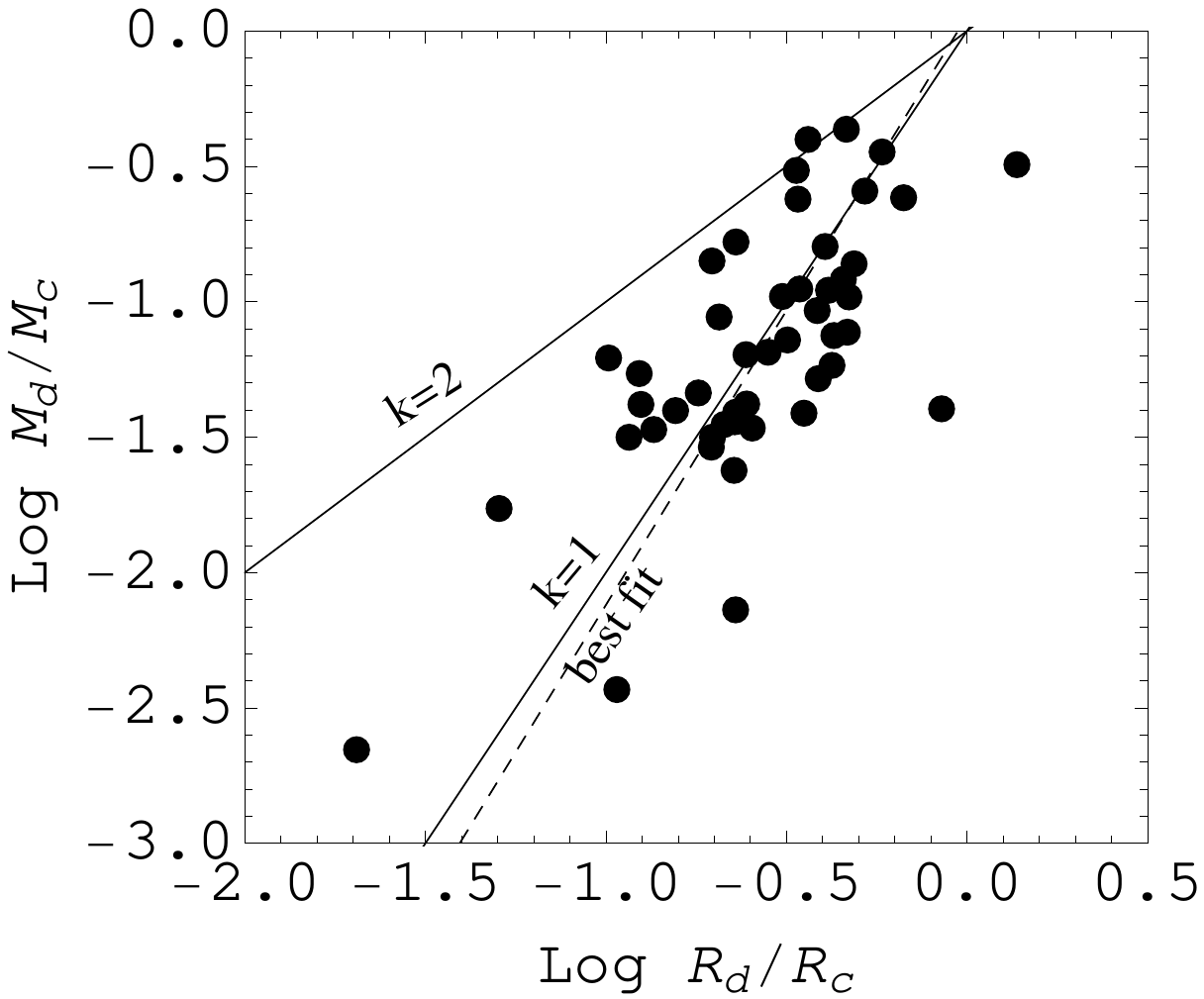}&\includegraphics[width=0.425\linewidth]{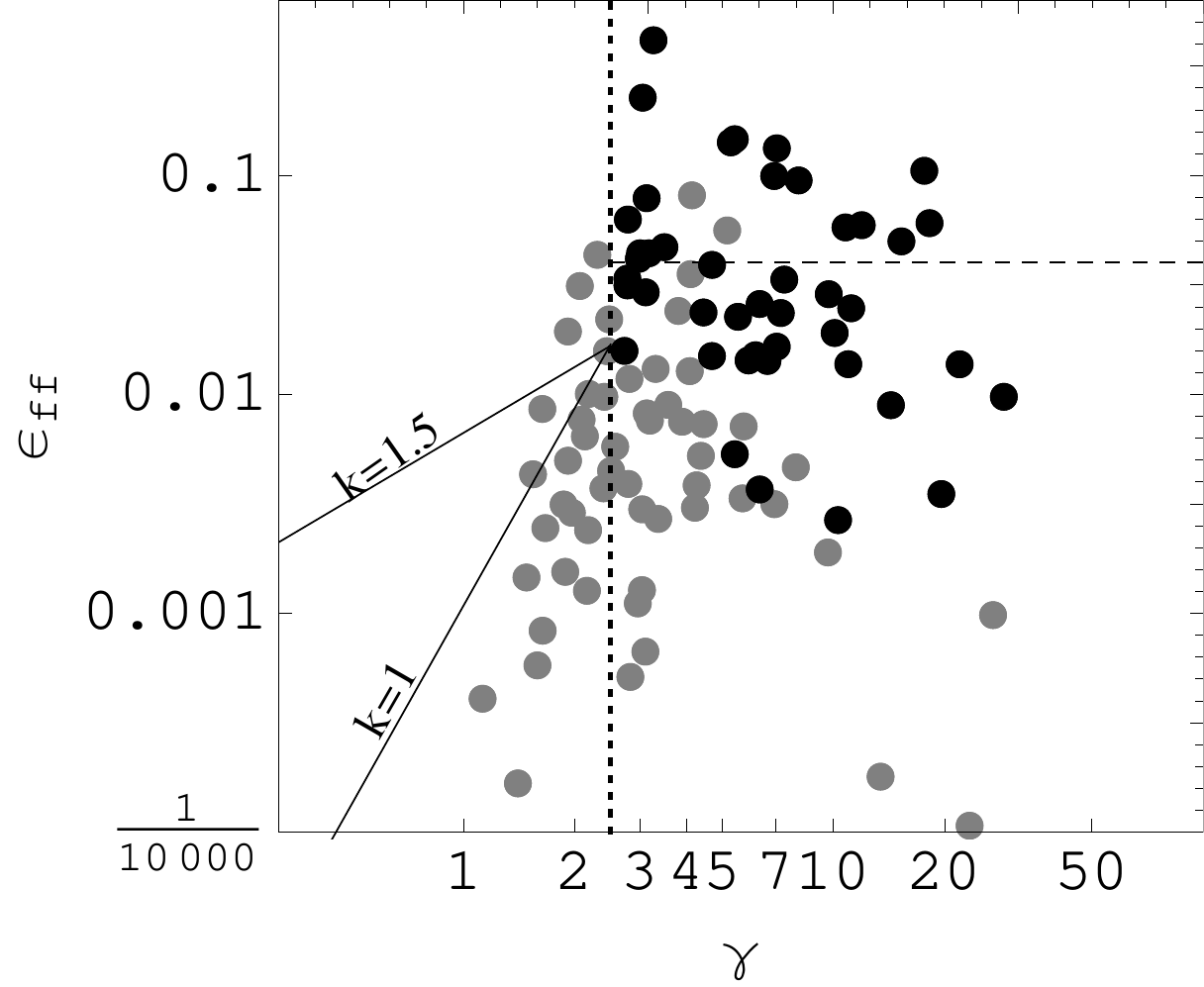}
\end{tabular}
\end{center}
\caption{(Left) Dense gas mass fraction $M_{\rm d}/M_{\rm c}$ vs.\ $R_{\rm d}/R_{\rm c}$ for the local clouds in the \citetalias{veh16} sample.  Two black lines show the relation predicted for the power-law density distribution $\rho\propto r^{-k}$ with $k=1$ or $k=2$.  The best-fit relation with slope corresponding to $k\sim$0.8 is plotted as a dashed line.  (Right) The efficiency per free-fall time $\epsilon_{\rm ff} = \mathrm{SFE} t_{\rm ff}$ above different densities in local clouds.  Measurements from high and low density tracers are shown in black and gray, respectively (see text and \citetalias{veh16}).  The vertical dotted line marks the collapse threshold $\gamma_{\rm coll}=2.5$ predicted in Section \ref{sec:collapsetime} that marks the division between gas that is coupled ($\gamma\ll\gamma_{\rm coll}$) and decoupled ($\gamma\gg\gamma_{\rm coll}$) from the galaxy potential.  The horizontal dashed line marks the average $\epsilon_{\rm ff}=0.04$ measured in the decoupled zone.  Two solid black lines show predictions for the dependence of $\epsilon_{\rm ff}$ on $\gamma$ in the coupled, weakly self-gravitating regime, as given by Eq.~\eqref{eq:epweak} assuming that $\epsilon=0.04.$  
  \vspace*{0.5cm} 
 }
\label{fig:MWeps}
\end{figure*}

\begin{figure*}[t]
\begin{center}
\begin{tabular}{cc}
\includegraphics[width=0.425\linewidth]{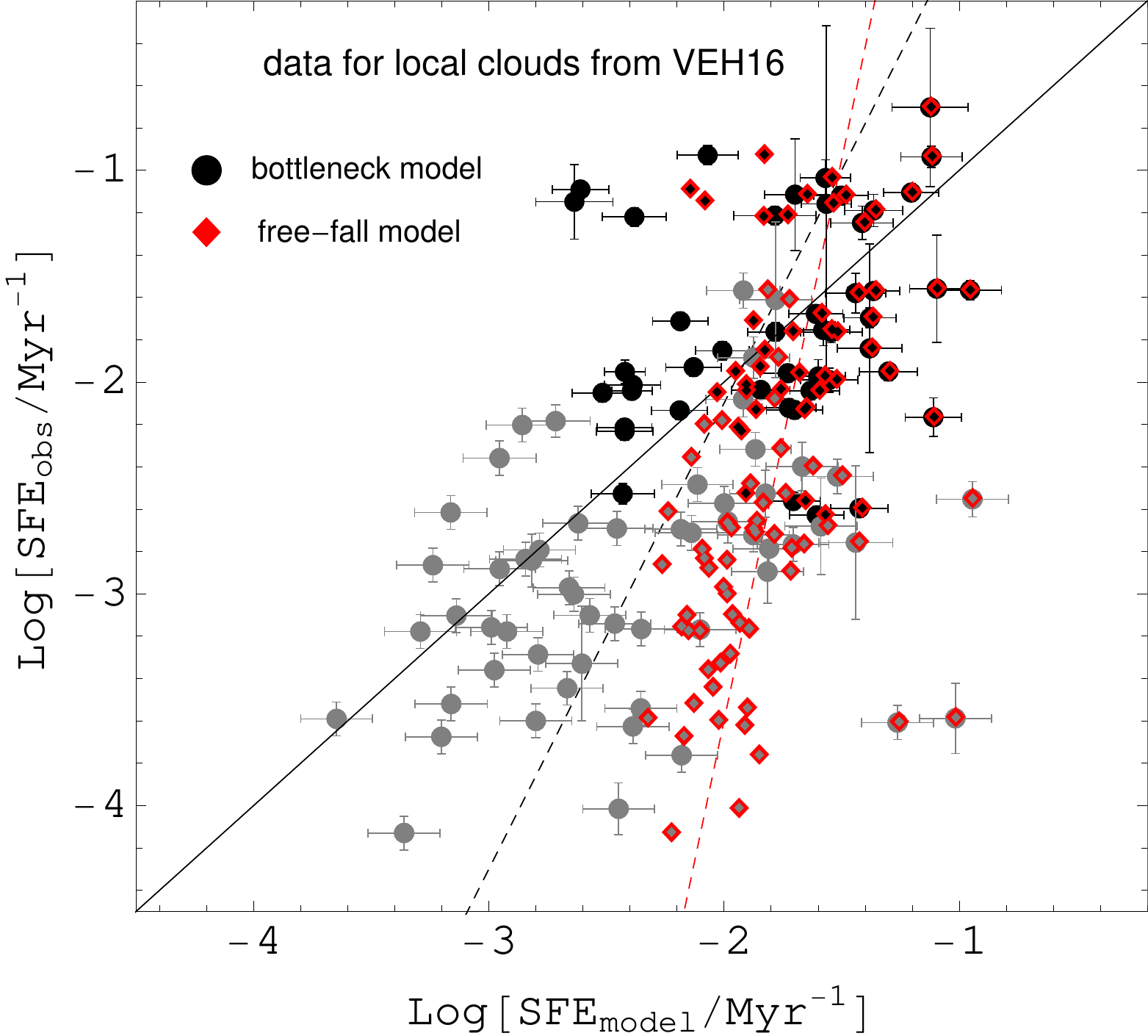}&\includegraphics[width=0.425\linewidth]{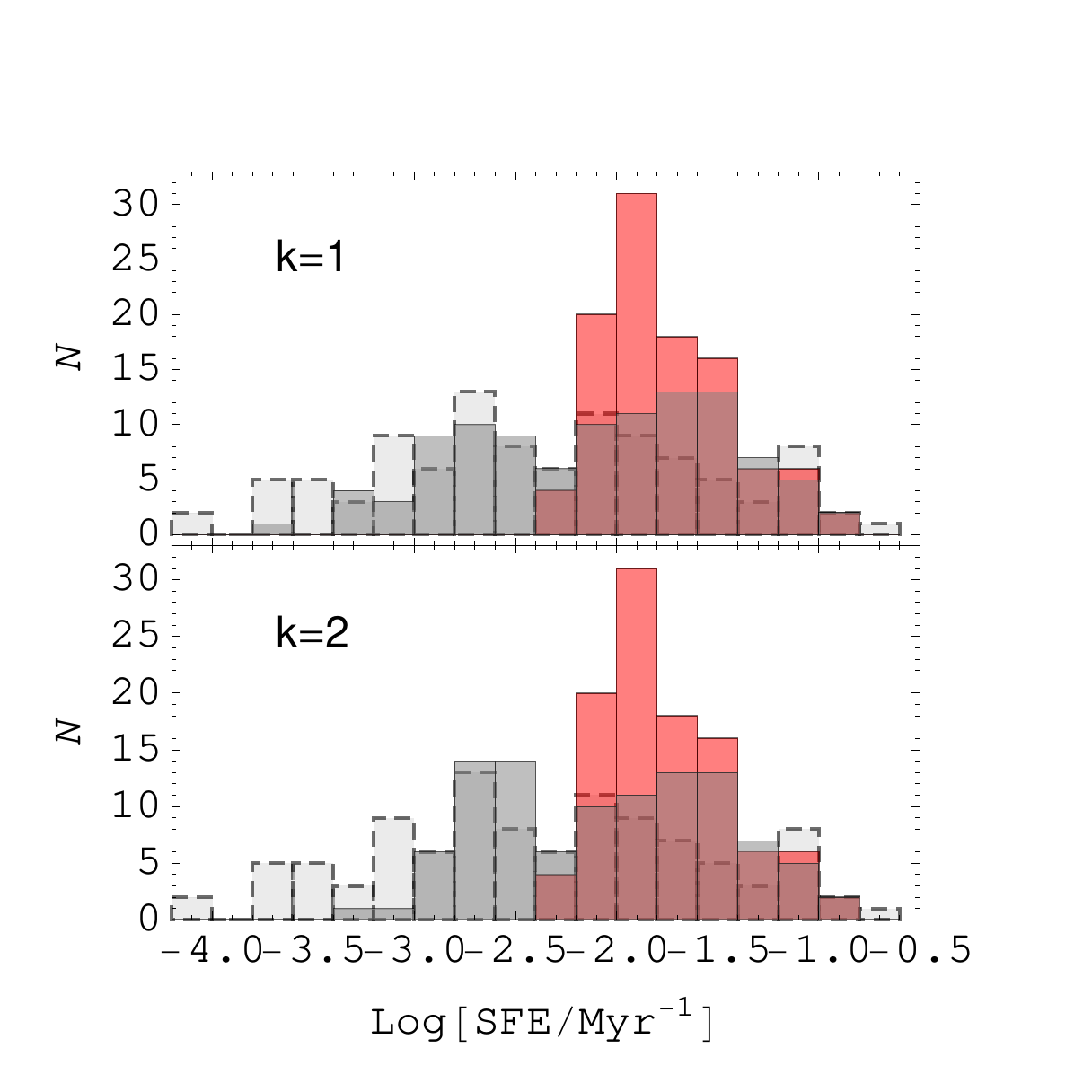}
\end{tabular}
\end{center}
\caption{(Left) Comparison between the observed SFE$_{\rm obs}$ and the SFE$_{\rm model}$ predicted by either the bottleneck model (black and gray circles) or the free-fall model (red outlined diamonds).  Measurements from two density regimes within the clouds are shown (low: gray and high: black).  Error bars are shown only on the bottleneck model predictions and reflect propagated uncertainties on either the measured mass and star formation rate (vertical) or the cloud mass and size (horizontal).  The black line indicates equality.  The gray (red) dashed line shows the best-fit relation to all points predicted by the bottleneck (free-fall) model. 
(Right) Histograms of observed and predicted SFE in local clouds.  In both panels, observed SFEs are shown in light gray with a black dashed boundary and the predictions of the free-fall model are shown in red.  Two predictions of the  bottleneck collapse model are highlighted (dark gray) assuming two different power-law indices: $k=1$ (top) and $k=2$ (bottom).  
\vspace*{0.5cm} 
 }
\label{fig:MWsfes}
\end{figure*}

\subsubsection{An empirical calibration of \texorpdfstring{$\epsilon$}{epsilon} from observations of Milky Way clouds}\label{sec:MWSFR}
In this section, we 
explore how SFEs vary throughout a sample of clouds in the disk of the Milky Way and use these observations to calibrate a value of $\epsilon$ that applies to approximately free-falling gas.  We consider $56$ clouds studied by \cite{veh16} (hereafter \citetalias{veh16}) with reliable SFRs (above the hard 5 $M_\odot {\rm Myr}^{-1}$ minimum advocated by \citetalias{veh16}).  We use two sets of properties tabulated by VEH16 measured either with a dense gas tracer (submillimeter dust continuum emission; \citealt{aguirre}) or with a lower density tracer ($^{13}$CO$(1-0)$ emission; \citealt{jackson,romanduval}) probing nearer to the cloud edge (see \citetalias{veh16} for details).  
These define a total of $103$ unique measures of the efficiency at different locations in the Milky Way.  

With this set of properties we also assemble $47$ measures of the dense gas mass ratio $f_{\rm d}=M_{\rm d}/M_{\rm c}$ (where available).  According to $\S$~\ref{sec:coupledSFE}, the internal density distribution affects the star formation rate per unit mass predicted for a given cloud.  For two equal mass clouds with a fixed threshold $\rho_{\rm coll}$, the cloud with the shallower density profile will appear to form fewer stars per unit mass (see also \citealt{tan06}; \citealt{burk18}; \citealt{parm}).  Thus our study of this set of local clouds begins with an examination of their internal structure.  The right panel of Figure~\ref{fig:MWeps} plots $f_{\rm d}$ vs. $R_{\rm c}/R_{\rm d}$ where, following \citet{meidt16}, we use $f_{\rm d}$ as a proxy for the density distribution.  In the case of a power-law $\rho\propto r^{-k}$, the relation between $f_{\rm d}$ and $R_{\rm c}/R_{\rm d}$ (and $\Sigma_{\rm c}/\Sigma_{\rm d}$) depends on the power-law index, i.e. $\log M_d/M_c =(3-k) \log R_d/R_c$ (see eq. \ref{eq:eqRatiosGen}).  From the slope of the best-fit linear relation in the figure, we infer $k$$\sim$0.8.  We therefore adopt $k$=1 in what follows, unless otherwise noted.  

The right panel of Figure~\ref{fig:MWeps} shows the variation in $\epsilon_{\rm ff}$ as a function of $\gamma$ on the outer measurement scale throughout the studied population, where $\epsilon_{\rm ff}$ is estimated from the measured $\mathit{SFE}$ using the free-fall time at the observed density, i.e., $\epsilon_{\rm ff}=\mathit{SFE} t_{\rm ff}$.  The estimated $\gamma$ (and $t_{\rm coll}$ and $\gamma_0$ assuming $k=1$) at a given density in a given cloud depends both on the properties of the gas and the cloud's (radial) location in the galaxy.  For each cloud we estimate the surface density $\Sigma_i$ at scale $R_i$ from the mean cloud surface density $\Sigma=M_i/(\pi R_i^2)$ derived from the gas mass $M_i$ inside the measured size $R_i$ using that, for our nominal density profile, $\Sigma_c=(3-k)/(4/k)\Sigma$.  We use the observed properties of the Milky Way's rotation curve \citep{reid} to estimate the epicyclic frequency $\kappa$ for each cloud's position (assigned by \citetalias{veh16}).  We then use this to estimate the vertical frequency $\nu$ following the approximation given in Paper~I, assuming a thin stellar disk scale height of $300$~pc in the Milky Way \citep{gilmore, rixbovy}.   Note that, by combining these estimates to yield a measure of $\sigma_{\rm gal}$, we do not account for local gas motions associated with non-axisymmetric structures in the disk.  Thus our estimates of $\gamma$ may be unrealistically high for this set of MW clouds, which populate the Sagittarius arm and an extrapolation of the arm spur identified by \citet{xuspur}.  

The local clouds in Figure~\ref{fig:MWeps} exhibit a wide range in $\epsilon_{\rm ff}$, highlighting cloud material in various states of collapse.  Local clouds tend to fall on either side of the $\gamma_{\rm coll}=2.5$ line, depending on density.  This line also tends to separate clouds into high and low efficiency star formation.  Toward the lower end of $\gamma_0$, a majority of the measured $\epsilon_{\rm ff}$ at low density are systematically lower than the overall level of the points in the figure.  This is consistent with the increased coupling of the gas to the galaxy potential predicted by the bottleneck model.  According to the model, for a fixed internal density distribution, observed star formation efficiencies $\epsilon_{\rm ff}$ fall below the value of $\epsilon$ when they reflect a contribution from the environment-dependent, non-collapsing cloud envelope (see Eq.~\eqref{eq:epweak}).  The relation between $\epsilon_{\rm ff}$ and $\gamma$ predicted in Eq.~\eqref{eq:epweak} in this case depends on the distribution of material.  The thin black line shows the relation assuming the value $k=1$ inferred from the right of Figure \ref{fig:MWeps}.  (For reference the relation assuming $k=1.5$ is also shown.) 

The greater majority of points sampling higher gas densities fall above $\gamma_0=2.5$.  As suggested by Figure~\ref{fig:MWeps}, these probe the fairly pervasive collapse predicted in molecular gas above a surface density $100$~M$_{\odot}$~pc$^{-2}$.  In the context of our model, at these densities, the gas should be fully decoupled from the galaxy and able to undergo approximately free-fall collapse.   

From Figure~\ref{fig:MWeps} we can infer that even in decoupled gas $\epsilon_{\rm ff}$ is well below a value of unity, which we take as an indication that collapse in this regime is regulated by non-gravitational factors, as discussed in $\S$~\ref{sec:decoupling}.  It also exhibits significant variation (as much $1$~dex), most of which is presumably due to temporal variations \citep{FeldmannGnedin11,lee16,grudic}, although differences in the way the material is distributed within clouds can also contribute (i.e. \citealt{tan06}; \citealt{burk18}; \citealt{parm}).  
Overall, however, the observations from the decoupled region of the plot (with $\gamma_0\gg2.5$) suggest a fairly regular $\epsilon_{\rm ff}=0.04$ on average, though with modest evidence for a decrease with increasing density (or $\gamma$).  For the purposes of investigating environmental variations in star formation efficiencies due to the galactic bottleneck, we will adopt this average value  in what follows, although it should be noted that additional variation is to be expected due to the spread in $\epsilon$ in collapsing gas.  According to eq. (\ref{eq:deltadPL}) in Appendix \ref{sec:appendixdeltad}, $\epsilon$ can be estimated from $\epsilon_{\rm ff}$ in the strongly self-gravitating regime, (when $t_{\rm coll}\approx t_{\rm ff}$) as $\epsilon=\delta_d^{-1}\epsilon_{\rm ff}$.  

This calibration of $\epsilon$ yields an important anchor for the model at the onset of collapse.  From the measured cloud properties in Figure~\ref{fig:MWeps} we find $\log \delta_d \epsilon/t_{\rm ff,coll} = log \delta_d\mathit{SFE}_{\rm coll}\,\mathrm{[Myr]} =-2.0\pm0.2$ on average, using the average free-fall time of $4$~Myr for material with $\gamma=2.5$ and our calibration $\epsilon_{\rm ff}$=0.04.  This is consistent with the average value determined by \citetalias{veh16} across the high density subset of the measurements considered here.   
The strong residual variation in the observed $\mathit{SFE}=\delta_d\epsilon/t_{\rm ff}$ found by \citetalias{veh16} would then be related to variations in free-fall time (and collapse time), as we explore in the next section.   In what follows, we will use this calibration as a constraint on $\mathit{SFE}_{\rm coll}\,\mathrm{[Myr]} \sim -2.0$, by adopting $k\sim$1 and $\delta_d\sim$1 (see Appendix \ref{sec:appendixdeltad}) appropriate for the clouds in this sample. 

\subsubsection{Predictions of free-fall collapse vs.\ inhibited collapse in coupled gas}\label{sec:MWSFRmodels}
Within the MW clouds studied in the previous section, the high density material appears to be decoupled from the background galactic potential and able to form stars at a fiducial free-fall rate, while the lower density material is observed to form stars with lower efficiency.  In this section we demonstrate how the bottleneck model provides a continuous description of star formation across these two regimes.  We compare the star formation efficiencies predicted with our collapse timescale to the efficiencies predicted in a model of universal free-fall collapse, as shown in Figure~\ref{fig:MWsfes}.  Both sets of predictions are anchored using our calibration for a universal average $\delta_d\epsilon=0.04$.  For the predictions of the bottleneck model plotted in the figure we adopt $k=1$ (see Figure~\ref{fig:MWeps}).  Here again we distinguish between high and low density tracers.  

There are several notable characteristics in the right panel of Figure~\ref{fig:MWsfes}.  First, the two predictions overlap at high density (as indicated by the overlap between the high density black circles and red-outlined black diamonds), which stems from the similarity between the collapse timescale and the free-fall time at these densities.  In this region of the plot, both predictions (by construction) also show good consistency with the observations.  The basic model of star formation proceeding in a free-fall time, with a universal $\epsilon$, is able to roughly capture systematic variation in observed SFEs.  Note that, in the context of the model, the residual scatter present in the figure would be attributable to deviations from the assumed universal efficiency  (and/or departures from the adopted model for internal cloud structure).  We therefore restrict the present discussion to the possibility that systematic variations in the measured SFEs can be described by environmental dependencies in the model.  

At lower density, the predictions diverge from each other (as highlighted by offset between low density gray circles and red-outlined gray diamonds).  The bottleneck model predicts SFEs a factor of $5{-}10$ lower than the free-fall model.  This leads to better agreement between the predictions of the bottleneck model and the observations.  The slope of the best-fit line between the observations and the predictions in the log-log plot in Figure ~\ref{fig:MWsfes} is ${\sim}7$ in the case of the universal free-fall model and ${\sim}2$ in the case of the bottleneck model with $k=1$.  Although the bottleneck predictions tend to appear more scattered than the free-fall predictions, they deviate less from one-to-one; the mean deviation of the bottleneck model from unity is a factor of $2$ lower than the free-fall prediction. 

As illustrated by the histograms of SFE in the right panels of Figure~\ref{fig:MWsfes}, the strength of the bottleneck model (dark gray) lies in its ability to cover the full range in observed SFEs (light gray), which extend nearly two orders of magnitude lower than those predicted by the free-fall model (red).  However, even in this case, the low-density SFEs still do not reach low enough to fully match the observations. There are several reasons that predictions for the low-density SFEs could be preferentially high.   The first is the possibility of systematic variation in the distribution of densities below the cloud scale, which is not presently accounted for. 
A factor of $2$ decrease in the $^{13}$CO conversion factor from the value assumed by \citetalias{veh16} (as indicated by observations and PDR models of the ratio W($^{13}$CO)/W($^{12}$CO) at low $A_V$ throughout the Perseus molecular cloud complex; \citealt{pineda}) would also bring the measured and predicted SFEs at low density into better agreement.  Underestimation of the SFE at low density might also reflect the importance of spiral arms on local motions, which are not accounted for by the present estimate of $\kappa$ in Eq.~\eqref{eq:siggalnet}.  (An underestimation of $\kappa$ would be expected to have larger consequences at low density when self-gravity and the energy in galactic motions are more comparable.)  

Despite these issues, Figure~\ref{fig:MWsfes} demonstrates that the bottleneck model captures one of the key features of the observations, namely the decreased efficiency of star formation at low density compared to higher density.  

\subsection{Environmental variations in star formation efficiencies}
\label{sec:environmentalvariations}
In this section, we use the bottleneck model to highlight scenarios in which the onset of collapse and star formation in molecular gas vary systematically with environment.  Our main aim here is to provide a sense for how much of the observed variation in SFE can be attributed to the galactic bottleneck.  
As emphasized earlier, the details of turbulence-regulated star formation could lead to deviations from either the universal $\epsilon$ or efficiency $\mathit{SFE}_{\rm coll}$ adopted at present by the model and/or our adopted description of internal cloud structure.  Without a detailed picture for how these quantities could vary systematically with environment, we assume that deviations from our adopted model will introduce scatter about the primary trends predicted by the model.  Observational constraints on this variation will be key to improving predictions of the model in the future.  

\subsubsection{Inefficient star formation in galaxy centers: application to the Central Molecular Zone}\label{sec:CMZ}
\begin{table*}
\begin{center}
\caption{Star formation rates in the Milky Way CMZ}\label{tab:CMZ}
\begin{threeparttable}
\begin{tabular}{rccc}
\hline
Region& Observed $\Sigma_{\rm SFR}$\tnote{a}& Predicted $\Sigma_{\rm SFR}$\tnote{b}, $k=2$ & Predicted $\Sigma_{\rm SFR}$, $k=1.5$\\
&(M$_\odot$ yr$^{-1}$ kpc$^{-2}$)& (M$_\odot$ yr$^{-1}$ kpc$^{-2}$) & (M$_\odot$ yr$^{-1}$ kpc$^{-2}$)\\
\tableline
230 pc-integrated& 0.2 & $0.12{-}0.2$ & $0.006{-}0.009$ \\
$1\fdg3$ cloud complex & 0.13&  $0.21{-}0.4$& $0.013{-}0.025$ \\
100 pc stream& 3.0 & $12.6{-}25.9$& $6.8{-}14.1$ \\
\hline
\end{tabular}
 \begin{tablenotes}
\item[a]Adopted from \cite{kruijssen14}.
\item[b]Derived with our model of dynamical regulation according to Eq.~\eqref{eq:gensfr} using the gas surface density and range in epicyclic frequencies tabulated for each region by \cite{kruijssen14} to calculate the collapse timescale (and the collapse mass fraction); see text for details.
 \end{tablenotes}
\end{threeparttable}

\end{center}
\end{table*}
In the main disk environment of the Milky Way, a large fraction of the material in clouds above densities probed by $^{13}$CO(1-0) emission is collapsing and forming stars near the free-fall rate, given the sizes and masses of the emitting regions. According to the strongly restricted collapsing mass fraction at small galactocentric radii in our model (see Figure \ref{fig:sfeplot}), galaxy centers, in contrast, are prototypical sites where molecular gas may be preferentially found to form stars inefficiently.  

We emphasize that, as expressed more precisely in Eq.~\eqref{eq:genericSFEc}, whether or not star formation is suppressed in the center environment depends on the exact distribution of gas densities, i.e., deviations from the basic axisymmetric exponential disk models adopted here, such as in the form of rings or at the locations of bar ends.  The gas in these regions can locally reach such high densities that it may be strongly self-gravitating even in the presence of a strongly varying central galactic potential so that star formation proceeds efficiently \citep[e.g.,][]{utomo17}.  To get a general sense for the degree to which our basic picture of star formation restricted by galactic motions applies, here we apply the framework of our model to the gas observed in the Central Molecular Zone (CMZ) of our own Galaxy, where star formation is observed to be strongly suppressed (\citealt{longmore}; \citealt{kruijssen14}).  

For molecular gas with surface density $\Sigma_{\rm gas}$ across some area $A$ that is forming stars at a rate $\dot{M}_{\rm star}$, we express the star formation rate surface density $\Sigma_{\rm SFR}=\dot{M}_{\rm star}/A$ as 
\begin{eqnarray}
\Sigma_{\rm SFR} = \delta_d\mathit{SFE}_{\rm coll}\frac{\gamma_{\rm coll}^{2(k-3)/k}}{t_{\rm coll,on}/t_{\rm ff,coll}} f_{\rm c}\Sigma_{\rm gas}\frac{M_{\rm coll}}{M_{\rm c}}\, , \label{eq:gensfr}
\end{eqnarray}
where $f_{\rm c}$ is the cloud fraction, $\mathit{SFE}_{\rm coll} = \epsilon/t_{\rm ff,coll}$ represents the efficiency with which free-falling gas forms stars at densities where gas begins to collapse, $\delta_d$ expresses that the integrated SFE depends on the internal density distribution \citep[see, e.g.][Appendix \ref{sec:appendixdeltad}] {burk18,parm},  $t_{\rm coll,on} = 2.4 t_{\rm ff}$ is the collapse time when the threshold $\gamma_{\rm coll}=2.5$ is reached (at which point $t_{\rm ff}=t_{\rm ff,coll}$), and $M_{\rm coll}$/$M_{\rm c}$ is the collapsing mass fraction of the gas regulated by the galaxy potential as described in $\S$~\ref{sec:massfrac}.  

Assuming that the gas contained in the CMZ is organized entirely into clouds ($f_{\rm c}=1$) with internal density profiles $\rho\propto r^{-k}$ with either $k$=1.5 or $k$=2 so that $M_{\rm coll}/M_{\rm c} = (\rho_{\rm coll}/\rho_{\rm c})^{2(k-3)/k}$, we can use the collapse fraction of the clouds to estimate the star formation rate there.  Using the properties tabulated for several well-studied regions in the CMZ by \citet{kruijssen14}, we estimate $M_{\rm coll}/M_{\rm c}$ by first determining the gas volume density $\rho_{\rm c}$ (from the tabulated surface density and vertical scale height $h$) and then $\rho_{\rm coll}$ according to Eq.~\eqref{eq:volthresh}.  For the galactic center environment, the potential gradient in the plane is expected to become comparable to the gradient in the vertical direction.  We thus assume that galactic motions are isotropic and replace $(\kappa^2+2\Omega^2+\nu^2)$ in the denominator of Eq.~\eqref{eq:volthresh} by $3\kappa^2$ using $\kappa$ measured by \citealt{kruijssen14} from the observed rotation curve\footnote{As the region of interest lies within the inner Lindblad resonance of the Milky Way's bar, $\Omega-\Omega_p < 2\kappa$, and we expect that the values of $\kappa$ assumed here (\citealt{kruijssen14}; based on background rotation) to be within a factor of $\sqrt{2}$ from the arguably more precise estimation of the epicyclic motions given in Paper~I that explicitly accounts for the effect of the bar.  Since this is within the accuracy quoted by \citet{kruijssen14}---given uncertainties in the circular velocity due to projection effects as well as uncertainty in the proximity and orientation of the region---we ignore this factor here.}).

The strongly restricted collapsing fractions $M_{\rm coll}/M_{\rm c}$ predicted by our model in this scenario imply very low star formation efficiencies within the material in the CMZ, specifically adopting the level $\log \mathit{SFE}_{\rm coll}\, \mathrm{[Myr]} =-2$ calibrated in local clouds in $\S$~\ref{sec:MWSFR} and our derived values of $\delta_d$ for each $k$ given in Appendix \ref{sec:appendixdeltad}.  Within the gas volume out to $230$~pc considered by \cite{kruijssen14}, we predict maximum star formation rates of $0.12{-}0.2$ M$_\odot$~yr$^{-1}$ adopting $k=2$ (given the tabulated range of $\kappa$), in good agreement with observations (on average $0.2$~M$_\odot$~yr$^{-1}$~kpc$^{-2}$ across the region; \citealt{kruijssen14} and  \citealt{longmore,barnes17}; see Table~\ref{tab:CMZ}).  Predictions are lowered by a factor of $\sim$ 20 adopting a shallower density profile with $k$=1.5 and would be lowered still further with $k=1$.  

We find similarly good agreement between the predictions of our model and the measured star formation rates in the cloud complex at $l=1\fdg3$, which is a prominent feature in the CMZ. Using the dense gas surface density of the `$1\fdg3$ cloud complex' tabulated by \cite{kruijssen14}, we estimate maximum star formation rates of $\Sigma_{\rm SFR} = 0.2{-}0.4$ M$_\odot$~yr$^{-1}$~kpc$^{-2}$ (again, for the tabulated range of $\kappa$ and adopting $k$=2). 

The model provides a weaker match to the `100~pc stream', where the densest clouds and most of the star formation activity in the CMZ are located.  In this zone, our model predicts star formation rates that overestimate the current observed SFR at that location by factors of $4{-}8$ when $k$=2 is adopted and $2{-}5$ even when the shallower profile with $k$=1.5 is adopted (see Table~\ref{tab:CMZ}).  However, this prediction should be considered in the context of models for star formation in the CMZ environment, which predict that the CMZ is currently experiencing a star formation lull during an episodic cycle \citep[e.g.,][]{kruijssen14,krumholz17}.  The incipient SFR may therefore be higher than the one currently observed.  Indeed, the model predicts the SFR expected given the current state of the gas, whereas SFR measurements probe the end result of previously existing gas clouds.

In general, our model provides a compelling interpretation of the low observed star formation rates in the (outer) CMZ that several other theoretical estimates (i.e., tabulated by \citealt{longmore}) fail to match by as much as a factor of $10$.  Our model better complements two lines of recent work that incorporate an increase in the density threshold for star formation in the CMZ as a result of high turbulent pressure \citep[e.g.,][]{rathborne14b,federrath16} and account for the impact of turbulence driven by shear-driven acoustic instabilities \citep[e.g.,][]{montenegro99,kruijssen14}, which have been found to be sufficient to explain the observed gas velocity dispersions in this region (\citealt{krumkruij15}; \citealt{krumholz17}).  In our model, the background galaxy is responsible both for in-plane shear associated with epicyclic motions (which also drives the acoustic instability) and vertical motions that together establish an effective gas pressure.  This elevated pressure, together with the high gas densities characteristic of a strong background potential, raises the density threshold for star formation in the CMZ. 

An obvious direction to improve the model would be to explicitly incorporate turbulence driven by acoustic instabilities \citep{montenegro99}, which should better capture the instantaneous dynamical state of the gas, e.g., in the outer `100~pc stream'.  At present, predictions more appropriately describe the initial conditions imprinted on the gas in this zone when it becomes self-gravitating \citep{molinari11,longmore13b,kruijssen15}. This event marks the onset of the evolution toward the actively star-forming phase in the cycle (for a recent review on the duty cycle in the CMZ, see \citealt{kruijssen17}).  

Despite omitting these details, our model is consistent with numerical simulations of the clouds orbiting on the `100~pc stream', which show that the observed structural properties and kinematics of the clouds (such as the velocity gradient of the Brick) can be attributed to shear \citep{kruijssen19}.  This consistency is an important validation of our model, which contributes a generalization of an essential part in this broader picture and emphasizes the role of the host galaxy on the dynamical state of the gas in other environments as well. 

\subsubsection{Variations in SFE throughout galaxies}\label{sec:sfedenseobs}
\begin{figure}[t]
\begin{center}
\includegraphics[width=1.0\linewidth]{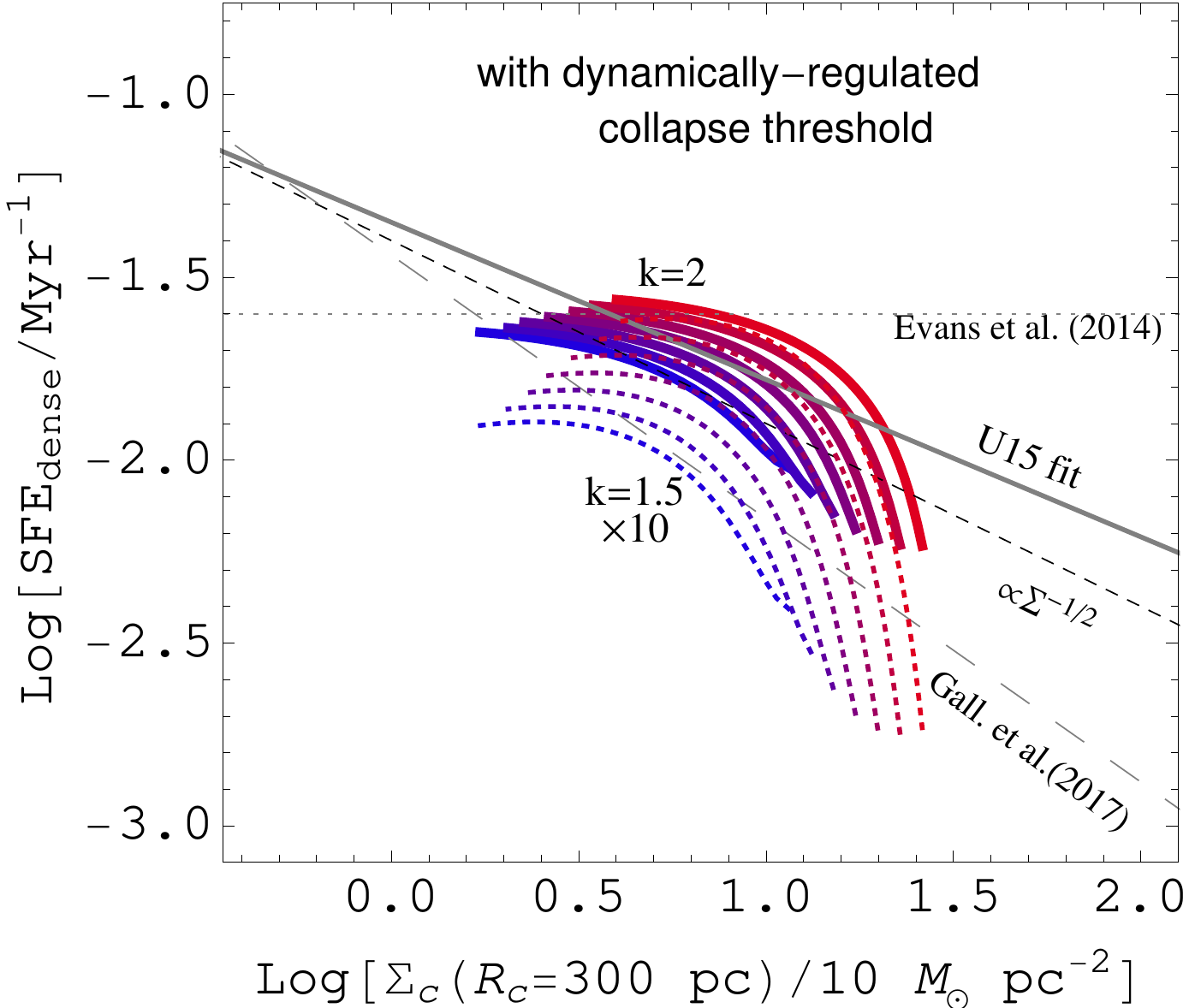}
\caption{
Variation in the dense gas star formation efficiency $\mathit{SFE}_{\rm dense}$ with the molecular gas surface density $\Sigma_{\rm H_2}$ on scale $R_c$= 300 pc.  Predictions from semi-empirical global galaxy models are shown as two sets of colored lines, assuming power-law density distributions with either $k=2$ (solid) or $k=1.5$ (dotted), with color-coding by galaxy stellar mass in the range $9.25<\log M/\mathrm{M}_\odot<10.75$ in steps of $0.25 \log M/\mathrm{M}_\odot$ (from low to high, blue to red). Curves extend out to $2.5R_{\rm e}$, enclosing the brightest molecular emission in typical nearby star-forming disks \citep{schruba}.  Predictions for $k$=1.5 are scaled up by a factor of 10.  The gray dotted horizontal line marks the efficiency $\log \mathit{SFE}\, \mathrm{[Myr]} = -1.6$ measured by \citet{evans14} in the high density star-forming gas in local clouds.  The observed relation between $\mathit{SFE}_{\rm dense}$ and $\Sigma_{\rm H_2}$ measured with an average beam size of $1.5$~kpc by Usero et al. (2015; U15) is shown as a gray solid line while the trend implied by the relation identified on similar scales in the inner regions of a similar set of galaxies by \cite{gallagher} is shown as a gray dashed line. The black dashed line highlights the overall generic trend predicted by our model described in the text.  The increasing limit to collapse in molecular gas imposed by the strengthening of cloud-scale orbital motions at smaller galactocentric radius (coinciding with an increase in the strength of the potential and $\Sigma_{\rm H_2}$) leads to a decrease in the star-forming fraction of the gas and thus a drop in $\mathit{SFE}_{\rm dense}$ with increasing $\Sigma_{\rm H_2}$. \vspace*{0.35cm}}
\label{fig:sfeplot}
\end{center}
\end{figure}

The previous sections demonstrate that star formation on the cloud scale is made inefficient by the suppression of star formation in the cloud envelope, where galactic motions keep the gas weakly self-gravitating.  The onset of free-fall collapse is limited to high density material, leading to more star formation per unit mass in dense gas compared to lower density gas, as exhibited by clouds in the Milky Way (see previous section).  

From this perspective, variations in the star formation efficiency on cloud-scales are largely related to variations in the amount of material in the weakly self-gravitating envelope or to variations in the amount of bound, strongly self-gravitating gas (see also \citealt{ostriker}).  However, even dense gas probes may exhibit varying levels of star formation efficiency, depending on the match between the threshold $\rho_{\rm coll}$ and, e.g., the critical density of the tracer (or how much of the non-collapsing envelope material is also traced).  

In our formalism, this is quantified as a deviation of the ratio $M_{\rm coll}$/$M_{\rm d}$ from unity.  Since $M_{\rm coll}$ depends on the relative strengths of gas self-gravity and the galactic potential, which vary from environment to environment, the star formation efficiency per unit time measured in the dense gas, $\mathit{SFE}_{\rm dense}$, may also reflect the decoupling of gas from the galaxy and exhibit characteristic variations with other globally varying gas properties.  

As an illustration of this behavior, below we highlight how $\mathit{SFE}_{\rm dense}$ is predicted to vary in our semi-empirical global galaxy models (see $\S$~\ref{sec:globalgalaxymodels}).  For this exercise we assume that the decoupled, collapsing gas forms stars with a universal efficiency per free-fall time, as discussed in $\S$~\ref{sec:anchor}.  This allows us to relate variations in the SFE on different scales within clouds to the influence of the galactic bottleneck. 

For each galaxy model, we estimate $\mathit{SFE}_{\rm dense}$ according to 
\begin{equation}
\mathit{SFE}_{\rm dense}= \frac{f_{\rm c} \mathit{SFE}_{\rm c}}{f_{\rm d}} \label{eq:sfedensemodel}
\end{equation}
where $f_{\rm d}$ is the dense gas mass fraction, $f_{\rm c}$ is the cloud fraction by mass, and $SFE_{\rm c}$ is the star formation rate in Eq.~\eqref{eq:sfmodel} per unit gas mass on some scale $R_{\rm c}$.
 
Given the values of $\gamma$ typical on cloud scales in our empirical cloud and galaxy models, the star formation proceeds in the weakly self-gravitating regime, reducing Eq.~\eqref{eq:sfedensemodel} to
\begin{eqnarray}
\mathit{SFE}_{\rm dense} &=& \frac{f_{\rm c} \delta_d\mathit{SFE}_{\rm coll}}{f_{\rm d}} \frac{\gamma_{\rm coll}^{2(k-3)/k}}{t_{\rm coll,on}/t_{\rm ff,coll}}\left(\frac{\rho_{\rm coll}}{\Sigma_c/(2R_{\rm c})}\right)^{(k-3)/k}\nonumber\\
 &=& \frac{f_{\rm c} \delta_d\mathit{SFE}_{\rm coll}}{f_{\rm d}}\frac{\gamma_{\rm coll}^{2(k-3)/k}}{t_{\rm coll,on}/t_{\rm ff,coll}} \nonumber \\
 & & \times \left(\frac{2\pi (3a_k/5) G\Sigma_{\rm c} R_{\rm c}^{-1}}{ \kappa^2+2\Omega^2+\nu^2}\right)^{(3-k)/k}
\label{eq:sfedense2}
\end{eqnarray}
In this expression, a fraction $f_c$ of the gas is assumed to be arranged into clouds with characteristic size $R_{\rm c}$, with the density on scales smaller than $R_{\rm c}$ distributed according to the nominal internal density distribution $\rho\propto r^{-k}$.  For the predictions in this section we consider power-laws with either $k=1.5$ or $k=2$.  

The collapsing mass fraction below scale $R_c$ represented by the factor in parentheses is estimated as in $\S$~\ref{sec:massfrac}, using the $\kappa$ and $\nu$ implied by the rotation curve (and stellar scale height) at a given galaxy mass together with our molecular gas surface density model $\Sigma_{\rm H_2}$.  To match the scales probed by the extragalactic observations of U15 and \cite{gallagher} we adopt $R_c = 300$~pc.  Our model for $\Sigma_{\rm H_2}$ provides a good match to the gas surface density measurements on this scale.  

The factor $\gamma_{\rm coll}^{2(k-3)/k}/(t_{\rm coll,on}/t_{\rm ff})$ in Eq. (\ref{eq:sfedense2}) is a constant factor that depends, for a given $k$, on the properties of collapse derived in $\S$~\ref{sec:collapsetime}.  In principle, the dense gas fraction $f_{\rm d}$ also follows from the assumed density distribution.  In practice, we estimate the cloud-scale $f_{\rm d}$ using the empirical relation between the dense gas fraction $\Sigma_{\rm dense}/\Sigma_{\rm H_2}$ and $\Sigma_{\rm H_2}/\Sigma_{\rm HI}$ found on ${\sim}1$~kpc scales by U15, which is better consistent with the nominal $k=2$ model \citep{meidt16}.  To assign a value of $\Sigma_{\rm dense}/\Sigma_{\rm H_2}$ at a given $\Sigma_{\rm H_2}$ we assume that $\Sigma_{\rm HI}=10$ M$_{\odot}$~pc$^{-2}$, appropriate for massive galaxies with approximate solar metallicity \citep{schruba18}.  To further anchor our predictions, we set the efficiency $\log\mathit{SFE}_{\rm coll}\,\mathrm{[Myr]} = -2\pm 0.03$ to the average value measured in $\S$~\ref{sec:MWSFR} in the local clouds studied by \citetalias{veh16}.  

Figure~\ref{fig:sfeplot} shows the variation of $\mathit{SFE}_{\rm dense}$ predicted by Eq.~\eqref{eq:sfedense2} as a function of molecular gas surface density $\Sigma_{\rm H_2}$ assuming $f_{\rm c}=0.5$.  Each line represents how $\mathit{SFE}_{\rm dense}$ and $\Sigma_{\rm H_2}$ are expected to vary in a galaxy with stellar mass in the range $9.25 < \log{M_{\star}/\mathrm{M}_\odot} < 10.75$.  For these predictions we assume that the distribution of recent star formation has the same filling factor as the dense gas.  

Two sets of lines the highlight predictions with either $k=1.5$ or $k=2$ assumed for the density distribution below $R_{\rm c}$.  Specific to each $k$ we adopt the values derived for $\delta_d$ in Appendix \ref{sec:appendixdeltad}.    
Since we expect the prediction assuming $k=2$ to be more internally consistent with our adopted prescription for $f_{\rm d}$, we will focus our discussion on this below, and comment on the prediction for $k=1.5$ at the end of the section.

\begin{table*}
\begin{center}
\caption{Variation in $\mathit{SFE}_{\rm dense}$ due to parameter choices.}\label{tab:params}
\begin{threeparttable}
\begin{tabular}{rcc}
Parameter & Fiducial Value & Spread in $\log \mathit{SFE}_{\rm dense}$ (dex)\\
\tableline
$\log \mathit{SFE}_{\rm coll}\,\mathrm{[Myr]} $ \tnote{a}& $-2\pm0.2$ & 0.2 \\
$f_{\rm d}(x=\Sigma_{\rm H_2}/\Sigma_{\rm HI})$\tnote{b} & 10$^{-1.46\pm0.04}x^{0.29\pm0.04}$ & 0.03 \\
measurement scale $R_{\rm c}$ [pc] & $300\pm100$ & 0.1 \\
clumping factor $c$\tnote{c} & $1-2$ & 0.16 \\
$\delta_d$\tnote{d} & 1.33 - 3.5 & 0.4  \\
\hline 
stellar mass ($\log M_{\odot}$) & $10\pm0.75$ & 0.17 \\
spiral $K$\tnote{e}& \nodata &0.23 
\end{tabular}
 \begin{tablenotes}
\item[a]Adopting the empirical value determined in $\S$~\ref{sec:MWSFR}
\item[b]Based on the \cite{usero} fit to observations of dense gas in nearby galaxies.
\item[c]This factor specifies the scaling between the surface density model on scale $R_c$ and the global molecular gas distribution implied by galaxy scaling relations (see Paper~I), i.e. $\Sigma_c$=$c\Sigma_{H2}$.  The adopted range of values is chosen so that the model matches observations made with a 0.5-1~kpc beam presented by U15 and \cite{gallagher}. Larger values would be required for smaller $R_c$.  
\item[d]The fiducial range spans approximate values for power-law profiles with index $k$ in the range 1-2 (see Appendix \ref{sec:appendixdeltad}).  
\item[e]This incorporates an increase in the local epicyclic frequency in the presence of an independently rotating spiral (or bar) pattern far from corotation, as given in Paper~I.  
 \end{tablenotes}
\end{threeparttable}

\end{center}
\end{table*}

\subsubsection{Characteristic decline of \texorpdfstring{$\mathit{SFE}_{\rm dense}$}{SFE(dense)} toward dense regions}\label{sec:trendssfe}
As illustrated by the trends in Figure~\ref{fig:sfeplot}, the galactic bottleneck model reproduces the  decline in $\mathit{SFE}_{\rm dense}$ with increasing $\Sigma_{\rm H_2}$ observed on large scales in nearby galaxies \citep[U15][]{gallagher, q19,jd19}.  The model further captures---with essentially no tuning---the low level of star formation observed overall.  Four factors: $\gamma_{\rm coll}^{2(k-3)/k}/(t_{\rm coll,on}/t_{\rm ff})$, $\mathit{SFE}_{\rm coll}$, $\delta_d$ and $f_{\rm d}$ directly determine the normalization of $\mathit{SFE}_{\rm dense}$ while another three indirectly influence the star formation rate (described in $\S$~\ref{sec:scatterSFEdense}).  The last three of these factors are chosen based on an empirical measurement or motivated by observations.  The first of these factors, on the other hand, is directly predicted by our model for collapse.  The closeness with which our predictions for $\mathit{SFE}_{\rm dense}$ match the low levels observed thus mostly serves as validation of this factor. As will be discussed further below, however, given our adopted $SFE_{\rm coll}$ and $f_d$, this match applies only to a restricted range of internal density power-law indices, since different values of $k$ can yield a wide range of SFEs in equal mass clouds \citep[see Appendix \ref{sec:appendixdeltad} and, e.g.,][]{burk18,parm}.  Note that, to match the observations, the factor $\delta_d$ for the $k$=2 case can not be much higher than the value $\delta_d$=3.5 assumed in the figure (see Appendix \ref{sec:appendixdeltad}), unless  $\mathit{SFE}_{\rm coll}$ - and $\epsilon$, in particular --  is not also lower than calibrated in $\S$ \ref{sec:MWSFR}.  

In the remainder of this section we focus on the systematic variation in SFE predicted by the bottleneck model.  The characteristic decline in $\mathit{SFE}_{\rm dense}$ with increasing $\Sigma_{\rm H_2}$ in the case of either $k=1.5$ or $k=2$ owes to the condition $\sigma_{\rm sg}/\sigma_{\rm gal}\approx1$ typical in the inner molecule-rich disks of galaxies (see $\S$~\ref{sec:motions}).  As a result, overall Eq.~\eqref{eq:sfedense2} with $k=2$ implies $\mathit{SFE}_{\rm dense}\propto SFE_{\rm coll} f_{\rm d}^{-1}$.   Since we have assumed the dense gas fraction $f_{\rm d}$ varies with galactic environment as $f_{\rm d}\propto\Sigma_{\rm H_2}^{1/2}$ (U15 and see \citealt{meidt16}), the predicted $\mathit{SFE}_{\rm dense}$ overall falls off roughly proportionally to $\Sigma_{\rm H_2}^{-1/2}$.  

However, the model also predicts curvature away from the $\Sigma_{\rm H_2}^{-1/2}$ line.  This stems from variation in the coupling of the dense gas to galactic environment encoded in the systematic (if modest) variation of the factor $\sigma_{\rm sg}/\sigma_{\rm gal}$, which follows from the faster rise in the strength of the external potential compared to $\Sigma_{\rm H_2}$ toward galaxy centers.  
This same behavior leads to the variation in $M_{\rm coll}/M_{\rm c}$ predicted on the cloud scale in Figure~\ref{fig:rsgplot0}.\footnote{It should be noted that the prominence of the curvature depends on the CO-to-H$_2$ conversion factor.  In Paper~I, we predicted that the CO-to-H$_2$ conversion factor varies with galactocentric radius, according to the change in the internal gas kinematics with $R_{\rm gal}$ resulting from variation in the balance between self-gravity and the external potential  (which also regulates the collapsing fraction).  The sense of the variation would tend to minimize the curvature away from the $\Sigma_{\rm H_2}^{-1/2}$ line in Figure~\ref{fig:sfeplot}.}  

As a result of this additional coupling to galactic environment in the model, the predicted relation between SFE and $f_{\rm d}$, and thus also $\Sigma_{\rm H_2}$, is environment-dependent.  In galaxy centers, the SFE falls off more rapidly with increasing $\Sigma_{\rm H_2}$ than in the main disk environment.  Observations suggest that this sort of environmental dependence may be recognizable.  The relation between $\mathit{SFE}_{\rm dense}$ and $\Sigma_{\rm H_2}$ fitted to the inner regions of nearby galaxies by \citet{gallagher} is steeper than the relation fitted by U15, who targeted regions sampling further out in the disk than the slightly higher resolution maps studied by \cite{gallagher} covering the inner $3{-}5$~kpc; of the $62$ regions sampled by U15, $1/3$ are located at radii beyond $4$~kpc.  It is worth noting, though, that strong bar and spiral features may tend to degrade the strength of the trends in Figure~\ref{fig:sfeplot} predicted in the case the axisymmetric disk models adopted there. 

The importance of the environmental coupling for the relation between SFE and $\Sigma_{\rm H_2}$ also underlines the sensitivity to the gas density distribution below scale $R_{\rm c}$, as highlighted in the Figure~\ref{fig:sfeplot}.  As previously noted, for a fixed threshold $\rho_{\rm coll}$, shallower profiles have a larger cloud envelope of decoupled material, reducing the star formation efficiency measured on scale $R_{\rm c}$.  The scaling factor $\delta_d$ is also lower in such cases \citep[see e.g.][and Appendix \ref{sec:appendixdeltad}]{tan06,parm}.  This is responsible for the offset toward lower SFEs predicted by the model with $k=1.5$ in Figure~\ref{fig:sfeplot}, which becomes more notable when paired with the exceptionally elevated collapse threshold characteristic of galaxy centers.  We emphasize, however, that the behavior highlighted in the figure is likely an exaggeration considering that a sharp decrease in cloud scale collapse fraction would normally be compensated by a matched decrease in dense gas fraction.  At present, however, the predictions for $k=1.5$ shown in Figure~\ref{fig:sfeplot} adopt the same empirical model for $f_{\rm d}$ used for the $k=2$ predictions, as well as the same $\epsilon$ (and $SFE_{\rm coll}$).

As will be addressed more in the next section, comparisons between the model and observations can be used to place constraints on the degree of environment-dependent variations possible in several of the factors that are assumed to be universal in the current set of model predictions (or, independent of processes on the outer cloud scale and beyond).  
Thus it is important to emphasize that predictions for $\mathit{SFE}_{\rm dense}$ to be compared to observations should be derived based on the observed structure of the galaxy potential and distribution of molecular gas, as well as the expected properties (size, surface density) of the cloud population, which are only approximated by the semi-empirical model of galaxy morphology and dynamics invoked here.  We expect that variations in cloud size (as predicted across galaxy disks in $\S$~\ref{sec:rsgpop1}, but not assumed in Figure~\ref{fig:sfeplot}) will restrict the curvature in $\mathit{SFE}_{\rm dense}$ vs.\ $\Sigma_{\rm H_2}$ to galaxy centers (high $\Sigma_{\rm H_2}$), for example. 

\subsubsection{Variations in \texorpdfstring{$\mathit{SFE}_{\rm dense}$}{SFE(dense)} due to parameter choices}\label{sec:scatterSFEdense}
Despite incorporating only the basic gravitational factors influencing the organization and kinematics of molecular gas, our model for the coupling of clouds to their environment captures the broad behavior of SFEs observed at or beyond the cloud scale across galaxy disks.  This would tend to suggest that additional factors that could introduce changes in, e.g., internal cloud structure, the onset of self-gravitation or $\epsilon$ are not strongly environment-dependent.  These factors may, however, be responsible for overall shifts toward lower or higher SFE or lead to scatter in the observational trends highlighted in Figure~\ref{fig:sfeplot} which would reflect the importance of the regulation of the star formation process on small scales.  Deviations between our model and  observations could provide a way to constrain the degree to which these factors influence extragalactic observations of molecular gas and star formation.  

\begin{figure}[t]
\begin{center}
\begin{tabular}{cc}
\includegraphics[width=0.9\linewidth]{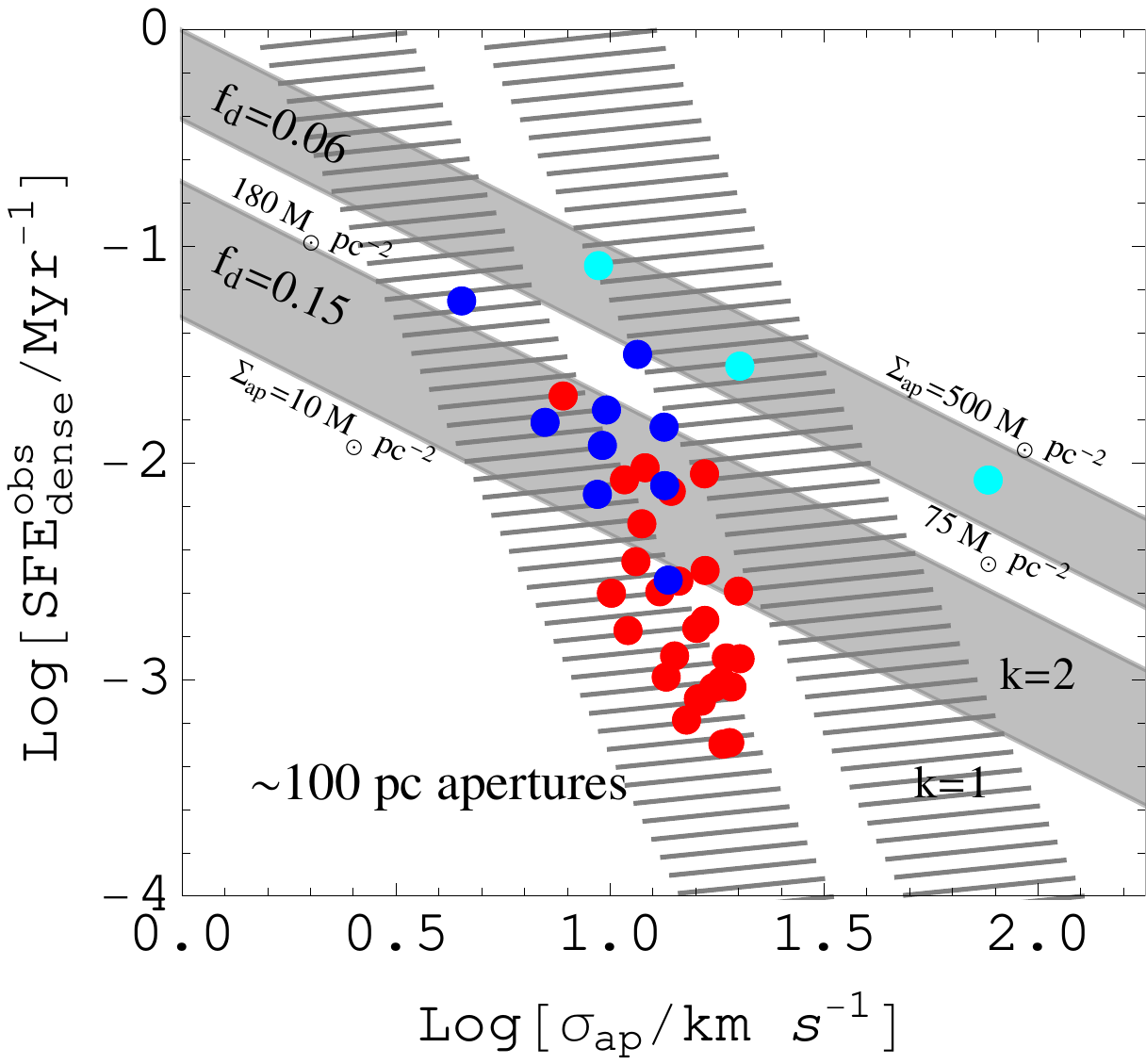}
\end{tabular}
\end{center}
\caption{Observed variation in the dense gas star formation efficiency $\mathit{SFE}_{\rm dense}$ with gas velocity dispersion $\sigma_{\rm ap}$ measured in ${\sim}100$~pc apertures throughout NGC~3627 (cyan; probing out to $R_{\rm gal}\sim3$~kpc), M31 (blue; out to $R_{\rm gal}\sim15$~kpc), and M51 (red; out to $R_{\rm gal}\sim3$~kpc) \citep[][and references therein]{q19}.  Gas kinematics on this scale provide a measure of motions in the galactic potential.  Four bands show the decrease in SFE with $\sigma_{\rm ap}$ predicted by Eq.~\eqref{eq:SFEvdisp} assuming either $k=2$ (solid gray) or $k=1$ (hashed gray), at two different (fixed) levels of the dense gas fraction $f_d$ and a range in gas surface density on $100$~pc scales, $\Sigma_{\rm H_2,ap} = 10{-}180$ M$_\odot~{\rm pc}^{-2}$ (bottom, left) and $\Sigma_{\rm H_2,ap} = 75{-}500$ M$_\odot~{\rm pc}^{-2}$ (top, right).  The lower/left trend for each $k$ is intended to span the observed ranges in M33 and M51 \citep{q19} while the upper/right trend is meant to match observations of the dense gas in NGC~3627 \citep{Murphy} and assumes the relevant aperture size of $300$~pc.  
 }
\label{fig:SFEdata}
\end{figure}

For this purpose, in Table~\ref{tab:params} we tabulate the variation in $\mathit{SFE}_{\rm dense}$ (at fixed galaxy mass) characteristic of the main disk environment of galaxies given the variation of all adopted parameters in Eq.~\eqref{eq:sfedense2} within their measured (or otherwise realistic) ranges.  This includes uncertainties in the assumed $\mathit{SFE}_{\rm coll}$, as well as realistic ranges in the scaling $\delta_d$, the clumping factor $c$\footnote{We emphasize that the clumping factor $c$ in the predictions in Figure \ref{fig:SFEdata} has been chosen to match the molecular gas surface density predicted by global galaxy scaling relations to extragalactic observations probing hundreds of parsecs.  We find that $c=1$ is sufficient but note that measured variations in the adopted galaxy scaling relations could accommodate the higher clumping factors measured in molecular gas by \cite{leroyclump}.} and cloud radius $R_{\rm c}$.  Note that, to generate realistic predictions with our `global galaxy models', the choice of $c$ here is coupled to the adopted $R_c$. Generally, however, a larger $R_c$ will lower the SFE, while a higher clumping factor will raise it. 

According to our model, variations in the rotation curve shape and normalization can 
introduce as much as ${\pm}0.17$~dex scatter in $\mathit{SFE}_{\rm dense}$ at fixed surface density throughout a survey of galaxies with stellar masses in the range $9.25<\log{M_{\star}/\mathrm{M}_\odot}<10.75$.  The relation fitted by U15 to measurements throughout the disks of $29$ galaxies with a similar range in galaxy mass exhibits slightly less scatter ($0.07$~dex; U15.  The data could vary less than predicted given changes in the CO-to-H$_2$ conversion factor (not accounted for in Figure~\ref{fig:sfeplot}) or as a result of the specific locations of the measurements within each galaxy and the true mass distributions and rotation curve shapes of the surveyed galaxies.  

Systematic variations in $\kappa$ due to bar or spiral perturbations in the stellar disk, which become stronger in more massive disks, could also impact the spread between galaxies of different masses.  As modeled in Paper~I, the increase in $\kappa$ reduces the collapsing mass fraction of clouds, which in turn further suppresses star formation.  Using our approximation for $\kappa$ far inside the corotation radius of a density spiral pattern, $M_{\rm sg}/M_{\rm c}$ is reduced by a factor $0.23$~dex, systematically shifting $\mathit{SFE}_{\rm dense}$ downward by this amount.  If stellar dynamical features are preferentially associated with more massive galaxies, this could reduce the range spanned by $\mathit{SFE}_{\rm dense}$.  However, if all of the dense gas in the surveyed galaxies arises from molecular gas that preferentially populates spiral arms and inner bar features, the $\mathit{SFE}_{\rm dense}$ measured  throughout all galaxies would shift downward.  As indicated by Table~\ref{tab:params}, the reduced $M_{\rm sg}$/$M_{\rm c}$ expected in the case that the epicyclic frequency in the presence of a bar or spiral perturbation (Paper~I) is more appropriate overall than $\kappa$ for an axisymmetric disk could still yield $\mathit{SFE}_{\rm dense}$ consistent with the observations given a value for $\epsilon$ at the high end of the observed range.   

\subsection{Variations in SFEs with velocity dispersion}\label{sec:SFEvdisppred}
The link between $\mathit{SFE}_{\rm dense}$ and gas surface density highlighted in the previous section emerges from the strong dependence of both quantities on galactocentric radius, given the way mass is distributed in galaxy disks and from the way gas is typically organized and distributed.  
Several other galaxy properties exhibit strong radial dependencies, including the stellar mass surface density and ISM pressure.  In this section we cast the environmental variation predicted for molecular gas SFEs specifically in terms of gas velocity dispersion, which is another property observed to vary strongly throughout the molecular gas disks of galaxies.  Most recently, velocity dispersions on the cloud scale have been tightly linked to molecular gas surface density \citep{sun}.  Together with the predictions in the previous section, this implies that the SFE should be found to decrease with increasing velocity dispersion.  The novelty of our model in this context is the ability to describe how this behavior depends on the spatial scale being probed (highlighted in $\S$~\ref{sec:SFEvdisp}) as the boundedness of gas changes.  

We demonstrate this here, focussing on the large-scale behavior of the relationship between SFE and $\sigma$, which is arguably the most straightfoward prediction of the galactic bottleneck model.  Compared to eq.~(\ref{eq:sfedense2}), which requires a reliable description of galactic motions (and thus accurate models for $\kappa$ and $\nu$)\footnote{Several approaches to observationally estimate $\kappa$ exist, e.g.\ using the rotation curve derived either directly from the observed velocity field or from the inferred distribution of mass (gas, stellar and dark matter).  However, without detailed modelling, these types of estimates do not incorporate the local enhancement to $\kappa$ in the presence of bar and spiral perturbations, as approximated in Paper I, and are otherwise subject to the sometimes large systematic uncertainties inherent in both types of rotation curve models due to the presence of non-axisymmetric features (e.g.\ \citealt{deBlok}; \citealt{meidt}).}, the expressions derived in $\S$~\ref{sec:SFEvdisp} rely on velocity dispersions at or near the cloud scale as the more direct probe of the galactic motions present on these scales within molecular gas.  

Figure~\ref{fig:SFEdata} illustrates the relation (given by Eq.~\eqref{eq:SFEvdisp}) predicted between $\mathit{SFE}_{\rm dense}$ and molecular gas velocity dispersion $\sigma_{\rm ap}$ measured by apertures probing ${\sim}100$~pc scales.  The molecular gas in this example is assumed to be organized into clouds with radii $R_{\rm c}=30$~pc and two different internal power-law distributions, $k=1$ or $k=2$, below $R_{\rm c}$.  All clouds in a given aperture are further assumed to have fairly uniform properties, sharing the same internal structure and gas surface density (as assumed in the derivation of Eq.~\eqref{eq:SFEvdisp}). 

The overall trend between $\log{\mathit{SFE}_{\rm dense}}$ and $\log{\sigma_{\rm ap}}$ is a linear relationship, with slope that depends on $k$.  For each value of $k$ two trends are shown, each with a unique value of $f_d$.   According to Eq. (\ref{eq:SFEvdisp}), $\mathit{SFE}_{\rm dense}\propto\sigma_{ap}^{-1}$ when $k=2$ and $\mathit{SFE}_{\rm dense}\propto\sigma_{\rm ap}^{-4}$ in the case that $k=1$.   In both cases, increasing gas velocity dispersion tracks growing strength in the galaxy potential that increasingly limits self-gravitation and collapse in molecular gas.  The result is a decrease in the star-forming fraction of the gas and thus a drop in $\mathit{SFE}_{\rm dense}$ with increasing $\sigma_{\rm ap}$. This drop is more severe for shallower density profiles.   

The level reached by $\mathit{SFE}_{\rm dense}$ in any environment (for a given $k$ and level of $f_{\rm d}$) also depends on gas surface density (as indicated by the width of each bar).  A larger amount of high density gas is expected to be collapsing than lower density gas in a dynamically similar location, increasing the amount of gas that gets converted into stars, as described in $\S$~\ref{sec:massfrac}.

\subsubsection{Comparison to observations}
Observations on ${\sim}100$~pc scales reveal similar behavior.  Individual data points in Figure~\ref{fig:SFEdata} represent measurements of $\mathit{SFE}_{\rm dense}$ from a number of regions throughout the inner disks of a handful of normal star-forming galaxies compiled and measured by \citet{q19}.  These assume the standard constant Galactic CO-to-H$_2$ conversion factor and HCN-to-dense gas conversion factor adopted by U15 to convert CO and HCN luminosities to surface densities.  All surface densities are corrected to the galaxy plane by including a factor $\cos(i)$.  The contribution from self-gravity to the observed motions in all of the 100~pc apertures is confirmed to be negligible.  

Overall, the measurements exhibit a clearly decreasing trend between $\mathit{SFE}_{\rm dense}$ and molecular gas velocity dispersion $\sigma_{\rm ap}$ on ${\sim}100$~pc scales \citep{q19}.  The decrease is similar to the behavior exhibited when turbulence on $100$~pc scales is driven by SNe \citep{padoan12,padoan17}.\footnote{Based on the star formation properties of cloud-scale simulations in which turbulence is driven by SNe explosions, \citet{padoanRev,padoan17} propose an empirical model for the efficiency per free-fall time that varies as $\exp{(-1.6 t_{\rm ff}/t_{\rm dyn})}$ where $t_{\rm dyn} = R\,\sigma^{-1}$ across a region of size $R$.}

In both the \cite{padoan17} and the bottleneck models, the scatter in the observations can be partially related to a residual dependence on gas surface density (represented by the width of the bands in Figure~\ref{fig:SFEdata}); higher density gas (with a shorter free-fall time) can form stars more efficiently in a given (turbulent) environment.  On the other hand, because $\Sigma_{\rm H_2}$ varies systematically together with $\sigma_{\rm gal}$ (and $f_{\rm d}$) throughout galaxies, the result is more likely steepening of the overall trend.  This could be responsible for the steeper-than-linear trend traced out by the measurements in the nearby galaxies M33 and M51 in the case that $k=2$, although a model with $k=1$ also provides a good qualitative match.  

Position in the $\mathit{SFE}_{\rm dense}$ vs.\ $\sigma$ parameter space is also sensitive to the dense gas fraction $f_{\rm d}$ of any given cloud or region (see Eq.~\eqref{eq:SFEvdisp}), which we expect to vary not only between galaxies but also within them (U15; \citealt{gallagher}); for a fixed $\rho_{coll}$, more of the dense gas in a cloud is likely to be collapsing when it represents only a small portion of the cloud.  In the context of the model, a higher dense gas fraction in M51 compared to M31 could lead to overall lower SFEs in the former galaxy, assuming that the gas in both systems follow a power-law distribution with $k=2$.  This might also be responsible for the high offset $\mathit{SFE}_{\rm dense}$ at fixed $\sigma_{\rm ap}$ in the strongly barred galaxy NGC 3627, which has overall lower dense gas fractions compared to the other systems.  

In practice, the shape of the trend between $\mathit{SFE}_{\rm dense}$ and $\sigma$ is driven by systematic variation in several other properties assumed in Eq.~\eqref{eq:SFEvdisp}, including the cloud fraction and the internal density distribution below scale $R_{\rm c}$.  For the present study, our constraints on these properties have been motivated by Galactic studies but in the immediate future PHANGS will reveal whether these vary systematically within and between galaxies.  Variations in these properties might thus either introduce scatter to the observed relation between $\mathit{SFE}_{\rm dense}$ and $\sigma_{\rm ap}$, or lead to a non-linear relationship. 

It is important to note that the dependence of $\mathit{SFE}_{\rm dense}$ on $f_{\rm d}$ can itself yield an apparent trend between $\mathit{SFE}_{\rm dense}$ and $\sigma_{\rm ap}$ when $f_{\rm d}$ and $\sigma_{\rm ap}$ vary tightly together throughout galaxy disks.  This might occur on small scales where the gas is already strongly self-gravitating, so that $\sigma_{\rm ap}$ decreases together with the exponentially declining gas surface density (e.g. \citealt{sun}).  Coupled with the exponential decrease in stellar surface density $\Sigma_\star$ and $f_{\rm d}$ toward larger $R_{\rm gal}$, $\sigma_{\rm ap}$ would appear to vary with $\mathit{SFE}_{\rm dense}$. In this regime, $\sigma_{\rm ap}$ is not tracing galactic motions (but the gas self-gravity), in which case Eq.~\eqref{eq:SFEvdisp} predicts that galactic motions have negligible impact on $\mathit{SFE}_{\rm dense}$ compared to the dependence on $f_d$.  

In practice, identifying whether $\mathit{SFE}_{\rm dense}$ varies most strongly with $f_{\rm d}$ or $\sigma_{\rm ap}$ may be difficult, as it depends sensitively on the measurement scale, and specifically whether the aperture samples the scale at which gas is decoupled from the galaxy so that it can collapse.  To avoid this ambiguity, we propose that a more straightforward test of the model is the relation between $\sigma_{\rm ap}$ and the star formation efficiency in molecular gas $\mathit{SFE} = f_{\rm d} ~\mathit{SFE}_{\rm dense}$ on cloud scales and larger, which is independent of $f_{\rm d}$ (see Eqs.\ \eqref{eq:SFEvdisp} and \eqref{eq:SFEvdispCloud1}). Such a test is now becoming possible with the collection of high and low density kinematic tracers on cloud scales across a variety of galactic environments being assembled by PHANGS.  These measurements will be key to characterizing the strength of the relation between $\mathit{SFE}$ and $\sigma_{\rm ap}$ across a range of spatial scales (and densities) and for distinguishing between different models for the development of turbulent motions. 

\section{Discussion}\label{sec:discussion}
\subsection{The molecular gas depletion time of galaxies}\label{sec:depletiontimes}
The bottleneck model divides inefficient star formation into two sources: i/a large fraction of non-star-forming gas, which is kept weakly self-gravitating by the role of galactic motions on large scales and ii/a lower-than-unity conversion efficiency $\epsilon$ in the gas that is decoupled from the galaxy and able to collapse, as a result of the turbulent properties of gas, the impact of feedback, and the influence magnetic forces (i.e.).  On its own, the latter inefficiency slows down the star formation process relative to the free-fall time by an order of magnitude, adopting our empirical calibration for $\epsilon$.  The bottleneck contributes an additional order of magnitude, easily making the difference between the long observed molecular gas depletion times $\tau_{\rm dep} = M_{\rm H_2} / \mathit{SFR} \approx2$~Gyr (\citealt{bigiel}; \citealt{leroy2013}) and the short ${<}10$~Myr free-fall times in molecular gas.  

The systematic trends in SFE throughout galaxies described by the bottleneck model have further implications for the way galaxies consume their gas.   At the same time as $\mathit{SFE}_{\rm dense}$ decreases with increasing $\Sigma_{\rm H_2}$ (as explored in $\S$~\ref{sec:sfedenseobs}), the dense gas fraction $f_{\rm d}$ is observed to increase $\propto \Sigma_{\rm H_2}^{1/2}$ (U15; \citealt{bigiel16}; \citealt{gallagher}).  This empirical trend has been associated with a dependence of the dense gas fraction on pressure in the ambient ISM \citep{usero,meidt16,gallagher,jd19}, which is empirically linked to the H$_2$-to-HI ratio.  As a result of these nearly reverse dependencies, our model predicts that the SFR per unit molecular gas mass is maintained at a roughly constant level, independent of $\Sigma_{\rm H_2}$ \citep{usero}.  This would lead to an approximately linear molecular gas Kennicutt-Schmidt star formation relation consistent with what is observed, and the corresponding rough universality of the molecular gas depletion time $\tau_{\rm dep}$ (\citealt{bigiel}; \citealt{leroy2013}).  

However, given that $\mathit{SFE}_{\rm dense}$ is predicted to vary in detail more fundamentally with (radially and azimuthally) varying dynamical quantities than with $\Sigma_{\rm H_2}$, the model more precisely predicts residual variation in $\tau_{\rm dep} = \mathit{SFE}^{-1}$.  The variation is qualitatively similar to the systematic variations in $\tau_{\rm dep}$ that have been shown by \cite{saintonge} and \cite{leroy2013} to correlate with global galaxy properties and local dynamical conditions \citep[see also][]{meidt, leroy2016, leroy2017m51, utomo17}.  

Since $\mathit{SFE}_{\rm dense}$ may decouple from $\Sigma_{\rm H_2}$ and $f_{\rm d}^{-1}$ in some environments, we favor the more straightforward link between $\mathit{SFE}$ and the dynamics of host galaxies given by Eq.~\eqref{eq:sfedense2} (and as re-expressed in $\S$~\ref{sec:SFEvdisp}) as a compliment to similar quantitative measures for the impact of dynamics on local gas stability already considered \citep[e.g.,][]{leroy08,meidt}.  

Environmental variations in $\tau_{\rm dep}$ emerge in our model from differences in galaxy dynamical structure, such as between the bulge and disk of a given galaxy, or within bars and spiral arms.  It is instructive to express Eq.~\eqref{eq:SFEweak} as 
\begin{eqnarray}
\mathit{SFE}_{\rm c}
=\frac{\delta_d\mathit{SFE}_{\rm coll}}{6}&\delta_\rho&\frac{t_{\rm orb}}{t_{\rm ff}}\left[(1+\beta)\left(1+\frac{R_{\rm gal}}{2 z_0}\right)+1\right]^{-1/2}
\label{eq:sfeETGs}
\end{eqnarray}
where $t_{\rm orb}=2\pi/\Omega$ is the orbital period, $\beta$ is the galactic rotation curve shear parameter, and $\delta_\rho$ is a factor of order unity that depends on the internal distribution of densities within the gas.   
This suggests that 
\begin{eqnarray}
\frac{\tau_{\rm dep}}{t_{\rm ff}}
\approx \left(\frac{\delta_d\mathit{SFE}_{\rm coll}}{6}\right)^{-1}t_{\rm orb}^{-1}\left[(1+\beta)\left(1+\frac{R_{\rm gal}}{2 z_0}\right)+1\right]^{1/2}\label{eq:tdepETGs}
\end{eqnarray} 
At fixed $t_{\rm orb}$, gas occupying a central bulge-dominated region where the rotation curve increases rapidly (i.e., large $\beta$) is predicted to form stars less efficiently than equivalent gas sitting in the disk of a galaxy, where rotational velocities typically flatten out (so that $\beta\approx0$).  The increased opposition to self-gravity possible in environments where dynamical feature like bars raise $\kappa$ (not explicitly incorporate into Eqs.\ \eqref{eq:sfeETGs} and \eqref{eq:tdepETGs}; see section $2.3.5$ in Paper~I) can also lead to less efficient star formation depending on the density of the material in the bar zone (i.e., how strongly self-gravitating it is).  Altogether, these environmental variations would lead to scatter  at fixed $t_{\rm orb}$ in the resolved star formation relation traced by sets of galaxies.  This is also a potential source of scatter in observed global star formation relations at fixed $t_{\rm orb}$ (e.g., \citealt{Daddi}; \citealt{Genzel}), as implied by the global version of Eq.~\eqref{eq:tdepETGs} adopting weighted averaging over all radially varying quantities.   

Likewise, differences in global galaxy dynamical structure, such as between early- and late-type galaxies, are predicted to introduce variations in the star formation rate supported by gas that may otherwise be  structurally, thermally, and chemically identical from galaxy to galaxy.  The molecular gas disks in more massive early-type galaxies with higher maximum rotation velocities (and shorter $t_{\rm orb}$) are predicted to form stars less efficiently than equivalent molecular gas in disky late-type galaxies (with longer $t_{\rm orb}$).  Note that, as described nicely by our empirically-motivated rotation curve model (see Paper~I), the inner rotational velocity gradient typically increases with galaxy stellar mass (Lang et al. subm.).  As a result, more of the molecular gas in massive galaxies tends to sit beyond the initial steep gradient, at radii where the rotation curve is flat.  This leads to lower overall $\beta_{\rm avg}$ measured at radii inside the location of the peak rotational velocity than measured in lower mass systems with more slowly rising rotation curves (for which $\beta_{\rm avg}$ inside the peak is overall higher; see \citealt{davis14}).  In terms of $\beta_{\rm avg}$ (which is to be distinguished from the local $\beta$ that applies at a particular radius), the depletion times predicted by Eq.~\eqref{eq:tdepETGs} are consistent with observations that suggest that gas is consumed more rapidly with increasing $\beta_{\rm avg}$ (\citealt{davis14}; Colombo et al. 2018).  

\subsection{Regulated star formation}
Starting with the picture of gas motions we developed in Paper~I, in this paper we describe how the galaxy acts as a bottleneck to star formation, limiting the rate at which gas can collapse to form stars.  In this picture, the turbulent motions in the molecular gas that regulate star formation are combined with an additional set of (three dimensional) motions that are gravitational in origin, rather than, e.g., purely stellar feedback-driven.  As such, gas self-gravity competes with orbital motions (either in a coherent state or once they have developed into turbulent motions), rather than being supported entirely by feedback-driven turbulence (c.f., \citealt{ostriker} and \citealt{ostrikerShetty}).  In this scenario, the galactic orbital motions acting within clouds set up an effective pressure that by definition balances the weight of the gas in  the galactic gravitational field, which frames those very (gravitational) motions. This makes the effective gas pressure in this case similar in magnitude to the feedback-driven turbulent pressure that must balance the midplane pressure set by the vertical gravitational field in feedback-regulated models.  An important difference, however, stems from the presence of gas motions (and the appearance of virialization) in the present model even in the absence of recent nearby star formation.  In this picture, star formation feedback would also not need to be as strong as predicted in feedback-only models to achieve the same appearance of pressure equilibrium.  

In our model, the galaxy itself helps regulate the star formation rate in cold, dense molecular gas. This regulation stems from the dynamical properties of galaxies, which are such that, on the (cloud) scale on which the cold gas is structured, the gas is maintained in a state where self-gravity is balanced almost entirely by motions in the galactic potential, signified by $Q\approx1$ \citep{meidt2018}.  Although the specific galaxy evolutionary factors that allow this regulatory quality to develop are yet to be shown, it clearly relies on the distribution of mass and angular momentum in galaxies, as well as the amount of dense gas and its characteristic exponential distribution \citep[e.g.,][]{elmstruck, struckelm}.  Given these factors, the gas that is able to form stars in galaxies is observed at densities for which self-gravity is mostly matched to the motions induced by the background galaxy potential.   

This idea of a `galactic equilibrium state' provides a useful context for understanding the progress of star formation over time, evolving together with galaxy morphology.  It also provides a compelling interpretation for extremes of star formation in terms of deviations from this `galactic equilibrium state.'  
For example, starbursts would be described as a consequence of gas build-up (following inflow driven by strong torquing) that leads to a substantial `excess above exponential' in the molecular gas distribution.  With enough gas confined to a small area, it can be strongly self-gravitating with $\sigma_{\rm sg} \gg \sigma_{\rm gal}$ even when the galaxy potential is locally strong.  As a result, more of it can collapse and get converted into stars much more rapidly.  Conversely, the suppression of star formation associated with a decrease in efficiency over time could follow from evolution in the underlying galaxy mass distribution while the gas disk remains exponential.  As the stars become more centrally concentrated, for example, the Coriolis force and tides strengthen and self-gravitation is strongly limited, thus leading to suppressed star formation in a scenario resembling that of `morphological quenching' \citep{martig}.

Compared to these extremes, the `equilibrium state' of an exponential gas disk embedded in a disk of stars and dark matter would appear to be the ideal configuration for sustained star formation.  This represents a compelling explanation for observations suggesting that the galaxies in which most new stars are formed, up to at least $z\sim2.5$, are rotationally-supported disks, not extreme, unrelaxed star-bursting~mergers (\citealt{forster09}; \citealt{vdW14}).   

To complete the picture of how galactic dynamics  help regulate star formation, we must also capture the interplay between gravity and star formation feedback not fully included in the present analytical model.  Numerical simulations that model the organization, structure, and kinematics of the gas in star-forming galaxies over a range of spatial scales are clearly a key component in this effort.  We note that, even without reaching below the cloud scale, simulations can easily incorporate the influence of the host galaxy on sub-grid star formation by starting with a realistic orbital energy distribution and interpolating the local large-scale gradient in the galaxy potential down to the cloud scale.  In this way it should be possible to capture the scales and densities where a given process predominantly regulates how molecular gas forms stars as a galaxy evolves.  

Observational constraints are also essential for testing the present model and improving some of the assumed inputs.  Measurements of SFEs at multiple densities matched to kinematic tracers in environments with well-measured rotation curves can show the conditions and environmental properties where the model is most applicable or where additional factors influence on the process of star formation.  Matched to this most basic set of constraints, an analysis of the organization and structure of the ISM to determine the typical scale at which the gas is arranged into clouds would help apply the model with greater accuracy.  Likewise, improved constraints on cloud density structure in a variety of environments will  show whether cloud structure varies systematically or introduces scatter in cloud-scale star formation relations. 

\section{Summary \& Conclusions}
In this paper we explore the implications of a new model that describes the impact of the dynamics of galaxies on the kinematics and star-forming ability of their molecular gas reservoirs.  We begin by describing the three dimensional gravitational motions of gas on cloud scales, as derived in the first paper of this series.  The model incorporates orbital motions in the gas---organized by the galaxy potential set up by the background distribution of stars and dark matter---as a contribution to the 3D internal motions of clouds.  These motions are large enough on the cloud scale to support the gas against its own self-gravity, thus presenting an obstacle to collapse and star formation. In this way, the model synthesizes the two main mechanisms proposed to regulate star formation in molecular gas: vertical pressure equilibrium (\citealt{ostriker} and \citealt{ostrikerShetty}) and shear and Coriolis forces in the disk plane as parameterized by Toomre $Q\approx1$ \citep[e.g.,][]{hunter, koyama, hopkins}.  The scale-dependent motions predicted by our model, which rival self-gravity on large scales within clouds, are continually established by the host galaxy.  Thus, they represent an obstacle to star formation that acts in addition to feedback-driven turbulence on large scales.  

With this gravitational support in mind, in this paper we examine the conditions (densities and spatial scales) under which gas decouples from the galactic potential so that it can reach a state in which self-gravity competes with the feedback from star formation to determine the efficiency of star formation.  We show that the gravitational motions predicted in realistic models of molecular cloud populations throughout galaxies imply that the galaxy potential becomes negligible compared to gas self-gravity only at high density (${\gtrsim}10^2~{\rm cm}^{-3}$) in the deep interiors (${<}10$~pc-scales) of clouds.  In addition to feedback, this limit to the onset of self-gravitation and free-fall collapse suggests that the dynamics of the host galaxy could offer a natural bottleneck to the star formation efficiency possible in individual molecular clouds.  

Indeed, in the scenario that directly relates the star formation rate to the rate at which collapsing gas decouples from the galactic potential, the galaxy takes on an important regulatory role: in our model, the contrast between self-gravity and the galactic potential contributes not only to overall inefficient star formation, but introduces systematic variations in the onset and the efficiency of star formation with environment.  
We present an expression for the star formation efficiency regulated by galaxy dynamics that varies both within and between galaxies according to the diversity in their gas distributions and dynamical properties.  

Our model specifically reproduces the recently observed decrease in the extragalactic dense gas star formation efficiency with increasing gas surface density toward galaxy centers \citep{usero,bigiel16,gallagher,jd19}.  The model can also describe the suppressed star formation in the CMZ of the Milky Way \citep{longmore}, as part of a continuum of increased coupling between molecular gas and the galaxy potential toward galaxy centers. However, the incipient SFR in the `100~pc stream' of the CMZ is predicted to be higher than currently observed, consistent with models predicting that star formation in the CMZ is episodic \citep[e.g.,][]{krumholz17}.  The close match between the low star formation rate predicted for the CMZ by our model and the observations lends support to the idea that the onset of star formation is not set by the turbulent properties of the gas alone, but also the stability and self-gravitation of the gas in the context of its dynamical environment (see also \citealt{kruijssen14}). 

By incorporating the influence of the background galactic potential, the galactic bottleneck model is also able to describe a decrease in the star formation efficiency of some of the densest material within  clouds located in environments with high turbulent motions at and beyond the cloud scale.  Thus the model offers a promising  avenue to explain recent observations that exhibit similar trends (\citealt{leroy2017m51}; \citealt{q19}), which are the reverse of most conventional theories of star formation.  Because the conditions that lead to star formation can be directly related to the global properties of galaxies, the picture of star formation described by our model is easy to integrate as a realistic `sub-grid' star formation prescription to implement in simulations of galaxy formation and evolution.

\acknowledgements
We would like to thank the anonymous referee for a number of suggestions that helped sharpen the  ideas presented in this paper.  S.E.M. acknowledges funding during part of this work from the Deutsche Forschungsgemeinschaft (DFG) via grant SCHI 536/7-2 as part of the priority program SPP 1573 “ISM-SPP: Physics of the Interstellar Medium”. E.R. acknowledges the support of the Natural Sciences and Engineering Research Council of Canada (NSERC), funding reference number RGPIN-2017-03987. J.M.D.K. and M.C. gratefully acknowledge funding from the German Research Foundation (DFG) in the form of an Emmy Noether Research Group (grant number KR4801/1-1). J.M.D.K. acknowledges support from the European Research Council (ERC) under the European Union’s Horizon 2020 research and innovation programme via the ERC Starting Grant MUSTANG (grant agreement No. 714907). E.S. acknowledges funding from the European Research Council (ERC) under the European Union’s Horizon 2020 research and innovation programme (grant agreement No. 694343). The work of J.P. and A.H. was partly supported by the Programme National “Physique et Chimie du Milieu Interstellaire” (PCMI) of CNRS/INSU with INC/INP co-funded by CEA and CNES, and by the Programme National Cosmology and Galaxies (PNCG) of CNRS/INSU with INP and IN2P3, co-funded by CEA and CNES.  F.B. acknowledges funding from the European Union's Horizon 2020 research and innovation programme (grant agreement No 726384-EMPIRE).  A.U. acknowledges support from the Spanish MINECO grants AYA2016-79006-P and ESP2015-68964-P.

\appendix
\section{Scaling factors for the SFE in gas with power-law density structure}\label{sec:appendixdeltad}

\begin{figure}[t]
\begin{center}
\begin{tabular}{c}
\includegraphics[width=0.45\linewidth]{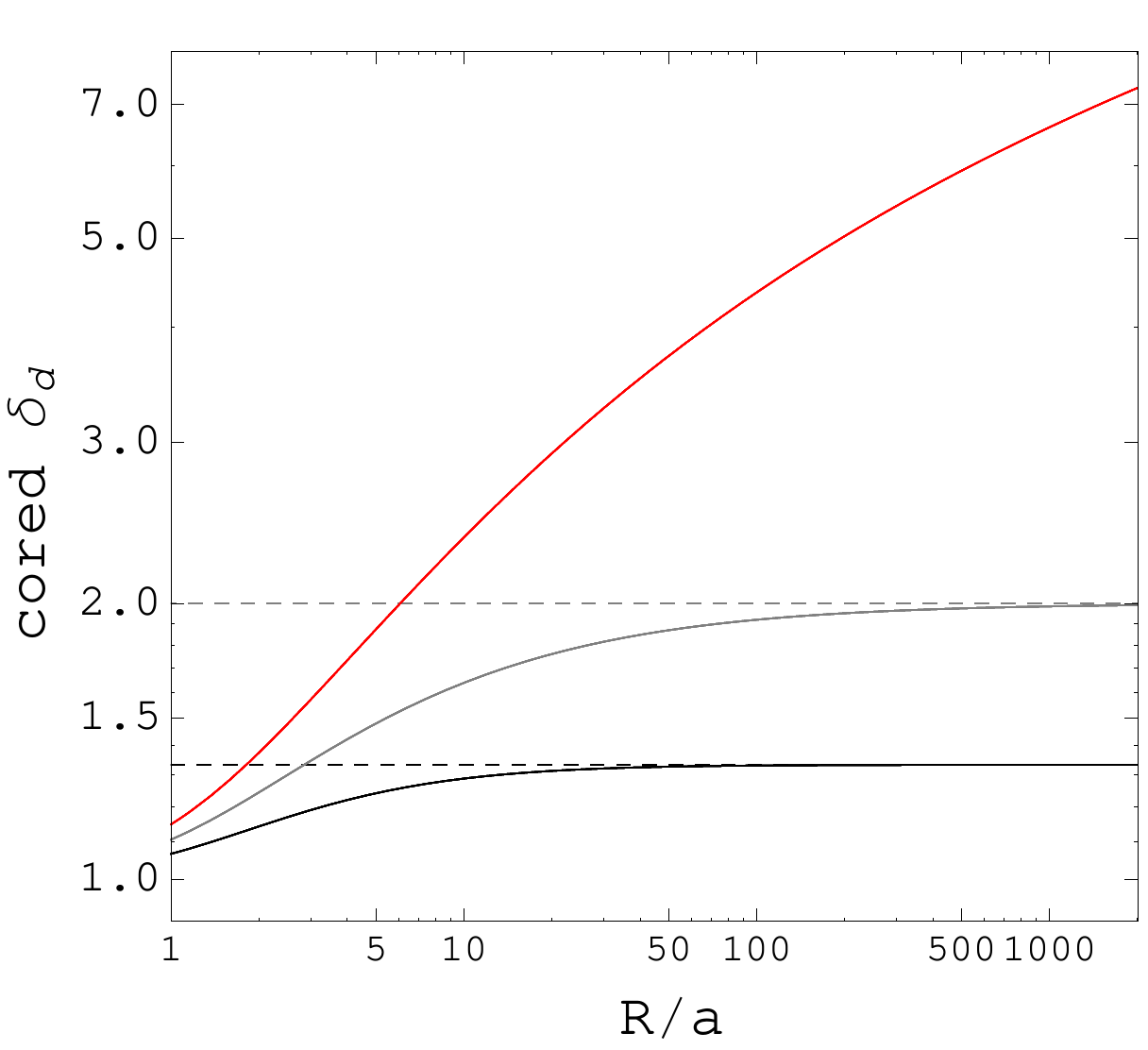}
\end{tabular}
\end{center}
\caption{Behavior of the factor $\delta_d$ for the cored power-law density profiles in eq. (\ref{eq:cored}) with index $k$=1 (black), $k$=1.5 (gray) and $k$=2 (red) as a function of the ratio of the cloud size $R_c$ to the core radius $a$.  Dashed black and gray horizontal lines show the value of $\delta_{d,PL}$ for a pure power law (which is finite only for $k$$<$2).
 }
\label{fig:deltad}
\end{figure}
For the power-law density distribution adopted for the material in clouds below scale $R_c$ in this paper (see $\S$ \ref{sec:internalstruc}), the collapse time (and the free-fall time) varies with location in a cloud.  Thus the integration that gives the total star formation rate in eq. (\ref{eq:sfmodel}) must be performed with $t_{\rm coll}$ inside the integral (see also \citealt{hennebelle}).  As shown previously (i.e. \citealt{tan06} and \citealt{parm}), in star formation models where the characteristic time is the free-fall time $t_{ff}\propto\rho^{-1/2}$, the power-law $\rho\propto r^{-k}$ yields a simple scaling factor $\delta_d$ between the total SFR and the SFR estimated at the cloud edge, i.e.  
\begin{eqnarray}
\dot{M}_{\rm stars}&=&\int_0^{R_c}\frac{\epsilon}{t_{\rm ff}}dM_c\nonumber\\
&=&\frac{\epsilon}{t_{\rm ff,c}}\frac{(2/3)(3-k)}{(2-k)}M_c \label{eq:deltadPL}
\end{eqnarray}
where $M_c$ =$(3-k)^{-1}4\pi R_c^{3}\rho_c$ is the mass in gas below scale $R_c$, $\rho_c$ is the density on scale $R_c$ and $t_{\rm ff,c}$ is the free-fall time at that density.  For this scenario we define $\delta_{d,PL}$=$(2/3)(3-k)/(2-k)$, in the case of a generic free-fall power-law (PL) model.  (The factor $\delta_d$ differs from that given by \citealt{tan06}, who assumed that $\epsilon$ also varies with density.) 

The integration with $t_{\rm ff}$ replaced by $t_{\rm coll}$ yields a slightly modified factor.  In the case that $R_c$ falls within gas with $\gamma$$\gtrsim\gamma_{\rm coll}$, we find 
\begin{equation}
\delta_{d}\approx\delta_{d,PL}\left(1-6.2\frac{(2-k)}{(6-k)}\right)
\end{equation}
using the approximation in eq. (\ref{eq:slowcoll}).

In this paper we consider both $k$=1 and $k$=2 profiles.  For the shallow $k$=1 density profile, $\delta_{d,PL}\sim$1.33, but for the $k$=2 model the SFR estimated in this way becomes infinite.  This is avoided when the central singularity at $r$=0 is replaced by an arguably more realistic constant density core.  We therefore recalculate the factor $\delta_d$ adopting the following density profile,
\begin{equation}
\rho=\frac{\rho_0}{\left(1+\left(\frac{r}{a}\right)^2\right)^{k/2}}. \label{eq:cored}
\end{equation}
This density distribution exhibits power-law behavior $\rho\propto r^{-k}$ everywhere except on the smallest scales, at the highest densities, where the profile flattens out inside a radius $a$.  

Figure \ref{fig:deltad} plots the scaling factor $\delta_d$ resulting from this density profile as a function of $R/a$.  For simplicity, in this calculation $t_{\rm ff}$ rather than $t_{\rm coll}$ has been adopted.    

For $k$=1 and $k$=1.5, $\delta_d$ roughly asymptotes to the approximation calculated from the un-cored power-law model $\delta_{d,PL}$ above.  Thus we will adopt $\delta_{d,PL}$ as a good approximation for $k$$<$2 throughout the rest of the paper.  For $k$=2, $\delta_d$ is now finite, but it spans a range of values 2.5$\lesssim$$\delta_d$$\lesssim$5 for 10$<$$R_c/a$$<$500.  Assuming core radii 0.1$<$$a$$<$1 pc\footnote{This is closer to the clump scale than the 0.01-0.1 pc sizes of the dense cores enclosed within a clump. Given that there are the many such individually star-forming cores in a clump, we choose the size of the density core $a$ to encompass the dense core-dominated region.}, we expect $\delta_d$=3.5$\pm$0.5 for typical clouds with sizes 10$<$$R_c$$<$80 pc.  We adopt this value for the $k$=2 distribution throughout the rest of the paper.

\section{The scales associated with the onset of collapse}\label{sec:appendix1}

Below we assemble predictions for the spatial scale  $R_{\rm coll}$ on which collapse is expected to occur throughout realistic cloud populations.  These provide either a direct estimate of $M_{\rm coll}/M_{\rm c}$ in a cloud with an internal density distribution $\rho\propto R^{-k}$ (see Eq.~\eqref{eq:eqRatios} or can be combined with observationally motivated models of cloud surface densities to generate realistic prescriptions of the volume density $\rho_{\rm coll}$ such as invoked in $\S$~\ref{sec:massfrac}.  

\subsection{Variations in the collapse scale throughout galaxies}
From the ratio of energies in Eq.~\eqref{eq:balance}, it follows generally that  
\begin{equation}
R_{\rm coll} = \frac{3^{1/2}\sigma_{\rm sg}}{\gamma_{\rm coll}(\kappa^2+2\Omega^2+\nu^2)^{1/2}}\label{eq:Rsg}\, ,
\end{equation}
where the small variation of $\kappa$ and $\nu$ across the cloud is ignored\footnote{With the approximation $\kappa\approx\sqrt{2}\Omega$, such as in the flat part of the rotation curve, the variation in $\kappa$ over one cloud radius can be written $\Delta\kappa\approx\sqrt{2}\Omega R_{\rm c}/R_{\rm gal}$ and so we expect fractional variation in $\kappa$ by an amount $R_{\rm c}/R_{\rm gal}$.} and $\sigma_{\rm sg}$ measures the strength of gas self-gravity at collapse.  

As in the main text, here the galactic motions within the cloud interior are assumed to be non-isotropic.  This choice is motivated by the properties of typical stellar disks, which define a potential that varies more rapidly over the extent of typical clouds in the vertical direction than in the plane.  Our preliminary inspection of cloud-scale molecular gas kinematics in galaxies with a range of inclinations (Paper~I) suggest that non-isotropy on the cloud scale is consistent with the observations.  

With this assumption, the cloud size estimated in Eq.~\eqref{eq:Rsg} is smaller than the Toomre length $\lambda_T=2\pi^2 G\Sigma/\kappa^2$, since generally $\nu \gg \kappa$ (see Paper~I). We can see this more easily by writing $R_{coll}$ in terms of the gas surface density at which collapse begins, i.e.,
\begin{equation}
R_{\rm coll}=\frac{2\pi (a_k/5) G\Sigma_{\rm coll} }{\gamma_{\rm coll}^2(\kappa^2+2\Omega^2+\nu^2)/3 } > \lambda_{\rm T} \label{eq:sgRsg}
\end{equation}

With this formulation, $R_{\rm coll}$ (and thus $\rho_{\rm coll}$) can be reconstructed from observables with knowledge of the host galaxy rotation curve shape (which yields an estimate of $\kappa$ and $\nu$ at all galactocentric radii) as long as the surface density $\Sigma_{\rm coll}$ measured on scale $R_{\rm coll}$ is also known.  

The collapse scale can be alternatively estimated given an arbitrary density on an arbitrary scale with an additional assumption for the distribution of densities below scale $R_{\rm c}$.  This is convenient for estimating $R_{\rm coll}$ in (extragalactic) clouds for which only global properties are measurable (i.e.\ total size, mass and surface density).  

For density profiles $\rho\propto r^{-2}$, for example, $\Sigma_{\rm coll} = R_{\rm c}/R_{\rm coll} \Sigma_{\rm c}$.  The collapse scale can thus be predicted given either the surface density $\Sigma_{\rm c}$ or volume density $\rho_{\rm c}$ on scale $R_{\rm c}$ according to 
\begin{eqnarray}
R_{\rm coll} &=& \left(\frac{2 \pi a_k/5 G\Sigma_{\rm c} R_{\rm c}^{k-1} }{\gamma_{\rm coll}^2(\kappa^2+2\Omega^2+\nu^2)/3}\right)^{1/k}\nonumber \\
&=& R_{\rm c}\left(\frac{4 \pi a_k/5 G\rho_{\rm c} }{\gamma_{\rm coll}^2(\kappa^2+2\Omega^2+\nu^2)/3}\right)^{1/k}\label{eq:Rcollout}
\end{eqnarray}

The left panel of Figure~\ref{fig:Rcoll} presents the collapse scale predicted by our semi-empirical galaxy models in the mass range $9.25<\log M/\mathrm{M}_\odot<10.75$, assuming two different density profiles below a fixed scale $R_{\rm c}=30$~pc and adopting our description for cloud-scale surface densities (i.e., $\Sigma_{\rm c,30pc}(R_{\rm gal}) = c \Sigma_{\rm H_2}(R_{\rm gal})$ with $c=2$; see $\S$~\ref{sec:globalgalaxymodels}).  (Note that doubling the clumping factor $c$ increases the predicted $R_{\rm coll}$ by a factor $\sqrt{2}\approx1.4$.)  This approximates the behavior expected at fixed spatial scale, such as probed by observations of the molecular gas distribution at fixed beam size when the beam size probes near the cloud scale.  

The right panel of the figure shows predictions adopting a fixed cloud mass, rather than a fixed scale, using the same cloud-scale surface density model $\Sigma_{\rm c}$ assumed in the left panel.  This is meant to illustrate how $R_{\rm coll}$ might vary throughout a cloud population, where the cloud size varies with location in the galaxy, according to $R_c=\sqrt{M_{\rm c}/\pi \Sigma_{\rm c}^{-1}(R_{\rm gal})}$.  

Both panels illustrate  the generic behavior in $R_{\rm coll}$ throughout typical galaxy disks, in which the background galaxy potential typically weakens with increasing $R_{\rm gal}$ faster than the gas self-gravity (as in our chosen empirical model; see $\S$~\ref{sec:globalgalaxymodels}).  This leads to a characteristic increase in $R_{\rm coll}$ from small to large galactocentric radius.  (Note that, in the right panel, the cloud scale itself drops off with $R_{\rm gal}$.)

The axisymmetric disk models assumed here undoubtedly oversimplify the true mass distributions of real galaxies, so we caution that the generic trends shown in Figure~\ref{fig:rsgplot0} may differ in detail from what would be predicted for a given observed rotation curve.  In $\S$~\ref{sec:CMZ} for instance, we use the variation of $\kappa$ in the center of the MW to more precisely predict the collapsing scale in the CMZ.  

\subsection{Relation to the cloud scale }\label{sec:rsgpop1}
\begin{figure*}[t]
\begin{center}
\begin{tabular}{cc}
\includegraphics[width=0.45\linewidth]{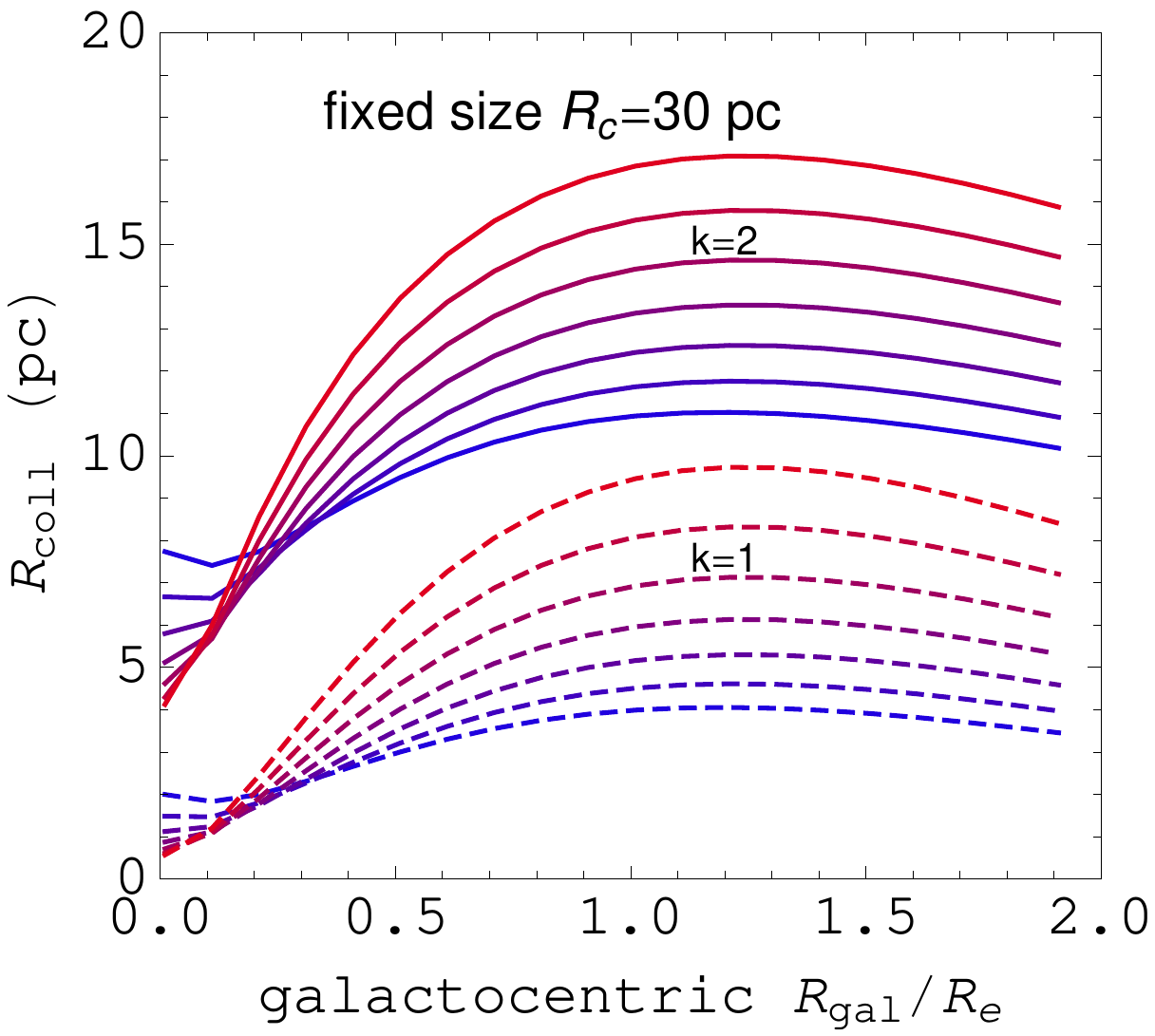}&\includegraphics[width=0.45\linewidth]{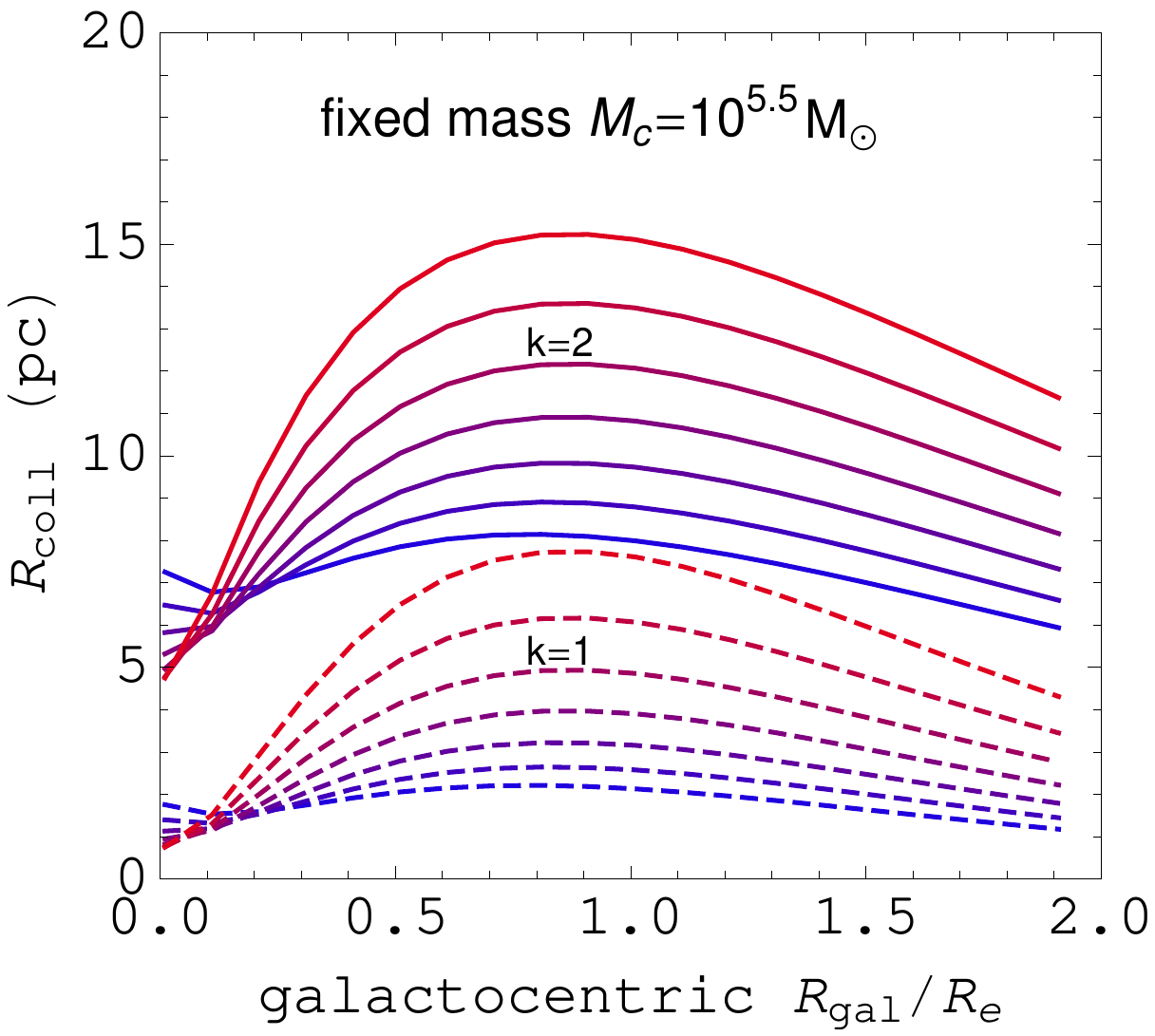}\\
\end{tabular}
\end{center}
\caption{Radial variation in the collapse scale $R_{coll}$ within clouds hosted by galaxies with different stellar masses in the range $9.25<\log M/\mathrm{M}_\odot<10.75$ (increasing from blue to red in steps of $0.5\log M/\mathrm{M}_\odot$; following from stellar and gas distributions suggested by global galaxy scaling relations as described in $\S$~\ref{sec:motions}).  Curves are chosen to span the brightest portion of typical molecular disks $0.01<R_{\rm gal}/R_{\rm e}<2.5$ (\citealt{schruba}; where galactocentric radius $R_{\rm gal}$ is given in terms of the disk scale length $R_{\rm e}$).  Each assumes a mass-dependent empirical cloud-scale surface density $\Sigma_{\rm c} = c \Sigma_{\rm H_2}$ where $c=2$ (see text).   Two sets of lines are shown, highlighting the choice of density distribution below a chosen scale $R_{\rm c}$, $k=2$ (solid) or $k=1$ (dashed).  Predictions in the left panel adopt a fixed $R_{\rm c}=30$~pc.  Predictions in the right panel are shown at fixed cloud mass, $M_{c} = 10^{5.5}~\mathrm{M}_\odot$.}
\label{fig:Rcoll}
\end{figure*}

From the collapse scale predicted in Figure \ref{fig:Rcoll} we infer that, in typical gas disks, only a fraction ($10{-}30$\%) of the mass in clouds is expected to collapse.  This is quantified more directly by the mass fractions $M_{\rm coll}/M_{\rm c}$ in Figure~\ref{fig:rsgplot0} in the main text, according to the relation between $M_{\rm coll}/M_{\rm c}$ and $R_{\rm coll}/R_{\rm c}$ in Eq.~\eqref{eq:eqRatiosGen} for the internal density distribution $\rho\propto r^{-k}$ (see $\S$~\ref{sec:massfrac}).  

It is worth noting that the formalism introduced here also presents a description for the sizes of clouds themselves, based on the view that clouds are by definition self-gravitating (and collapsing) objects, i.e., precisely objects with sizes $R_{\rm coll}$.  In such a cloud-based description of the organization of the molecular gas, the entirety of clouds would undergo collapse.  Our expressions for the star formation efficiency in terms of the collapsing fraction as presented in  $\S$~\ref{sec:coupledSFE} would then need to be re-expressed as a dependence of the SFE in a given aperture on the fraction of the gas in an aperture in the form of clouds.  

We find the more generic formulation of collapse fraction in terms of the density within a given volume of gas preferable, however, given that it is less sensitive to the definition of cloud size, which can be adifficult property to assess observationally, depending on the precise definition assumed and the strategy used to identify clouds (see discussion by \citealt{hughesI}).\footnote{In crowded areas, cloud edges assigned in relation to the surrounding material (according to some prescribed intensity contrast), i.e. using dendrograms, would tend to yield smaller sizes than those measured by extrapolating cloud properties to infinite sensitivity, as performed by the commonly employed mode of techniques like CPROPS \citep{CPROPS}.  Yet other techniques may define the cloud edge as specifically the location where the cloud becomes bound, which may or may not agree with other definitions of the cloud edge for the same objects.  Homogeneous treatment of comprehensive molecular gas surveys that probe to the relevant cloud scale throughout galaxies with a variety of morphological and kinematic properties (i.e., A.~K.~Leroy et al., in prep.) will shed new light on the distributions of cloud properties and how they emerge from the local (environmental) conditions.  The possibility of radial variation in cloud sizes, for one, is ideally tested alongside potential changes in cloud mass spectra as probes of the mechanisms that can lead to cloud growth and destruction (E.~Rosolowsky et al., in prep).} 

\section{Scale-dependence of the inverse relation between SFE and gas velocity dispersion}\label{sec:appSFEveldisp}
In this section, we cast the expressions in section $\S$~\ref{sec:sfedenseobs} into a more explicit relation between $\mathit{SFE}_{\rm dense}$ and gas velocity dispersion measured on relatively large  scales across some aperture (or beam), sampling at or above the cloud (${\sim}60$~pc) scale.  

\subsubsection{Large aperture (\texorpdfstring{${>}60$~pc}{greater than 60 pc}) trends}
With our assumed model of the way gas is typically structured and organized, we can use velocity dispersions measured on relatively large-scales to approximate the magnitude of galactic motions on the cloud scale and below.  Likewise, aperture measures of the gas surface density on cloud scales and beyond roughly constrain the surface density of the dense, potentially star-forming gas situated on scales near $R_{\rm coll}$. 
 
We consider a scenario in which the gas in an aperture with radius $R_{\rm ap}$ is structured into non-cloud and cloud components, designated by mass in terms of the cloud mass fraction $f_c$, and assume that the cloud component $f_{\rm c} \Sigma_{\rm ap}\pi R_{\rm ap}^2$ is arranged into a number of clouds $N_{\rm cl} = R_{\rm ap}^2/R_{\rm c}^2$ with similar sizes $R_{\rm c}$ and surface densities.  
In this case $\Sigma_{\rm ap} R_{\rm ap} \approx f_{\rm c}^{-1}\Sigma_{\rm c} R_{\rm c} N_{\rm cl}^{1/2}$ where $\Sigma_{\rm c}$ is the average surface density of clouds in the aperture. (Note that, with this formulation, the clumping factor we use in the main text to model the cloud surface density from an assumed disk surface density profile acts as the factor $f_{\rm c} N_{\rm cl}^{-1/2}$.)
 
With this assumed gas structure, we can also use the observed motions on the aperture scale $\sigma_{\rm ap}$ to approximate the motions on the scale of the gas that is beginning to collapse.  Since we are interested in examining the influence of the galactic bottleneck, we consider the case in which the kinematics of molecular gas on or near cloud scales are dominated by the combination of galactic motions and motions associated with self-gravity, either due to collapse or originating with feedback that keeps the cloud in equilibrium.  (Note that for this exercise we neglect the possibility of super-virial feedback-driven motions or the influence of magnetic forces.)  
In this case we write
\begin{equation}
\sigma_{\rm ap}^2\approx\sigma_{\rm gal}^2(1+\gamma^2),
\end{equation}
using $\gamma$ defined in $\S$~\ref{sec:forcebalance}.  For the gas in the molecular disks of typical nearby main sequence galaxies, $\gamma \lesssim 1$ on cloud scales and larger (see Paper~I) so that $\sigma_{\rm ap}^2\approx\sigma_{\rm gal}^2$.  Using the approximate relation between $\nu$ and $\kappa$ given in Paper~I and the expression for the line-of-sight projection of 3D galactic motions there we find
\begin{equation}
\sigma_{\rm gal,los}^2\approx\kappa^2 R_{\rm ap}^2\left(1+\frac{R_{\rm gal}}{2z_0} \cos^2{i}\right),
\end{equation}
in the main disk environment where rotation curves are approximately flat (so that $\kappa^2=2\Omega^2$).  Here $i$ is the inclination of the galaxy with respect to the line of sight. 

From here we estimate the velocity dispersion recovered in the aperture as 
\begin{equation}
\sigma_{\rm ap}\approx\kappa \left[1+\left(\frac{R_{\rm gal}}{2z_0}\right)\cos^2{i}\right]^{1/2} R_{\rm c} (R_{\rm ap}/R_{\rm c}). 
\end{equation}
In terms of the galactic motions on scale $R_{\rm c}$
\begin{equation}
\sigma_{\rm gal}\approx\kappa R_{\rm c}\left(2+\frac{R_{\rm gal}}{2z_0}\right)^{1/2}, 
\end{equation}
we write
\begin{equation}
\sigma_{\rm ap}\approx\sigma_{\rm gal} \frac{\left(1+\frac{R_{\rm gal}}{2z_0}\cos^2{i}\right)^{1/2}}{\left(2+\frac{R_{\rm gal}}{2z_0}\right)^{1/2}} N_{\rm cl}^{1/2}, 
\end{equation}
This reduces to $\sigma_{\rm ap}\approx(\sigma_{\rm gal}\cos{i})N_{\rm cl}^{1/2}$ at large galactocentric radius and approaches $\sigma_{\rm ap}\approx(\sigma_{\rm gal}/\sqrt{2})N_{\rm cl}^{1/2}$ toward galaxy centers.

Altogether, we can write the dense gas SFE on scales larger than the typical cloud size as \begin{eqnarray}
\mathit{SFE}_{\rm dense} &=& f_{\rm d}^{-1} \frac{\delta_d\mathit{SFE}_{\rm coll}}{2.4\gamma_{\rm coll}^{2(3-k)/k}}\left(\frac{\sqrt{2\pi a_k/5 G\Sigma_{\rm ap}R_{\rm ap}}}{\sigma_{\rm ap}}\right)^{2(3-k)/k}\nonumber\\
&\times& \left(f_{\rm c}^{1/2}N_{\rm cl}^{1/2}\frac{\left[1+\left(\frac{R_{\rm gal}}{2z_0}\right)\cos^2{i}\right]^{1/2}}{\left[2+\frac{R_{\rm gal}}{2z_0}\right]^{1/2}}\right)^{2(3-k)/k}\label{eq:SFEvdisp}
\end{eqnarray}

\subsubsection{Small aperture (\texorpdfstring{${\lesssim}60$~pc}{less than 60 pc}) trends}

When the aperture samples at or near the typical cloud size, the denominator of the term in the first set of parentheses in Eq.~\eqref{eq:SFEvdisp} should account for the increased contribution from gas self-gravity.  For example, in the case of $k=2$, on cloud scales we write
\begin{eqnarray}
\mathit{SFE}_{\rm dense} &=& \frac{\delta_d\mathit{SFE}_{\rm coll}}{2.4\gamma_{\rm coll}}\frac{\sqrt{2\pi a_k/5 G\Sigma_{\rm ap}R_{\rm ap}}}{\left(\sigma_{\rm ap}^2-\sigma_{\rm sg}^2\right)^{1/2}}\nonumber\\
&\times& f_{\rm c}^{1/2}f_{\rm d}^{-1} \frac{\left[1+\left(\frac{R_{\rm gal}}{2z_0}\right)\cos^2{i}\right]^{1/2}}{{\left[2+\frac{R_{\rm gal}}{2z_0}\right]^{1/2}}},\label{eq:SFEvdispCloud1}
\end{eqnarray}
using also that $N_{\rm cl}\approx1$ in this case. 

Note that the factor $f_{\rm d}$ is necessary to use the gas surface density and the line-of-sight velocity dispersion in an aperture to approximate the density and motions on smaller scales.  However, when the dense gas can be observed directly and the measurement aperture approaches the size of the region within clouds typically occupied by the dense gas, $\sigma_{\rm ap}$ directly probes the motions of the high density material and the expression for $\mathit{SFE}_{\rm dense}$ simplifies further.  Thus $f_{\rm d}=1$ (and $N_{\rm cl}=1$) so that, when $k=2$, 
\begin{equation}
\mathit{SFE}_{\rm dense} = \frac{\delta_d\mathit{SFE}_{\rm coll}}{2.4\gamma_{\rm coll}}\frac{\sqrt{2\pi a_k/5 G\Sigma_{\rm ap}R_{\rm ap}}}{\left(\sigma_{\rm ap}^2-\sigma_{\rm sg}^2\right)^{1/2}}
\label{eq:SFEvdispDense}
\end{equation}
again removing an estimation of the contribution from motions due to self-gravity on the measurement scale (estimated from the observed gas surface density) from the line-of-sight velocity dispersion.

\subsubsection{Variations in SFE with virial parameter \texorpdfstring{$\alpha_{\rm vir}$}{alpha(vir)}}
The above relations for $\mathit{SFE}_{\rm dense}$ expressed in terms of gas velocity dispersion make it clear that the SFE at any density predicted by our model depends on the boundedness of the gas as measured by the virial parameter $\alpha_{\rm vir}=5\sigma^2R/(GM)$ \citep{bertoldimckee}.  The dependence is in the same sense as suggested by the first cloud-scale study undertaken in M51, where higher $\alpha_{\rm vir}$ (lower gas boundedness $b=\Sigma/\sigma^2\propto\alpha_{\rm vir}^{-1}$) is linked to lower rates of star formation per unit mass \citep{leroy2017m51}. 

Starting with Eq.~\eqref{eq:SFEvdisp}, we write the molecular gas SFE measured in an aperture with radius $R_{\rm ap}$ sized at or near the cloud scale as
\begin{eqnarray}
\mathit{SFE} &=& \frac{\delta_d\mathit{SFE}_{\rm coll}}{2.4\gamma_{\rm coll}^{2(3-k)/k}}\frac{1}{\left(\alpha_{\rm vir}a_k^{-1}-1\right)^{(3-k)/k}}\nonumber\\
&\times& f_{\rm c}^{(3-k)/k}\left[1+\left(\frac{R_{\rm gal}}{2z_0}\right)\cos^2{i}\right]^{(3-k)/k}.\label{eq:SFEvdispCloud}
\end{eqnarray}
Here $\alpha_{\rm vir}=5\sigma_{\rm ap}^2/(\pi G\Sigma_{\rm ap} R_{\rm ap}$) in the notation used in the previous section and the last term on the right corrects cloud-scale non-isotropic motions down to the isotropic motions predicted on smaller scales where star formation occurs.  

Equation (\ref{eq:SFEvdispCloud}) suggests that the SFE will decrease as roughly $\alpha_{\rm vir}^{-1/2}$ when $k=2$ until the minimum value of $\alpha_{\rm vir}$ is approached, which we expect to be near $\alpha_{\rm vir,min}=1$ when $\sigma_{\rm ap}$ predominantly reflects motions due to self-gravity and the background galactic potential, as in the current model.  

In terms of boundedness $b=\Sigma_{\rm ap}/\sigma_{\rm ap}^2$ (in the notation used above), 
\begin{eqnarray}
\mathit{SFE} &=& \frac{\delta_d\mathit{SFE}_{\rm coll}}{2.4\gamma_{\rm coll}^{2(3-k)/k}}\left[\frac{b(2\pi a_k/5 G R_{\rm ap})}{1-b(2\pi a_k/5 G R_{\rm ap})}\right]^{(3-k)/k}\nonumber\\
&\times& f_{\rm c}^{(3-k)/k}\left[1+\left(\frac{R_{\rm gal}}{2z_0}\right)\cos^2{i}\right]^{(3-k)/k}.\end{eqnarray}
and SFE increases as roughly $b^{1/2}$ until a maximum $b_{\rm max}=(2\pi a_k/5 G R_{\rm ap})^{-1}$ is reached. 

\section{Glossary}
\begin{table*}
\begin{center}
\caption{Definitions of symbols used in the context of the galactic bottleneck model}\label{tab:paramTable}
\begin{threeparttable}
\begin{tabular}{p{0.1\linewidth} p{0.8\linewidth}}
Symbol&Description\\
\tableline
$\kappa$ & frequency of in-plane epicyclic motions in the galactic potential\\
$\nu$& frequency of vertical epicyclic motions in the galactic potential\\
$R_{\rm gal}$& galactocentric radius\\
$V_{\rm c}$&galaxy circular velocity \\
$\Omega$&galaxy angular velocity $V_{\rm c}/R_{\rm gal}$\\
$\beta$ & logarithmic derivative of the circular velocity $\partial (\ln V_{\rm rot})/\partial(\ln R_{\rm gal})$\\
$R_{\rm c}$ & arbitrary scale within the interior of a cloud \\
$\Sigma_{\rm c}$ & gas surface density at scale $R_{\rm c}$ within a spherically symmetric cloud \\
$\rho_{\rm c}$ & gas volume density at scale $R_{\rm c}$ within the cloud \\
$M_{\rm c}$ & the mass in gas above $\rho_{\rm c}$ within the cloud \\
$\rho_{\rm d}$ & the volume density of gas above a characteristic `dense' threshold \\
$\Sigma_{\rm d}$ & surface density of dense gas above a volume density $\rho_{\rm d}$\\
$M_{\rm d}$ & the mass in gas above $\rho_{\rm d}$ within a cloud \\
$f_{\rm d}$ & dense gas (mass) fraction $M_{\rm d}/M_{\rm c}$ \\
$k$ & index of the spherically symmetric power-law density profile $\rho\propto R^{-k}$\\
$\sigma_{\rm sg}$ & velocity dispersion associated with gas self-gravity, originating either with collapse or as a result of feedback-driven turbulence in equilibrium with self-gravity \\
$\gamma$ & the (density and/or scale-dependent) ratio of gravitational energies in molecular clouds\\
$\gamma_{\rm coll}$ &the value of $\gamma$ specifically at the onset of collapse\\
$\rho_{\rm coll}$ & gas volume density at which gravitational energies are out of balance by the factor $\gamma_{\rm coll}$ \\
$\kappa_{\rm sun}$ & frequency of in-plane epicyclic motions in the Solar Neighborhood \\
$\alpha_{\rm vir}$& the virial parameter of the gas, i.e. the ratio of cloud kinetic energy to the cloud potential \\
$f_{\alpha}$ & the ratio of the energy in feedback driven turbulent motions to the cloud potential\\
$\Sigma_{\rm H_2}$ & large-scale molecular surface density \\
$c$ & clumping factor $\Sigma_{\rm c}/\Sigma_{\rm H_2}$ between $\Sigma_{\rm H_2}$ and cloud-scale surface density $\Sigma_{\rm c}$\\
$f_{\rm c}$ & the cloud mass fraction \\
$R_{\rm ap}$ & the arbitrary size of an observational aperture \\
$N_{\rm cl}$ & number of clouds in an aperture \\
$\sigma_{\rm ap}$ & velocity dispersion measured in an aperture sampling above or near cloud scales\\
$\dot{M}_\star$ & star formation rate \\
$\epsilon$ & dimensional conversion efficiency from gas to stars during the star formation process\\
$\epsilon_{\rm ff}$ & the star formation efficiency per free fall time at a given density\\
$\mathit{SFE}$ & (large-scale) molecular gas star formation efficiency $\dot{M}_\star/M_{\rm H_2}$ (with inverse time units) \\
$\mathit{SFE}_{\rm c}$ & cloud-scale molecular gas star formation efficiency $\dot{M}_\star/M_{\rm c}$  \\
$\mathit{SFE}_{\rm dense}$ & star formation efficiency in `dense gas' above $\rho_{\rm d}$\\
$\delta_d$ & density profile-dependent factor to scale up the $SFE$ measured from properties on scale $R_c$ to the integrated $SFE$ out to scale $R_c$\\
$t_{\rm epic}$ & epicyclic period $2\pi/\kappa$ \\
$t_{\rm ff}$& the local free-fall time\\
$\tau_{\rm orb}$ & orbital period $2\pi/\Omega$ \\
$\tau_{\rm dep}$ &molecular gas depletion time $M_{\rm H_2}$/$\dot{M}_\star = \mathit{SFE}^{-1}$\\
$M_\star$ & galaxy stellar mass \\
$i$ & galaxy inclination\\
\tableline
\end{tabular}
\end{threeparttable}

\end{center}
\end{table*}


\end{document}